\documentclass[preprint]{aastex63}
\usepackage{amssymb,amsmath,amsopn}
\usepackage{bm}
\usepackage{mathbbol}
\usepackage{mathtools} 
\usepackage{enumitem}
\usepackage{hyperref}
\usepackage[utf8]{inputenc}
\usepackage{newunicodechar,graphicx} 
\usepackage{color}
\usepackage{soul}

\DeclareRobustCommand{\okina}{%
  \raisebox{\dimexpr\fontcharht\font`A-\height}{%
    \scalebox{0.8}{`}%
  }%
}
\newunicodechar{ʻ}{\okina}

\newcommand{\vect}[1]{\boldsymbol{\mathbf{#1}}}
\newcommand{\Matrix}[1]{\mathbb{#1}}
\newcommand{\fnc}[1]{\textsf{#1}} 

\newcommand{\lare}{\fnc{LaRe3D}}
\newcommand{\charc}{\fnc{CHAR}}
\newcommand{\ddbc}{\fnc{DDBC}}
\newcommand{\nrbc}{\fnc{NRBC}}
\newcommand{\wmsd}{\fnc{wMSD}}
\newcommand{\xb}{\ensuremath{\vect{x}_b}}
\newcommand{\incoming}{\ensuremath{\mathcal{I}}} 
\newcommand{\outgoing}{\ensuremath{\mathcal{O}}} 
\newcommand{\residual}{\ensuremath{\mathcal{R}}} 

\newcommand{\widebar}[1]{\ensuremath{\overline{#1}}}

\newcommand{\charderiv}{\ensuremath{\vect{L}}}

\newcommand{\charderivz}{\ensuremath{\charderiv_z}}
\newcommand{\charderivi}{\ensuremath{\charderiv_\incoming}}
\newcommand{\charderivo}{\ensuremath{\charderiv_\outgoing}}
\newcommand{\charderivzi}{\ensuremath{\charderiv_{z\incoming}}}
\newcommand{\inhomoterms}{\ensuremath{\vect{D}}}
\newcommand{\mhdstate}{\ensuremath{\vect{U} } }
\newcommand{\gtstate}{\ensuremath{\vect{G} } }
\newcommand{\svdrhs}{\ensuremath{\vect{f}}}
\newcommand{\minnormsoln}{\ensuremath{\svdrhs^*}}

\newcommand{\tstartname}{O1}
\newcommand{\tendname}{O2}
\newcommand{\tstart}{\ensuremath{t_{\tstartname}}} 
\newcommand{\tend}{\ensuremath{t_{\tendname}}}     
\newcommand{\dtdata}{\ensuremath{\Delta t}}       
\newcommand{\dtsim}{\ensuremath{\delta t}}        
\newcommand{\dtsample}{\ensuremath{\Delta t_\text{crit}}}        
\newcommand{\totsteps}{\ensuremath{\Gamma}}

\newcommand{\ts}{\ensuremath{\eta}}           
\newcommand{\tspj}{\ensuremath{t_{\tstartname+\ts}}}        
\newcommand{\tspjpone}{\ensuremath{t_{\tstartname{}+\ts+1}}}        
\newcommand{\Uj}{\ensuremath{\vect{U}_{\tstartname{}+\ts}}}

\newcommand{\dUdtest}{\ensuremath{\dot{\mhdstate}}} 
\newcommand{\tcycleend}{\ensuremath{t_{\tstartname{}+\Gamma}}}        

\newcommand{\observedstate}{\ensuremath{\widehat{\mhdstate}}}
\newcommand{\targetstate}{\ensuremath{\widebar{\mhdstate}}}
\newcommand{\targettimederiv}{\ensuremath{\widebar{\dot{\mhdstate}}}}

\newcommand{\UBtstartobs}{\ensuremath{\observedstate(\tstart)}}  
\newcommand{\UBtendobs}{\ensuremath{\observedstate(\tend)}}  
\newcommand{\UBtime}[1]{\ensuremath{\mhdstate(#1,\xb)}}  

\newcommand{\evalmat}{\ensuremath{\Matrix{\Lambda}}}
\newcommand{\xevalmat}{\ensuremath{\evalmat_x}}
\newcommand{\yevalmat}{\ensuremath{\evalmat_y}}
\newcommand{\zevalmat}{\ensuremath{\evalmat_z}}
\newcommand{\evalmati}{\ensuremath{\evalmat_\incoming}}      
   
\newcommand{\smat}{\ensuremath{\Matrix{S}}}
\newcommand{\smati}{\ensuremath{\smat_\incoming}}
\newcommand{\smato}{\ensuremath{\smat_\outgoing}}
\newcommand{\smatinv}{\ensuremath{\smat^{-1}}}
\newcommand{\smatinvi}{\ensuremath{\smatinv_\incoming}}
\newcommand{\smatinvo}{\ensuremath{\smatinv_\outgoing}}
\newcommand{\smatzi}{\ensuremath{\smat_{z\incoming}}}

\newcommand{\amat}{\ensuremath{\Matrix{A}}}            
\newcommand{\amatp}{\ensuremath{\amat_+}}              
\newcommand{\amatm}{\ensuremath{\amat_-}}              
\newcommand{\amati}{\ensuremath{\amat_\incoming}}      

\newcommand{\pmat}{\ensuremath{\Matrix{P}}}     
\newcommand{\pmatp}{\ensuremath{\pmat_+}}       
\newcommand{\pmatm}{\ensuremath{\pmat_-}}       
\newcommand{\pmatpm}{\ensuremath{\pmat_\pm}}    
\newcommand{\pmati}{\ensuremath{\pmat_\incoming}}      
\newcommand{\pmato}{\ensuremath{\pmat_\outgoing}}      
\newcommand{\pmatio}{\ensuremath{\pmat_{\incoming,\outgoing}}}   

\newcommand{\rightxevecmat}{\ensuremath{\smat_x}}
\newcommand{\leftxevecmat}{\ensuremath{\rightxevecmat^{-1}}}
\newcommand{\rightyevecmat}{\ensuremath{\smat_y}}
\newcommand{\leftyevecmat}{\ensuremath{\rightyevecmat^{-1}}}
\newcommand{\rightzevecmat}{\ensuremath{\smat_z}}
\newcommand{\leftzevecmat}{\ensuremath{\rightzevecmat^{-1}}}

\newcommand{\normalize}[1]{\ensuremath{#1_N}}
\newcommand{\lnorm}{\normalize{L}}
\newcommand{\rnorm}{\normalize{\rho}}
\newcommand{\pnorm}{\normalize{P}}
\newcommand{\vnorm}{\normalize{V}}
\newcommand{\tnorm}{\normalize{T}}
\newcommand{\bnorm}{\normalize{B}}

\newcommand{\rhat}{\ensuremath{\widehat{\vect{r}}}}
\newcommand{\phihat}{\ensuremath{\widehat{\vect{\phi}}}}
\newcommand{\thetahat}{\ensuremath{\widehat{\vect{\theta}}}}
\newcommand{\xhat}{\ensuremath{\widehat{\vect{x}}}}
\newcommand{\yhat}{\ensuremath{\widehat{\vect{y}}}}
\newcommand{\zhat}{\ensuremath{\widehat{\vect{z}}}}
\newcommand{\nhat}{\ensuremath{\widehat{\vect{n}}}}
\newcommand{\del}{\nabla}

\newcommand{\curl}{\nabla\times}
\renewcommand{\div}{\nabla\cdot}

\newcommand{\bigo}{\ensuremath{\mathcal{O}}}
\providecommand{\abs}[1]{\lvert#1\rvert}

\DeclareMathOperator{\sign}{sign}

\newcommand{\unit}[1]{\ensuremath{\, \mathrm{#1}}}
\newcommand{\figref}[1]{Figure~\ref{#1}}

\newcommand*{\horzbar}{\rule[.5ex]{2.5ex}{0.5pt}}

\makeatletter
\def\env@matrix{\hskip -\arraycolsep 
  \let\@ifnextchar\new@ifnextchar
  \array{*{\c@MaxMatrixCols}c}}
\makeatother

\defcitealias{Tarr:2023}{Paper I}
\defcitealias{Kee:2023}{Paper II}

\begin{document}
\title{
Simulating the photospheric to coronal plasma using magnetohydrodyanamic characteristics I: data-driven boundary conditions}
\correspondingauthor{Lucas A. Tarr}
\email{ltarr@nso.edu}
\author[0000-0002-8259-8303]{Lucas A. Tarr}
\affiliation{National Solar Observatory, 22 Ohi\okina{}a Ku St, Makawao, HI, 96768}

\author[0000-0001-7322-5401]{N. Dylan Kee}
\affil{National Solar Observatory, 22 Ohi\okina{}a Ku St, Makawao, HI, 96768}

\author[0000-0002-4459-7510]{Mark G. Linton}
\affil{U.S. Naval Research Laboratory, 4555 Overlook Ave SW, Washington, DC 20375}
\author[0000-0003-1522-4632]{Peter W. Schuck}
\affil{NASA Goddard Space Flight Center, 8800 Greenbelt Rd, Greenbelt, MD 20771}
\author[0000-0002-6936-9995]{James E. Leake}
\affil{NASA Goddard Space Flight Center, 8800 Greenbelt Rd, Greenbelt, MD 20771}

\begin{abstract}
We develop a general description of how information propagates through a magnetohydrodynamic (MHD) system based on the method of characteristics and use that to formulate numerical boundary conditions (BCs) that are intrinsically consistent with the MHD equations.
Our formulation includes two major advances for simulations of the Sun.
First, we derive data-driven BCs that optimally match the state of the plasma inferred from a time series of observations of a boundary (e.g., the solar photosphere).
Second, our method directly handles random noise and systematic bias in the observations, and finds a solution for the boundary evolution that is strictly consistent with MHD and maximally consistent with the observations.
We validate the method against a Ground Truth (GT) simulation of an expanding spheromak.  
The data-driven simulation can reproduce the GT simulation above the photosphere with high fidelity when driven at high cadence.  
Errors progressively increase for lower driving cadence until a threshold cadence is reached and the driven simulation can no longer accurately reproduce the GT simulation.
However, our characteristic formulation of the BCs still requires adherence of the boundary evolution to the MHD equations even when the driven solution departs from the true solution in the driving layer.
That increasing departure clearly indicates when additional information at the boundary is needed to fully specify the correct evolution of the system.
The method functions even when no information about the evolution of some variables on the lower boundary is available, albeit with a further decrease in fidelity.

\end{abstract}

\section{Introduction}\label{sec:introduction}
How is the three-dimensional (3D) state of the MHD plasma in the solar atmosphere advanced in time given (i) the initial state of the atmosphere and (ii) observations of the lower (photospheric) boundary at the later time?
This is a problem at the heart of solar physics research.
First, all energy in the chromosphere/corona originates in the convection zone and must transit the photosphere. In plain language, all of the energy in photosphere-to-corona models ``originates'' at the boundary!
Thus, correctly solving for the MHD evolution of the boundary is of paramount importance for correctly solving for the evolution within the the volume of interest.
Second, photospheric emissions originate in a relatively narrow layer between optically thick and optically thin radiative enviornments.
Thus, the photospheric observables derived from these emissions can be assumed to originate from a narrow height range in the solar atmosphere\----key for boundary driving.
Third, the photosphere represents the narrow boundary layer between the pressure dominated convection zone physics and magnetically dominated coronal physics.
Thus, photospheric ``boundary'' driving of photosphere-to-corona models of the solar atmosphere is an attractive approach for understanding the evolution of the corona because of practicality\----the photospheric observations are the most localized (and perhaps the most reliable), and simplicity\----the physics of the pressure-dominated convection zone is lumped into the boundary conditions for theoretical studies.

For these reasons, the rigorous time-advance of MHD models given only the time-evolution of the boundary is of paramount importance for modeling coronal activity such as the evolution of active regions, of coronal heating, of flare energy release, of coronal mass ejections, and for space weather prediction.
Ideally, the solution to this problem---i.e., the values of the mass density $\rho$, internal energy density $\epsilon$, magnetic field vector $\vect{B}$, and velocity vector $\vect{v}$ at every location in the photosphere-to-corona volume at the later time---should be exactly the same as the case in which a more extended problem is solved wherein we
model the entire universe; that is, from the solar core, through the convection zone, corona, heliosphere, and into interstellar space.
The boundary driving must properly encode the influence of the rest of the universe on our patch of solar atmosphere, and the influence of our patch of solar atmosphere on the rest of the universe.
The daunting complex nature of this task is concealed by its typical plain-language summary: we need to set the boundary conditions ``correctly.''
This paper rigorously addresses ``correct'' in the MHD context.

The good news is that the photosphere is nearly continuously observed at decent temporal cadence and spatial resolution by a variety of spectroscopic and polarimetric instruments.
Combined with a mix of models and assumptions, these observations can be used to infer a good fraction of the MHD state vector, $\mhdstate:=(\rho,\epsilon,\vect{v},\vect{B})^T$, in the photosphere on a routine basis, especially within active regions, which are the focus of this study.
The Helioseismic and Magnetic Imager \citep{Scherrer:2012,Schou:2012} instrument onboard the Solar Dynamics Observatory \citep{Pesnell:2012} regularly infers vector magnetic field estimates of $\vect{B}$ for the earth-facing side of the Sun at a 12 minute cadence and one arcsecond resolution ($\approx 730\unit{km}$ on the Sun).
Optical flow methods such as DAVE4VM \citep{Schuck:2008} or PDFI\_SS \citep{Fisher:2020} can provide estimates of plasma velocity $\vect{v}$.
Thermodynamic variables such as density $\rho$ and internal energy density $\epsilon$ (or, equivalently, the temperature $T$ or pressure $P$) can be derived by combining model atmospheres with observations of the continuum intensity, multiple spectral lines \citep{Maltby:1986}, or using empirically derived relations \citep{Maltby:1986,Jaeggli:2012, Borrero:2019}.
For smaller spatial patches of the Sun, higher resolution and higher sensitivity spectropolarimetric, multi-line observations are providing ever better constrained estimates of the photospheric state vector \citep{Yadav:2019, QuinteroNoda:2021,Vissers:2021,daSilvaSantos:2022}, especially in combination with spatially-coupled inversion approaches based on machine learning techniques \citep{AsensioRamos:2017, AsensioRamos:2019} and the newest generation of solar telescopes, such as the Daniel K. Inouye Solar Telescope \citep{Rimmele:2020}.

This wealth of observational data has made photosphere-to-corona simulations an appealing approach for studying the solar atmosphere, but such simulations come with unique challenges in terms of their boundary conditions.
Indeed, even in the best case scenario where the entire MHD state vector may be \emph{estimated} on the boundary from observations, this estimated state vector will contain bias and noise and, in general, will not be completely consistent with the internal state of the photosphere-to-corona model of the solar atmosphere \citep{Borrero:2019}.
This same issue arises in purely theoretical models where the model boundary is driven in a particular way without regard to the internal state of the model, in which case the boundary driving can over-determine the MHD model leading to an improper solution in the volume of interest.
Thus, the question arises, how best to use \emph{estimated} boundary data given the MHD state of the volume of interest?

Photosphere-to-corona simulations occupy a special space in plasma physics modeling in terms of what boundary conditions are appropriate to the situation.
This is in stark contrast to many other subdisciplines of plasma physics, where researchers are most often interested in the evolution of a plasma that is energized in its interior, usually by instabilities, and the boundary plays a secondary role.
To give several examples:
\begin{enumerate}[label={(\roman*)}]
  {\item Local homogeneity and local isotropy\----the volume of interest can be considered a subvolume of a much larger turbulent system  \citep[]{Kolmogorov1941b} so that periodic boundary conditions, which pass energy out of the volume from one boundary and inject it in on the opposite boundary, may be applicable.}
  {\item Local plasma confinement\----the volume of interest is either isolated from the rest of the universe or the boundary is strictly controlled by external electric circuits \citep[][\S4.6.1]{Goedbloed:2004} and therefore conducting boundary conditions, which reflect energy back into the volume, may be appropriate.}
  {\item Spatially and temporally localized phenomena\----the dynamics of interest are transient, localized in space, and considered independent of the boundary conditions (e.g., some reconnection scenarios \citep{Nitta:2001}, bouyant flux ropes \citep{Leake:2022}, or interacting plasma wave packets \citep{Verniero:2018}).  Therefore, any numerically stable boundary condition may be used provided all analysis of the interesting physics is carried out before the boundaries can learn about the evolution of the plasma and act back on the region of interest.}
\end{enumerate}
The last example (iii) makes explicit the idea that some information propagates from a localized region of space, interacts with a distant boundary (artificial or not), and then some other, possibly different, information propagates back from the boundary into the localized subvolume of interest.

None of the above boundary conditions are appropriate for photosphere-to-corona simulations because the corona is driven by energy that originates in the convection zone and is transported through the photosphere.
From the perspective of a photosphere-to-corona simulation, the energy that ultimately structures the corona originated in the simulation boundary.
Therefore, for photosphere-to-corona simulations, the boundary and its treatment have elevated importance.

Despite the wealth of observational data mentioned previously, it remains an open question how to correctly evolve an MHD simulation consistent with an observed time series of the MHD state vector on the lower boundary.
This is true regardless of whether the entire state vector is known or only a subset.
Numerous attempts to solve related problems in the field of solar physics stretch back decades \citep{Schmidt:1964, Altschuler:1969,Nakagawa:1980,Alissandrakis:1981,Low:1985,Bogdan:1986,Yang:1986,Nakagawa:1987,Wu:1987,Aly:1989,Sakurai:1989,Wiegelmann:2004,Mackay:2006,Wu:2006,Valori:2007,Wheatland:2007,Jiang:2011,Cheung:2012,Malanushenko:2012,Aschwanden:2013,Yeates:2014,Fisher:2015,Aschwanden:2016, Goldstraw:2018,Zhu:2018,Pomoell:2019,Zhu:2019, Boocock:2019,Price:2020,Lumme:2022,Mathews:2022,Yeates:2022}.
In brief, methods to estimate the fully 3D state of the solar corona can be grouped into static methods and evolving methods.  
Static methods construct an estimate of the coronal state based on boundary observations at a single instance in time, and evolving methods use a time series of observations at a boundary to iteratively modify the volume state.
Many of the methods just cited operate in the limit of negligible pressure gradient forces and therefore only attempt to determine the coronal magnetic field.
On the other hand, the most sophisticated methods are fully dynamic: they numerically integrate the full MHD equations such that the state at one time is causally related to the state at the next time through the combination of Maxwell's equations and Newton's laws from which the MHD equations are derived.

In the present context we use the term ``data-driven simulation'' to mean a dynamically evolving simulation that directly incorporates estimates for the primitive variables (the elements of the MHD state vector $\mhdstate$) into the simulation boundary.
We are agnostic as to where those estimates come from: they could, for instance, be derived from observations, extracted from another simulation, or imposed according to some analytic model.
The problem is challenging because it is possible to prescribe the boundary values in a way that is fundamentally inconsistent with the MHD equations.
In fact, this will almost certainly be the case unless deliberate precautions are taken.
For example, most published data driving approaches hold some subset of the primitive variables constant in time while directly setting others \citep{Bourdin:2013, Galsgaard:2015, Hayashi:2018,Jiang:2020,Kaneko:2021, Guo:2021, Inoue:2023}; typically, some subset of the velocity and magnetic field vectors are time-varying while the remaining velocity and magnetic field variables, the density, and the pressure are held constant.
Many variations on this approach exist, and while the results may appear reasonable, in general \emph{it cannot and does not evolve the boundary in a way that is consistent with the MHD equations.}
The reason is that the evolution of the primitive variables are not independent from each other but instead are coupled together by the MHD equations.
Stated another way, an arbitrary perturbation to the MHD state vector is typically not a dynamical solution to the MHD equations; only displacements that can be expressed as combinations of eigenmodes of the MHD equations are dynamically allowed \citep[see, e.g., the Lagrangian formulation of MHD in][]{Dewar:1970, Dewar:2015}.
As such, at present, none of the published data-driven MHD approaches appear to drive their boundaries in a way that are guaranteed to be fully consistent with the MHD equations.
While this may be well recognized, it is often not stated explicitly, which is why we have emphasized it here.

The inherent challenge of data-driven photosphere-to-corona simulations and the shortcomings of the current methods were well illustrated in \citet{Toriumi:2020}.
They tested the fidelity of multiple coronal reconstruction methods by having each method attempt to reproduce the coronal magnetic field from an \textit{ab initio} flux emergence simulation that spanned the convection zone to the corona.
Each method was given synthetic ``observational'' data consisting of the time series of the MHD state vector extracted from a horizontal slice through the emergence simulation's photosphere.
\citet{Toriumi:2020} found that all of the methods they tested for reconstructing the coronal field from photospheric observations---static extrapolation, magnetofrictional, or data-driven MHD---struggled to accurately reproduce the ground truth field on a variety of metrics, including the extent of field expansion, the total magnetic energy, and the energy in excess of the potential field (although a magnetofrictional method was able to reproduce the relative magnetic helicity and asymptotically match the total magnetic energy).
Qualitatively and quantitatively, the structure of field lines differed substantially between the ground truth solution and all reproductions, even when a metric like the total magnetic energy matched the ground truth value to a factor of two or so (see especially Figures 3 and 5 of \citet{Toriumi:2020}).

\citet{Jiang:2020} presented a follow-up to \citet{Toriumi:2020} in which they performed additional tests of one of the MHD based data-driven simulations\footnote{The simulations used a variation of the code described in \citet{Jiang:2016}.} to study the effect of using data extracted at different heights from ground truth simulation as a boundary for the data-driven simulation.
They tested their data driving approach using layers extracted between the photosphere (which had a large Lorentz force imbalance) up to the low corona (where the Lorentz force was approximately zero, a so-called ``force free'' solution).
They found much better agreement when driving from layers with smaller residual Lorentz forces.
However, it is unclear why an MHD-based data driving approach was unable to reproduce the self-consistently generated MHD state of the ground truth simulation, regardless of the size of the Lorentz force; Lorentz forces are clearly an inherent and necessary part of the MHD equations.

This is a puzzling and troubling result that is worth emphasizing: for the test case of \citet{Toriumi:2020} and \citet{Jiang:2020} the ``observational'' data are perfect, the layer used for driving is pulled directly from a self-consistent MHD simulation, and yet none of the methods were able to fully reproduce the coronal field when driven from the photosphere (the material plasma variables were not considered in that study).
The unavoidable conclusion is that, at a fundamental level, none of the tested methods are fully consistent with the MHD equations \emph{in the driven photospheric boundary layer}.
At some level these methods must not correctly encode how information crosses the driving boundary.
This emphasizes that if the boundary is not evolved correctly then the interior will not be a valid solution to the problem at hand, and may not be a valid solution to any MHD system.

In the present work we derive boundary conditions for all primitive variables that are necessarily consistent with the MHD equations, up to the level that the numerical scheme matches the analytic equations.
To do so we express the evolution of the boundary using the characteristic form of MHD \citep[see, e.g.,][]{Jeffrey:1964, Goedbloed:2010}.
The characteristic form of MHD ultimately encodes how information (i.e., disturbances) propagate locally through the plasma.
This is possible because ideal MHD is a hyperbolic system\footnote{MHD is not strictly hyperbolic: for certain states of the system, two or more of the eigenmodes may become degenerate.  However, those cases may typically be resolved, see \citet{Roe:1996}.} and therefore has propagating solutions, each of which describe the directional propagation of information \citep{Goedbloed:2010}.
It is therefore possible to decompose the dynamics of the boundary into a portion that is due only to information originating within the simulation and a portion that is due only to information originating from outside the simulation, i.e., in the external universe.
The observed dynamics of the boundary are due to the superposition of these two sources of propagating information through the boundary plus any information propagating within the boundary itself, i.e., in the transverse direction.
The transverse portion, arising from transverse derivatives, can be calculated using only data within the boundary, which is already known in the simulation.
Then, supposing the time evolution of the boundary is known, and because the information originating inside the simulation and within the boundary is known, it follows that information originating from outside the simulation can be solved for and used to consistently set the boundary conditions correctly.
This is the fundamental idea behind our data-driven boundary condition (\ddbc).

In summary, the problem of data driving can be stated as this: How can the MHD state vector at time $t$ in the volume be evolved to time $t+\dtdata$, given a new target state vector on the lower boundary at time $t+\dtdata$ and a requirement that the evolution is fully consistent with MHD?
We stress again that, due to incomplete or uncertain knowledge of the boundary layer, it may not be possible to simultaneously satisfy both the estimated boundary evolution and the MHD equations, and in such cases we side with the MHD equations.
With that in mind, the goal of this paper is to describe both the mathematical background and numerical implementation needed to evolve an MHD simulation from $t$ to $t+\dtdata$ in a way that is fully consistent with the MHD equations in the volume and maximally consistent with the new, prescribed MHD state vector on the lower boundary at $t+\dtdata$. 
This solution to the problem may then be repeated $N$ times to drive the MHD system from $t$ to $t+N\,\dtdata$.

The structure of the rest of the paper is as follows.
In \S\ref{sec:characteristic-mhd} we recast the MHD equations into characteristic form and discuss how this describes the flow of information, or equivalently, the propagation of specific disturbances, through the system.
In \S\ref{sec:characteristic-bounds} we give an overview of using the characteristics to set boundary conditions and show how the time evolution of any boundary can be decomposed into the effect of information that passes through that boundary from each side.
In \S\ref{sec:ddbcs} we show how a time series of observations of a boundary can be used to solve for the information propagating in one of the normal directions, thus formally solving the problem of correctly setting boundary conditions in the continuous case.
In \S\ref{sec:fvddbc} we describe the representation of characteristic based MHD in a finite-volume numerical scheme, \charc{}, and the implementation of data-driven boundary conditions in that code, which we call \charc-\ddbc.
In \S\ref{sec:testcase} we describe our validation suite, which centers on a ``ground-truth'' simulation of an expanding spheromak implemented the Lagrangian Remap 3D MHD code \citep[\lare{}:][]{Arber:2001}.
We emphasize here that part of our validation involves coupling the \charc-\ddbc{} code to the \lare{} code, which have unrelated numerical schemes; \charc-\ddbc{} is general-purpose and can be coupled to any MHD code.
We describe the results of the validation simulations in \S\ref{sec:results}, provide some additional discussion in \S\ref{sec:discussion}, and summarize in \S\ref{sec:conclusions}.
Several appendices follow that serve to keep the present work self-contained.

As a final note in this Introduction, we stress that our characteristic method can be used to formulate arbitrary boundary conditions in addition to data-driven boundary conditions.
This will be explored in subsequent papers in this series.
In particular, a problem that is intricately related to the present one is how to set up boundary conditions that allow information to freely leave a simulation volume, a so-called (and perhaps unfortunately named) \emph{nonreflecting boundary condition} (\nrbc).
For that topic, we refer the reader to \citet{Kee:2023}, hereafter called \citetalias{Kee:2023}.
Put succinctly, if the present \citetalias{Tarr:2023} is about correctly getting information into a simulation volume, \citetalias{Kee:2023} is about correctly getting information out of the simulation volume.
As it turns out, the former stands on much more solid ground than the latter.

\section{Rewriting MHD in characteristic form}\label{sec:characteristic-mhd}
The key to formulating data-driven boundary conditions consistent with the internal state of an MHD simulation is to transform the traditional Eulerian description of MHD into the \emph{characteristic form} which directly describes the flow of information through the plasma as specific types of propagating modes.
The characteristic form is particularly useful for writing and analyzing boundary conditions for three reasons: 
\begin{enumerate}[nosep]
    {\item By identifying each \emph{type} of information (each mode), the characteristics encode precisely how the plasma at a given location responds to each mode propagating through that location.}
    {\item By identifying the direction and speed at which each mode propagates relative to the plasma velocity $\vect{v}$, the characteristic form distinguishes between information that propagates out of the simulation and information that propagates into the simulation, through the boundary, from the external universe.}
    {\item By expressing the boundary conditions in terms of the characteristics, they are intrinsically consistent with the MHD equations themselves. \newline This final reason has as a corollary
      that:
    \begin{enumerate}[label=\arabic{enumi}\alph*.,nosep]
        \item Any physical assumptions about the dynamics in the external universe inherent in the boundary conditions must be made explicit in terms of the allowed characteristic modes.
    \end{enumerate}
    }
\end{enumerate}

The first step towards formulating boundary conditions in characteristic form is to write the MHD equations in that form without any reference to a boundary, i.e., for an arbitrary location in space\footnote{From a data driving perspective, this initial formulation could be used for internal data driving and/or data assimilation in general MHD models.}.
While little of the information in this section is completely novel---among many other references, the characteristics of perturbative MHD are detailed in the text by \citet{Jeffrey:1964} and the full theory for numerical simulations was applied to a 1D Roe-type upwind differencing scheme in \citet{Brio:1988}---we include it here to keep the present text self-contained and to highlight aspects that are important for developing data-driven boundary conditions for 3D MHD.

The ideal MHD equations that describe conservation of mass, momentum, and energy, plus the ideal induction equation, are:
\begin{subequations}
  \label{eq:mhd}
  \begin{gather}
    \frac{\partial\rho}{\partial t} + \div(\rho\vect{v})=0\\
    \frac{\partial\vect{v}}{\partial t} + \vect{v}\cdot\del\vect{v} + \frac{1}{\rho}\del P - \frac{1}{\rho}\vect{j}\times\vect{B} - \vect{g}=\vect{0}\label{eq:momentum}\\
    \frac{\partial\epsilon}{\partial t} + \vect{v}\cdot\del\epsilon + \frac{P}{\rho}\div\vect{v}=0\\
    \label{eq:induction}
    \frac{\partial\vect{B}}{\partial t} +(\vect{v}\cdot\del)\vect{B} - (\vect{B}\cdot\del)\vect{v} + \vect{B}(\div\vect{v}) = \vect{0}.
    \intertext{The system is closed using the ideal equation of state relating pressure $P$ to density and internal energy density,}
    P = (\gamma-1)\rho\epsilon.
  \end{gather}
\end{subequations}
A general equation of state, with pressure a general function of density and internal energy $P(\rho,\epsilon)$, could also be used and would be better suited to a photospheric plasma.  However, the ideal equation of state leads to slightly simplified notation and so we adopt it here. All of the following work carries over with minor changes for the general case.
The primitive variables are density $\rho$, specific internal energy density $\epsilon$, the velocity vector $\vect{v}^T = (v_x,v_y,v_z)$, and the magnetic vector $\vect{B}^T=(B_x,B_y,B_z)$.
The electric current density is $\vect{j}=\curl\vect{B}/\mu_0$, where $\mu_0$ is the permeability of free space.

The equations are nondimensionalized by choosing normalization constants for the magnetic field, density, and length: $B=\bnorm B^*$, $\rho=\rnorm \rho^*$, and $L=\lnorm L^*$, where symbols subscripted with $N$ have units and starred quantities are unitless.
The normalizations for all other quantities are defined in terms of these three, the values of which are given in Appendix \S\ref{sec:normalization} where we describe the details of the numerical simulations and their initial conditions.
Gradients are given by $\nabla = \nabla^*/\lnorm$, and for the remaining variables we have $\vnorm = \bnorm/\sqrt{\mu_0\rnorm}$, $\tnorm =\lnorm/\vnorm$, $\pnorm=\bnorm^2/\mu_0$, $\epsilon_N = \bnorm^2/\mu_0\rnorm$, $g_N = \lnorm g_R/\vnorm^2$ (with $g_R$ the value of gravity at the solar surface in SI units), and $j_N = \bnorm/\mu_0\lnorm$.
The final of these means that the current density in normalized units has the form $\vect{j}^* = \nabla^*\times\vect{B}^*$ so that the permeability $\mu_0$ drops out of the equations.
Thus, the MHD equations in Equation \eqref{eq:mhd} are the same in both normalized and unnormalized (or starred and unstarred) form, with the single difference that $\vect{j}$ is replaced with $\curl\vect{B}$ in the momentum equation \eqref{eq:momentum}, without $\mu_0$.
From here on we drop reference to the starred variables and all expressions are taken to be normalized unless stated explicitly.

The total MHD state vector written in primitive variables is $\vect{U}^T = (\rho,\epsilon,v_x,v_y,v_z,B_x,B_y,B_z)$.
The MHD equations can then be rewritten in matrix form as
\begin{equation}
  \label{eq:mhd-matrix}
  \partial_t \vect{U} + \Matrix{A}_{x}\cdot\partial_x\vect{U} + \Matrix{A}_{y}\cdot\partial_y\vect{U}+ \Matrix{A}_{z}\cdot\partial_z\vect{U} + \vect{D} = \vect{0},
\end{equation}
where, for instance, the coefficient matrix $\Matrix{A}_z$ is given by
\begin{equation}
  \Matrix{A}_z = \begin{bmatrix}
      v_z &   0 &   0 &   0 & \rho               & 0        & 0        & 0 \\
      0   & v_z &   0 &   0 & (\gamma-1)\epsilon & 0        & 0        & 0  \\
      0   &   0 & v_z &   0 & 0                  &-B_z/\rho & 0        & 0\\
      0   &   0 &   0 & v_z & 0                  & 0        &-B_z/\rho & 0 \\
      \frac{(\gamma-1)\epsilon}{\rho} & (\gamma-1) & 0 & 0 & v_z & B_x/\rho & B_y/\rho  & 0\\
      0 & 0 & -B_z & 0    & B_x & v_z & 0 & 0\\
      0 & 0 &   0  & -B_z & B_y & 0   & v_z & 0\\
      0 & 0 &   0  & 0    & 0 & 0 & 0 & v_z
  \end{bmatrix}.
\end{equation}
The coefficient matrices $\Matrix{A}_x$  and $\Matrix{A}_y$ are similarly defined, and their full expressions are given in Appendices \ref{sec:Ax} and \ref{sec:Ay}.
The vector \inhomoterms{} may contain any inhomogeneous terms one wishes to include, such as gravity, resistive terms from a generalized Ohm's law, volumetric heating terms, and so on. 
We have included gravity as an example, in Equations \eqref{eq:mhd}; thus, in this paper, the inhomogeneous term is simply $\inhomoterms{} = \vect{g}=-g\zhat$, where $g=1$ is the value of gravity near the solar surface in normalized units.

The matrices $\amat_n$ are individually diagonalizable \citep{Roe:1996}, but not simultaneously so outside of special cases.\footnote{Notation: $n$ will index the spatial directions $x,y,z$, and $\sigma$ and $\zeta$ will index the variables of our system of equations, e.g. $[\amat_z]_{\sigma\zeta}$.  Thus, the first element of the MHD state vector is $U_{\sigma}=\rho$ with $\sigma=1$.  Scalars are written in standard font face, vectors in bold face, and matrices in a double-struck face.}
This is not a concern for us: we ultimately wish to formulate a boundary condition and are free to diagonalize the system in the direction normal to that boundary.
From here on, we will mostly focus on $\zhat$ as an example, but the analysis carries over to all directions by cyclic permutation of $\{x,y,z\}$.
Looking ahead towards the application of the formalism to data driving, the diagonalization will eventually be applied individually to each face of a computational cell to construct a discrete, finite volume method (see Appendix \ref{sec:transverse} for a discussion of our choice of a finite volume over a finite difference approach to the characteristics).

The primitive variables (and the more typical conserved variables, for that matter) do not constitute a good set of variables for describing how information propagates through the plasma.
To find a suitable set, we can rotate the state-space coordinate system in Equation~(\ref{eq:mhd-matrix}) describing the MHD equations in the standard way by diagonalizing each of the $\amat$ matrices with a similarity transformation.
As will be clear shortly, this transformation essentially separates out the different types of information that can propagate in a given direction.
Using \zhat{} as an example, the similarity transformation for $\amat_z$ is
\begin{equation}\label{eq:diagonalization}
  \amat_z = \rightzevecmat\zevalmat\leftzevecmat\quad \Longleftrightarrow \quad \zevalmat = \leftzevecmat\amat_z\rightzevecmat,
\end{equation}
where \zevalmat{} is the diagonal matrix of eigenvalues of $\amat_z$ and the \leftzevecmat{} and \rightzevecmat{} matrices contain the eigenvectors of $\amat_z$.
The left eigenmatrix $\leftzevecmat$ has the left eigenvectors $\vect{l}_{z,\sigma}^T$ as each row while the right eigenmatrix $\rightzevecmat$ has the right eigenvectors $\vect{r}_{z,\sigma}$ as each column, where the index $\sigma$ runs from 1 to 8, corresponding to the 8 primitive variables (or any transformed variables derived from them).
Full expressions for the left and right eigenvectors in the \zhat{} direction are given in Appendix \ref{sec:Az}.

The eigenvalues $\lambda_{z,\sigma}$ and left and right eigenvectors satisfy the usual eigenequations: $\vect{l}_{z,\sigma}^T\cdot\amat_z = \vect{l}_{z,\sigma}^T\lambda_{z,\sigma}$ and $\amat_z\cdot\vect{r}_{z,\sigma} = \lambda_{z,\sigma}\vect{r}_{z,\sigma}$.
The eigenvectors are biorthonormal so that $\vect{l}_{z,\sigma}^T\cdot\vect{r}_{z,\zeta} = \delta_{\sigma\zeta}$, the Kroneker delta, and $\leftzevecmat\rightzevecmat=\rightzevecmat\leftzevecmat=\Matrix{I}$, the identity matrix.
Complete descriptions of the left (\leftzevecmat) and right (\rightzevecmat) eigenmatrices are given in Appendix \ref{sec:Az}; the corresponding matrices for the $\xhat$ and $\yhat$ directions are given in Appendices \ref{sec:Ax} and \ref{sec:Ay}, respectively.

The eigenvalues of $\amat_z$ are found by solving the standard equation
\begin{gather}
  \fnc{Det}(\Matrix{A}_z-\lambda\Matrix{I})=0,
  \intertext{or, with the determinant expanded out,}
  \Bigl[(v_z-\lambda)^2\Bigr]\Bigl[b_z^2-(v_z-\lambda)^2\Bigr] \Bigl[(v_z-\lambda)^4-(a^2+b^2)(v_z-\lambda)^2+b_z^2a^2\Bigr]=0.
\end{gather}
This equation has eight solutions\footnote{The eight solutions are unique when $\amat_z$ is full rank, but in certain limits the eigenvalues become degenerate.  For example, when the magnetic field is reduced to zero the 8 MHD solutions collapse onto the 5 hydrodynamic solutions.  See \citet{Roe:1996} for a full discussion of the eigenstructure of MHD equations in various limits.}:
\begin{subequations}\label{eq:mhd-evals}
  \begin{gather}
    \lambda_1 = v_z,    \label{eq:eval1}\\
    \lambda_2 = v_z,    \label{eq:eval2} \\
    \lambda_3 = v_z-c_a,\label{eq:eval3} \\
    \lambda_4 = v_z+c_a,\label{eq:eval4}\\
    \lambda_5 = v_z-c_s,\label{eq:eval5}\\
    \lambda_6 = v_z+c_s,\label{eq:eval6}\\
    \lambda_7 = v_z-c_f,\label{eq:eval7}\\
    \lambda_8 = v_z+c_f, \label{eq:eval8}
    \intertext{where}
    c_a = \abs{b_z}\label{eq:ca},\\
    c_f^2 = \frac{1}{2}(a^2+b^2) + \frac{1}{2}\sqrt{(a^2+b^2)^2 - 4a^2b_z^2}\label{eq:cf},\\
    c_s^2 = \frac{1}{2}(a^2+b^2) - \frac{1}{2}\sqrt{(a^2+b^2)^2 - 4a^2b_z^2}\label{eq:cs},\\
    a^2 = \gamma(\gamma-1)\epsilon, \quad b_n = B_n/\sqrt{\rho} \text{ for $n\in(x,y,z)$}, \quad b^2 = \sum b_n^2.
  \end{gather}
\end{subequations}
The eigenvalues $\lambda_\sigma$ in Equations \eqref{eq:mhd-evals} constitute the components of the diagonal eigenvalue matrix in Equation \eqref{eq:diagonalization}, $\Matrix{\Lambda}_z=\fnc{diag}(\vect{\lambda}_z)$.
The eigenvalues have units of velocity and are expressed in terms of the bulk velocity in the diagonalization direction $v_z$, the sound speed $a$, the total and projected Alfv\'en speeds $b$ and $c_a$ (with projected components $b_n$ in each direction $n\in(x,y,z)$), the fast magnetosonic speed $c_f$, and the slow magnetosonic speed $c_s$.
Note that the Alfv\'en, fast, and slow speeds are modified from the standard 1D expressions by incorporating a projection onto the diagonalization direction: only $b_z$ appears in the definitions of the projected Alfv\'en speed $c_a$, or in the $4a^2b_z^2$ term under the radical in Equations \eqref{eq:cf} and \eqref{eq:cs}.

The eigenvalues themselves are propagation speeds, i.e., the speed at which each possible type of information described by the MHD equations flows through the plasma in the diagonalization direction: $\lambda_{1,2}$ are advection in $\zhat{}$, $\lambda_{3,4}$ the Alfv\'en speeds, $\lambda_{5,6}$ the slow magnetosonic speeds, and $\lambda_{7,8}$ the fast magnetosonic speeds, all calculated relative to the background velocity in the diagonalization direction, in this example $v_z$.
These characteristic speeds are equivalent to those reported by, e.g., \citet{Roe:1996}, although ours has one extra eigenvalue $v_z$; this is because \citet{Roe:1996} solved the 1-D system with $\nabla\cdot\vect{B}=0$ enforced by construction, while we choose to enforce the solenoidal constraint with an additional eigenmode; we discuss this further at the end of this section, after Equations \eqref{eq:mhd-chars}.

The eigendecomposition of the $\Matrix{A}_z$ matrix described in Equation \eqref{eq:diagonalization} can also be carried out for $\Matrix{A}_x$ and $\Matrix{A}_y$, as given in in Appendices \ref{sec:Ax} and \ref{sec:Ay}, respectively.
Doing so leads to essentially the same set of expressions for the eigenvalues and eigenvectors in each case, up to cyclic permutations of the variables in $x$, $y$, and $z$.
The fact that the eigenvalues and vectors for each direction only match up to a cyclic permutation of the projection direction is the reason that the MHD equations \eqref{eq:mhd-matrix} are not diagonalizable in all directions simultaneously: the eigensystem of one matrix does not match those of the others.
Nonetheless, the eigendecomposition can be carried out in each direction separately.

Substituting the diagonalized forms for all the coefficient matrices $\amat_x,\ \amat_y,\text{ and } \amat_z$ into Equation \eqref{eq:mhd-matrix}, we rewrite the MHD equations as
\begin{gather}\label{eq:mhd-matrix-expanded}
  \partial_t \vect{U} + 
  \rightxevecmat\Matrix{\Lambda}_x\leftxevecmat\cdot\partial_x\vect{U} +
  \rightyevecmat\Matrix{\Lambda}_y\leftyevecmat\cdot\partial_y\vect{U}
  + \rightzevecmat\Matrix{\Lambda}_z\leftzevecmat\cdot\partial_z\vect{U} + \inhomoterms{}= \vect{0}.
  \end{gather}
Grouping the derivative term in each direction  $(\partial_n\mhdstate{})$ with its respective eigenvalue matrix $(\evalmat_n)$ and left eigenvector matrix $(\smatinv_n)$ defines the characteristic derivative vector $\charderiv_n$ in each direction: 
\begin{subequations}\label{eq:charderiv}
    \begin{gather}
        \charderiv_x=\xevalmat{}\smatinv_x\cdot\partial_x\mhdstate\label{eq:charderivx},\\
        \charderiv_y=\yevalmat{}\smatinv_y\cdot\partial_y\mhdstate\label{eq:charderivy},\\
        \charderiv_z=\zevalmat{}\smatinv_z\cdot\partial_z\mhdstate\label{eq:charderivz}.
    \end{gather}
\end{subequations}
Substituting these $\charderiv_n$ into Equation \eqref{eq:mhd-matrix-expanded}, the MHD equations become
\begin{gather}
  \partial_t \mhdstate + 
  \rightxevecmat\cdot\charderiv_x + 
  \rightyevecmat\cdot\charderiv_y +
  \rightzevecmat\cdot\charderiv_z +\inhomoterms = \vect{0}. \label{eq:mhd-char-full}
\end{gather}
Equation \eqref{eq:mhd-char-full} is equivalent to the full system of MHD equations \eqref{eq:mhd} or \eqref{eq:mhd-matrix}, but rewritten in characteristic form.
Note that we have not yet chosen a boundary direction; this is still a description of a plasma whose properties are known at all spatial locations.

The physical meaning of each element in Equation \eqref{eq:mhd-char-full} can be understood even before the full expressions are worked out.
Each component ($\sigma$) of one of the characteristic derivatives, $L_{n,\sigma}$, corresponds to a specific combination of information (derivatives of primitive variables) that propagates with a given speed and direction relative to the diagonalization direction, \nhat; for example, a fast mode disturbance propagating in $\zhat{}$.
To see this, observe that: 
\begin{enumerate}[label=(\roman*),nosep]
    {\item the derivatives in Equation \eqref{eq:charderiv} are changes in the state of the plasma from point to point in space,}
    {\item a single row $\sigma$ of a left eigenvalue matrix $\smatinv_n$, which is the left eigenvector $\vect{l}^T_\sigma$, couples together a set of those changes through the dot product, \newline
    and}
    {\item the coupled set of changes is multiplied by a single eigenvector speed $\lambda_{n,\sigma}$ (recall the matrix $\evalmat{}_n$ is diagonal).}
\end{enumerate}
The result of the calculation is an object which represents a specific set of coupled changes in the primitive variables that propagate at a particular speed in a particular direction, i.e., it represents a specific type of information propagating through the plasma.

We emphasize that one component $\sigma$ of a characteristic derivative vector in one direction (e.g., $L_{n,\sigma}$) is a \emph{scalar} quantity that encodes how \emph{all} components of the MHD state vector are simultaneously modified due to the $\sigma$ mode propagating in the $\nhat$ direction.
That coupled change between all the primitive variables is the embodiment of the dynamical constraints imposed by the MHD equations: out of all arbitrary variations to the MHD state vector $\mhdstate{}$ one might consider, only the eight described by the eight components of the characteristic derivative $\charderiv_n$ in the \nhat{} direction are allowed by, and consistent with, the MHD equations.

A superposition of multiple modes is, of course, allowed, and the right eigenvectors $\vect{r}_\sigma$ describe the relative contribution of each mode to the time update of each primitive variable.
To see this, look again to Equation \eqref{eq:mhd-char-full}. 
The right eigenmatrix $\smat_n$ multiplies its associated characteristic derivative vector.
Each column of the right eigenmatrix $\smat_n$ is a right eigenvector $\vect{r}_\sigma$, and, carrying though the multiplication between $\smat_n$ and $\charderiv_n$, the elements of a right eigenvector give the relative contribution of each mode to the time-update of the primitive variables.

Summarizing the above few paragraphs, the specific combination of weights of primitive variable derivatives prescribed by a left eigenvector $\vect{l}_{n,\sigma}$ defines a mode, the scalar value of the characteristic derivative $L_{n,\sigma}$ defines the strength of the mode, and each right eigenvector $\vect{r}_{n,\zeta}$ describes the response of a primitive variable to each mode\footnote{Throughout, we will often refer to ``the $L_{n,\sigma}$ mode'' or ``$L_{n,\sigma}$ characteristic'' when, strictly speaking, the characteristic derivative $L_{n,\sigma}$ is the value of the weighted sum of primitive variable derivatives associated with the propagation of the $\sigma$ mode evaluated at a certain spatial location.  In practice, the modes and the derivatives associated with them are matched 1-to-1, so we will typically use the less verbose terminology.}.
The interpretation of Equation \eqref{eq:mhd-char-full} that follows from this discussion is that the temporal update to the MHD state vector $\mhdstate{}$ at some location $\vect{x}$ is due to the combined effect of all information propagating through that location from each type of eigenmode in each direction.
The eigendecomposition performed at Equation \eqref{eq:diagonalization} separated the various types of information that can propagate through the plasma at $\vect{x}$ from the effect that each type of propagating information has on the plasma at location $\vect{x}$.
This decomposition will eventually allow us to identify the information propagating into a subvolume of plasma from the external universe, and thereby enable us to enforce only allowed changes to the plasma at a simulation boundary.
Note that, if the supplied information is not exactly self-consistent then the boundary condition will not satisfy the Cauchy conditions and hence will be over-determined \citep[][Chapter 6]{Morse:1953a}.
In that case, we must either depart from the known physics of our universe or search for the closest boundary state that \emph{does} obey MHD; our solution for the latter case is in \S\ref{sec:fvddbc}.

We turn now to the details of analyzing how information propagates in a specific direction with the eventual goal of determining the way in which information can cross a boundary and what effect it has on the plasma at the boundary.
To make contact with photosphere-to-corona simulations, we take the \zhat{} direction as an example and group the transverse terms (involving $\partial_x,\partial_y$) and inhomogeneous terms (\inhomoterms{}) in Equation \eqref{eq:mhd-char-full} into a single vector $\vect{C}$:
\begin{align}
  \label{eq:c}\vect{C} = \Matrix{A}_{x}\cdot\partial_x\vect{U} + \Matrix{A}_{y}\cdot\partial_y\vect{U}+\vect{D} = \rightxevecmat\cdot\charderiv_x + \rightyevecmat\cdot\charderiv_y +\inhomoterms.
\end{align}
Full expressions for the vector $\vect{C}$ are given in Appendix \ref{sec:transverse}.
Note that $\vect{C}$ lives in the space of the MHD primitive variables, i.e., each element $C_\sigma$ contributes to the update of one of the primitive variables.
Substituting $\vect{C}$ into Equation \eqref{eq:mhd-char-full}, we finally arrive at
\begin{equation}
    \partial_t\mhdstate{} + \rightzevecmat\cdot\charderivz + \vect{C} = \vect{0}.\label{eq:mhdbc}
\end{equation}
Equation \eqref{eq:mhdbc} is the primary equation we will focus on for the remainder of this paper.
It is still just an equivalent form of the MHD equations but now written in a way that is useful for describing boundary conditions in the \zhat{} direction.
In particular, it consolidates all our knowledge about the \emph{flow of information in the \zhat{} direction} into the vector \charderivz.

Consider an arbitrary plane at constant $z=z_b$, which we suggestively label $\vect{x}_b = (x,y,z_b)$.
If the state of the system is known everywhere in the plane $\vect{x}_b$ at some time $t$ then every component of the combined transverse and inhomogeneous term $\vect{C}(t; \vect{x}_b)$ can be calculated in the plane at that time.
On the other hand, the characteristic derivative vector \charderivz{} defined in Equation \eqref{eq:charderivz} contains derivatives in direction normal to the plane and therefore inherently requires information from outside of the plane in order to fully specify it.
At the moment, the information propagating in each direction through the plane is still bundled together in the vector \charderivz{}, but these two pieces will soon be separated out.
If the arbitrary plane is actually a boundary, so that the MHD state vector is known everywhere on one side (the interior) but not the other (the exterior), then the components of \charderivz{} that describe inward propagating modes will represent independent information coming into the system.

Having discussed the meaning of each element of the MHD equations when written in characteristic form, either as Equation \eqref{eq:mhd-char-full} or \eqref{eq:mhdbc}, we turn now to the mathematical details of those elements. 
The following discussion still applies to the continuous case and makes no assumptions about boundaries or numerical discretization of the system.
We will return to the application of the characteristic derivatives to boundary conditions (and data-driven boundary conditions in particular) in Section \ref{sec:characteristic-bounds}.

Each component of the characteristic derivative\footnote{To reduce notational clutter, for the rest of this section elements of an eigensystem that do not explicitly refer to a direction will refer to the \zhat{} direction.  When either the \xhat{} or \yhat{} directions are needed, they will be explicitly labeled with a subscript.  For example, $L_1$ is taken to mean $L_{z,1}$, and the fast mode speed labeled in the most complete fashion is $c_{z,f}$.} \charderiv{}, $L_\sigma$, is a weighted sum of the normal derivatives of the primitive variables multiplied by a characteristic speed $\lambda_\sigma$. The components of the left eigenvector are the weights for each primitive variable derivative for the $\sigma$ mode:
\begin{gather}\label{eq:li-index}
L_\sigma=\lambda_\sigma(\vect{l}_\sigma^T\cdot\partial_z\vect{U}) = \lambda_\sigma \sum_\zeta S_{\sigma\zeta}^{-1} \partial_z U_\zeta.
\end{gather}

The eight components of $\charderiv$ (from which the left eigenvectors can be read off, cf. Appendix \ref{sec:Az}) are
\begin{subequations}\label{eq:li}
  \begin{align}
    \text{\textsl{div}B} && L_1 & = v_z\biggl[B_z^\prime\biggr] \label{eq:l1z}\\
    \text{entropy} && L_2 & = v_z\biggl[ \frac{1-\gamma}{\gamma \rho}\rho^\prime +\frac{1}{\gamma\epsilon}\epsilon^\prime \biggr] \label{eq:l2z}\\
    \text{reverse Alfv\'en} && L_3 & = \frac{v_z-c_a}{2}\biggl[-\beta_yv_x^\prime + \beta_x v_y^\prime - \frac{\beta_ys_z}{\sqrt{\rho}}B_x^\prime +\frac{\beta_xs_z}{\sqrt{\rho}}B_y^\prime \biggr]\label{eq:l3z} \\
    \text{forward Alfv\'en } && L_4 & = \frac{v_z+c_a}{2}\biggl[\beta_yv_x^\prime - \beta_x v_y^\prime - \frac{\beta_ys_z}{\sqrt{\rho}}B_x^\prime +\frac{\beta_xs_z}{\sqrt{\rho}}B_y^\prime \biggr]\label{eq:l4z} \\
    \text{reverse slow} && L_5 & = \frac{v_z-c_s }{2}\biggl[\frac{\alpha_s}{\gamma\rho}\rho^\prime +\frac{\alpha_s}{\gamma\epsilon}\epsilon^\prime - \frac{\beta_x\alpha_f c_f}{a^2}s_zv_x^\prime - \frac{\beta_y\alpha_f c_f}{a^2}s_zv_y^\prime - \frac{\alpha_s c_s}{a^2}v_z^\prime -\frac{\beta_x\alpha_f}{\sqrt{\rho}a}B_x^\prime - \frac{\beta_y\alpha_f}{\sqrt{\rho}a}B_y^\prime \biggr]\label{eq:l5z} \\
    \text{forward slow} && L_6 & = \frac{v_z+c_s}{2}\biggl[\frac{\alpha_s}{\gamma\rho}\rho^\prime +\frac{\alpha_s}{\gamma\epsilon}\epsilon^\prime + \frac{\beta_x\alpha_f c_f}{a^2}s_zv_x^\prime + \frac{\beta_y\alpha_f c_f}{a^2}s_zv_y^\prime + \frac{\alpha_s c_s}{a^2}v_z^\prime -\frac{\beta_x\alpha_f}{\sqrt{\rho}a}B_x^\prime - \frac{\beta_y\alpha_f}{\sqrt{\rho}a}B_y^\prime \biggr]\label{eq:l6z} \\
    \text{reverse fast} && L_7 & = \frac{v_z-c_f }{2}\biggl[\frac{\alpha_f}{\gamma\rho}\rho^\prime +\frac{\alpha_f}{\gamma\epsilon}\epsilon^\prime + \frac{\beta_x\alpha_s c_s}{a^2}s_zv_x^\prime + \frac{\beta_y\alpha_s c_s}{a^2}s_zv_y^\prime - \frac{\alpha_f c_f}{a^2}v_z^\prime +\frac{\beta_x\alpha_s}{\sqrt{\rho}a}B_x^\prime + \frac{\beta_y\alpha_s}{\sqrt{\rho}a}B_y^\prime \biggr]\label{eq:l7z}\\
    \text{forward fast} && L_8 & = \frac{v_z+c_f}{2}\biggl[\frac{\alpha_f}{\gamma\rho}\rho^\prime +\frac{\alpha_f}{\gamma\epsilon}\epsilon^\prime - \frac{\beta_x\alpha_s c_s}{a^2}s_zv_x^\prime - \frac{\beta_y\alpha_s c_s}{a^2}s_zv_y^\prime + \frac{\alpha_f c_f}{a^2}v_z^\prime +\frac{\beta_x\alpha_s}{\sqrt{\rho}a}B_x^\prime + \frac{\beta_y\alpha_s}{\sqrt{\rho}a}B_y^\prime \biggr]\label{eq:l8z},
  \end{align}
\end{subequations}
where the prime denotes differentiation in the $z$-direction, i.e., $\partial_z q = q^\prime$, for $q$ an arbitrary variable.
Each subequation is labeled by the type of mode (or disturbance) it represents.
The $\textsl{div}B$ and entropy modes each stand alone, while the Alfv\'en-, slow-, and fast-mode families each include a pair of related modes. 
The related modes in a given family have the same coefficients on each primitive variable derivative, but unique signs.

The above equations make use of the auxiliary variables
\begin{subequations}
  \label{eq:aux-var}
  \begin{align}
      s_z &= \sign(b_z) \\ 
      b_\perp^2 & = b_x^2+b_y^2 \\
      \beta_x &= \frac{b_x}{b_\perp}\label{eqn:Betax} \\
      \beta_y &= \frac{b_y}{b_\perp}\label{eqn:Betay} \\
      \alpha_f^2 &= \frac{a^2-c_s^2}{c_f^2 - c_s^2}\\
      \alpha_s^2 &= \frac{c_f^2-a^2}{c_f^2 - c_s^2},
  \end{align}
\end{subequations}
where $\alpha_f^2+\alpha_s^2=1$.
The prefactor in front of the bracketed term on the right hand side of each of Equations \eqref{eq:li} is proportional to the eigenvalue $\lambda_\sigma$ of the eigenmode and has units of velocity (see Eq. (\ref{eq:mhd-evals})).

\begin{figure}
    \centering
    \includegraphics{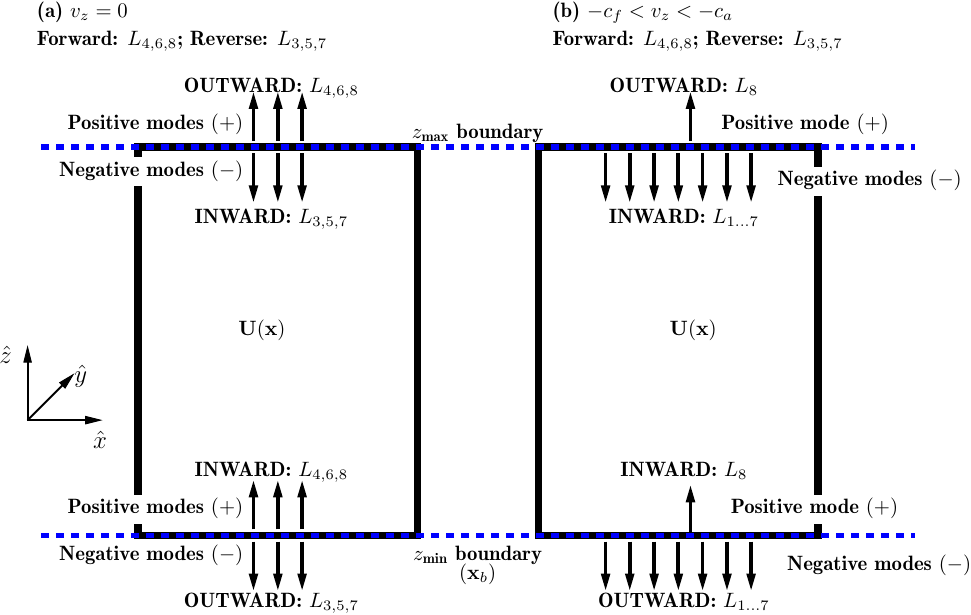}
    \caption{Diagram summarizing the relationship between the three types of propagation, defined relative to: (i) the bulk velocity, (ii) the coordinate axes, and (iii) the boundary of a subvolume of the plasma.  Two examples are shown for the  velocity $v_z$ in the \zhat{} direction.  The modes are labeled by their characteristic derivatives $L_\sigma$.  (a) $v_z = 0$, for which the forward-propagating modes (relative to the bulk velocity $v_z$) coincide with the positive propagating modes (relative to unit direction $\zhat$), and the reverse-propagating modes coincide with the negative propagating modes.  In example (b), the bulk velocity is directed downward with a magnitude exceeding all but the fast mode speed $c_f$.  The forward and reverse propagating modes are the same as in case (a), but now the identification of the (grid-) positive and negative modes has changed, due to the different bulk velocity.  Finally, the identification of \emph{inward} versus \emph{outward} propagating modes, defined relative the boundary-normal directions, are also indicated.
    \label{fig:axes-propagation}}
\end{figure}

There are three ways of defining the direction of propagation for each mode that need to be considered, as depicted diagrammatically in Figure \ref{fig:axes-propagation}.
First, the sign of the eigenvalue of the $\sigma$ mode, $\fnc{sign}(\lambda_\sigma)$, determines the direction of propagation for that mode relative to the diagonalization direction (i.e., the coordinate axes), with \emph{positive} (+) eigenvalues propagating in $+\zhat$ and \emph{negative} $(-)$ eigenvalues in $-\zhat$.
Second, the $+$ or $-$ sign in the expression for each eigenvalue in Equations \eqref{eq:eval1}-\eqref{eq:eval8} determines if that mode propagates in the \emph{forward} or \emph{reverse} direction relative to the bulk velocity $v_z$.
Eigenmodes 1 and 2 are purely advective (with $\lambda = v_z$), and hence do not propagate relative to the bulk velocity.
The final way of defining the propagation direction is relative to the boundary of a subvolume of interest.
For instance, at the bottom of a simulation, the positive modes propagate \emph{inward} (\incoming{}) towards the simulation interior and the negative modes propagate \emph{outward} (\outgoing{}) from the simulation towards the exterior universe.
Thus, at the top boundary of a simulation, the roles of the positive (negative) propagating modes are instead identified as outward (inward) propagating.

Figure \ref{fig:axes-propagation} presents two examples that help to clarify the three types of propagation that we will repeatedly refer to in this paper.
Panel (a) presents the case of a static plasma with $v_z=0$, and Panel (b) the case where the bulk velocity is downward with a magnitude in between the fast magnetosonic speed and the Alfv\'en speed, that is, $-c_f<v_z<-c_a$.
The \emph{forward} and \emph{reverse} propagating modes are labeled at the top of each panel in the Figure.
Because there is a 1-1 correspondence between characteristic derivatives and the modes that define them, we simply use the names of the derivatives to refer to the modes.
Note that the characteristic modes identified as forward- or reverse-propagating are always the same regardless of the normal bulk velocity, so they do not change between Panels (a) and (b). 
This is because the characteristics are ultimately defined in a co-moving frame with the plasma: all the eigenvalues in \eqref{eq:mhd-evals} are Doppler shifted relative to $v_z$.

Next, the terms \emph{positive} and \emph{negative} refer to propagation relative to the coordinate axes, or equivalently, the grid axes in a simulation, and the number and identification of these modes changes depending on the bulk plasma velocity.
For the static case $v_z=0$ in Panel (a), the positive and negative modes precisely coincide with the forward and reverse modes.
In this case, the purely advective modes $L_1$ and $L_2$ are non-propagating.
In contrast, for the example of the large magnitude downward velocity in Panel (b), only the $\sigma=8$ mode can propagate in the positive direction while all the other modes are negatively propagating relative to the normal coordinate axis.
In the extreme limit where the bulk velocity exceeds the fastest characteristic speed, $\abs{v_z}> c_f$, all modes will propagate in a single coordinate direction, e.g., $\pm\zhat$.

Finally, the identification of the \emph{incoming} and \emph{outgoing} propagating modes in each panel is simply a renaming of the positive and negative propagating modes with the direction of a boundary taken into account.
However, this is still necessary nomenclature for properly describing boundary conditions, because the outward propagating modes represent calculable information communicated from the known plasma properties to the external universe while the inward propagating modes represent the injection of new, previously unknown information into the region of space where the state of the plasma is known.

The type of information propagated through the plasma by each mode is labeled in each of Equations \eqref{eq:li}.
In order, the characteristic derivatives describe: the conservation of magnetic flux with the fluid flow ($L_1$), advection of entropy with the fluid flow ($L_2$), reverse and forward propagation of Alfv\'enic disturbances (relative to the fluid flow; $L_3$ and $L_4$), reverse and forward propagation of slow disturbances (relative to the fluid flow; $L_4$ and $L_6$), and reverse and forward propagation of fast disturbances (relative to the fluid flow; $L_7$ and $L_8$).
Note that, while each of these modes is related to the waves in linearized MHD, they are much more general, but also local (waves are less general but non-local).
Because the characteristic description is exactly equivalent to other forms for full MHD, any static configuration or dynamic process allowed by the full, nonlinear MHD equations is expressible in terms of the characteristic modes, although carrying out such a description may be difficult.

As expressed in Equations \eqref{eq:charderiv} and \eqref{eq:mhd-char-full}, the characteristic derivatives are defined in terms of left eigenvectors and then multiplied by the right eigenvectors in order to determine the time update to the primitive variables in the vector \mhdstate{}.
Full expressions for the left and right eigenvectors are given in Appendix \ref{sec:Az}.  
These are equivalent to those given by \citet{Brio:1988} and \citet{Roe:1996}, both of whom used the same set of auxiliary variables \eqref{eq:aux-var} but different primitive variables.
These auxiliary variables basically define the normalization of the eigenvectors.
The eigenvectors of any system can be arbitrarily normalized, and the normalization given above in terms of the auxiliary variables is preferable for numerical computation because, as \citet{Roe:1996} show in their \S2, the eigenvectors of an MHD coefficient matrix $\amat$ remain well behaved in various limits when written in terms of the auxiliary variables~(\ref{eq:aux-var}).
In particular, the apparent divergence in~(\ref{eqn:Betax})\--(\ref{eqn:Betay}) as $b_\perp\rightarrow 0$ remains bounded, and setting $\beta_x = \beta_y \rightarrow 1/\sqrt{2}$ in that case preserves the orthonormality of the eigenvectors.

The characteristic derivatives in a direction (e.g., \charderivz{}) have a complicated relationship to the spatial derivatives of the primitive variables in that same direction ($\partial_z\mhdstate$).
The coefficients of the normal derivatives in Equations \eqref{eq:li} do define a transformation between the two through the transformation matrix $\Matrix{E}$\footnote{Appendix \ref{sec:LtoU} provides the inverse matrix $\Matrix{E}^{-1}$ as well as the derivatives of the primitive variables in terms of the characteristic derivatives.}
\begin{gather}
    \charderivz = \zevalmat\smatinv_z\partial_z\mhdstate \equiv \Matrix{E}\partial_z\mhdstate,\label{eq:uprime2l}
    \intertext{which has an inverse transformation}
    \partial_z\mhdstate = \Matrix{E}^{-1}\charderivz = \smat_z\zevalmat^{-1}\charderivz\label{eq:l2uprime},
\end{gather}
where $\evalmat^{-1}_{z,\sigma\sigma} = \frac{1}{\lambda_\sigma}$ for $\lambda_\sigma\ne0$ and 0 otherwise (i.e., it is the pseudoinverse).
For any valid 3D MHD field the transformation matrix $\Matrix{E}$ should be full rank\footnote{Alternatively, the transformation matrix may gracefully collapse onto a subspace of the equations when any eigenvalues and eigenvectors overlap.  This happens, for instance, in the hydrodynamic limit of $\vect{B}\rightarrow \vect{0}$.  Again, see \citet{Roe:1996} for a full description of these limits.}; hence, if either the characteristic derivatives or primitive variable derivatives are given then the other set can be calculated from the first.
The fact that the transformation matrix $\Matrix{E}$ is not a diagonal matrix makes clear that varying a single component of a characteristic derivative vector produces corresponding variations in multiple primitive variables, and vice versa.
This interdependence is why the characteristics are useful for setting boundary conditions: each mode encodes a coupled, physically admissible variation of the entire MHD state vector; or in the other direction, out of all possible variations of the MHD state vector, those expressible in terms of characteristic derivatives are the physically allowed variations.
However, because the characteristics represent the directional flow of information, it is not immediately obvious how one should set up a discrete version of the transformation. 
The characteristic derivatives should presumably be cast in terms of one-sided derivatives (i.e., upwind and downwind), while this is not necessary for derivatives of the primitive variables.
We will return to that discussion in \S\ref{sec:fvddbc}.

Carrying out the matrix multiplication in Equation \eqref{eq:mhdbc} between the right eigenvector matrix \rightzevecmat{} and the characteristic derivatives \charderivz{} allows the MHD equations to be expressed directly in terms of the characteristic derivatives $L_\sigma$:
\begin{subequations}
  \label{eq:mhd-chars}
  \begin{align}
    &    \partial_t \rho -\rho [L_2] + \alpha_s\rho[L_5+L_6] + \alpha_f\rho[L_7+L_8]+  C_\rho = 0\label{eq:rhochar} \\
    &    \partial_t \epsilon + \epsilon[L_2] +\alpha_s\frac{a^2}{\gamma}[L_5+L_6]+\alpha_f\frac{a^2}{\gamma}[L_7+L_8] + C_\epsilon = 0 \\
    &    \partial_t v_x - \beta_y[L_3-L_4]-\alpha_f\beta_xc_fs_z[L_5-L_6] + \alpha_s\beta_xc_ss_z[L_7-L_8] + C_{v_x} = 0 \\
    &    \partial_t v_y + \beta_x[L_3-L_4] - \alpha_f\beta_yc_fs_z[L_5-L_6] + \alpha_s\beta_yc_ss_z[L_7-L_8]+ C_{v_y} = 0 \\
    &    \partial_t v_z - \alpha_sc_s[L_5-L_6] - \alpha_fc_f[L_7-L_8]+ C_{v_z} = 0 \\
    &    \partial_t B_x -\beta_y\sqrt{\rho}s_z[L_3+L_4] - \alpha_f\beta_x\sqrt{\rho} a[L_5+L_6] + \alpha_s\beta_x\sqrt{\rho}a[L_7+L_8] + C_{B_x} = 0 \\
    &    \partial_t B_y +\beta_x\sqrt{\rho}s_z[L_3+L_4] - \alpha_f\beta_y\sqrt{\rho}a[L_5+L_6] +\alpha_s\beta_y\sqrt{\rho}a[L_7+L_8]+ C_{B_y} = 0 \\
    &    \partial_t B_z +[L_1] + C_{B_z} = 0.
  \end{align}
\end{subequations}
This expanded form of the MHD equations is exactly equivalent to any set of Equations \eqref{eq:mhd}, \eqref{eq:mhd-matrix}, \eqref{eq:mhd-matrix-expanded}, \eqref{eq:mhd-char-full}, or \eqref{eq:mhdbc}.
The subscripted $C$ in Equations \eqref{eq:mhd-chars} are the components of the vector of transverse and inhomogeneous terms $\vect{C}$, defined in Equation \eqref{eq:c}. 
Full expressions for each component are given in Appendix \ref{sec:transverse} (the subscript notation for $C$ emphasizes that each component contributes to the time update of a single primitive variable).

The Equation \eqref{eq:mhd-chars} form of the MHD equations makes it abundantly clear how the time variation of each primitive variable depends on the characteristic derivatives and the flow of information they imply.
For instance, the continuity equation \eqref{eq:rhochar} shows that the density at some location will change because fluid with a different entropy is advected to that location (via $L_2$), or because slow or fast disturbances propagate to that location (via any of $L_{5,6,7,8}$). 
Similar interpretations hold for the other primitive variables.
From the Eulerian perspective of a primitive variable at a given location, it does not matter in which direction any of the modes propagate: the variable simply responds to all the modes that contribute to its update, with the strength of the response given by the value of each mode's characteristic derivative multiplied by the corresponding component of the right eigenvector.
On the other hand, calculating the value of a characteristic derivative does depend on the direction of propagation for the mode it represents: the derivative should be calculated in the upwind direction so that the changes in the plasma represented by a characteristic derivative are propagated at the characteristic speed in the correct direction for the associated mode.

Recall that we defined three flavors of propagation direction when discussing Figure \ref{fig:axes-propagation}: forward/reverse, positive/negative ($+/-$), and inward/outward ($\incoming/\outgoing$), which refer to propagation relative to the bulk velocity, the coordinates (or numerical grid), and a boundary, respectively.
In each of Equations \eqref{eq:mhd-chars} the forward-- and reverse--propagating pair of modes in each family (i.e., the slow, Alfv\'en, and fast modes relative to the bulk velocity) are grouped within square brackets\footnote{The $\nabla\cdot\vect{B}$ preserving mode, $L_1$, and the entropy mode, $L_2$, are purely advective modes and do not come in pairs.}.
Referring to the definitions of eigenvalue speeds in \eqref{eq:mhd-evals}, depending on the value of the bulk velocity $v_z$, either or both of the modes grouped in a single set of brackets can propagate in the positive or negative direction (i.e., relative to the coordinate axes).
Then, once a boundary $\vect{x}_b$ is specified, for instance taking the $z=0$ plane as the $z_\text{min}$ boundary of our volume of interest, the positive and negative propagating modes in turn define the inward and outward propagating modes relative to that boundary.

\begin{deluxetable}{ccl||ccc}
  \tablecaption{\label{tab:incoming}Characteristic derivatives (second and fourth columns) in the $\zhat$ direction ordered by boundary--normal velocity $v_z$ (first column), for the bottom boundary ($z_\text{min}$) of some volume. The horizontal line marks $v_z=0$, with $v_z\leq0$ above the line and $0\leq v_z$ below the line.  For reference, the fourth column shows the outgoing characteristic derivatives, which are simply the reverse ordering of the inward characteristics. $P$ refers to the number of inward propagating modes and $Q$ to the number subequations in \eqref{eq:mhd-chars} that the incoming characteristic derivatives participate in.  The slow, Alfv\'en, and fast speeds are all defined to be non-negative and are ordered $0\leq c_s\leq c_a \leq c_f$; see \eqref{eq:mhd-evals}.  Note that whenever the bulk velocity matches a mode's propagation velocity the corresponding mode is nonpropagating and, technically, neither incoming nor outgoing.  However, in that case the mode has zero amplitude (see Eqns.~\eqref{eq:li}) and thus no effect on the system.  Practically, this means we can include the equals sign on both sides of each division (`$\lesssim$' instead of `$<$') without loss of generality. }
  \tablehead{
    \colhead{Velocity} & \colhead{Incoming (\incoming) $L_\sigma$} &
    \colhead{Dependent Derivatives} & \colhead{Outgoing (\outgoing) $L_\sigma$} & \colhead{P} & \colhead{Q}}
  \startdata
  $v_z \leq -c_f$           &  -  & - & $L_7,L_3,L_5,{L_1},{L_2},L_6,L_4,L_8$ & 0 & 0\\
  $-c_f\leq v_z \leq -c_a$& $L_8$ & $\rho^\prime,\epsilon^\prime, v_x^\prime, v_y^\prime,v_z^\prime,B_x^\prime,B_y^\prime$ & $L_7,L_3,L_5,{L_1},{L_2},L_6,L_4$ & 1 & 7\\
  $-c_a\leq v_z \leq -c_s$& $L_4,L_8$ & $\rho^\prime,\epsilon^\prime, v_x^\prime, v_y^\prime,v_z^\prime,B_x^\prime,B_y^\prime$ & $L_7,L_3,L_5,{L_1},{L_2},L_6$ & 2 & 7 \\
  $-c_s\leq v_z \leq 0$   & $L_6,L_4,L_8$ & $\rho^\prime,\epsilon^\prime, v_x^\prime, v_y^\prime,v_z^\prime,B_x^\prime,B_y^\prime$ & $L_7,L_3,L_5,{L_1},{L_2}$ & 3 & 7 \\
  \hline
  $0\leq v_z \leq c_s $   & ${L_1},{L_2},L_6,L_4,L_8$ & ${\rho^\prime,\epsilon^\prime}, v_x^\prime, v_y^\prime,v_z^\prime,B_x^\prime,B_y^\prime,{B_z^\prime}$ & $L_7,L_3,L_5$ & 5 & 8 \\
  $c_s\leq v_z \leq c_a$  & $L_5,{L_1},{L_2},L_6,L_4,L_8$ & ${\rho^\prime,\epsilon^\prime}, v_x^\prime, v_y^\prime,v_z^\prime,B_x^\prime,B_y^\prime,{B_z^\prime}$ & $L_7,L_3$ & 6 & 8 \\
  $c_a\leq v_z \leq c_f$  & $L_3,L_5,{L_1},{L_2},L_6,L_4,L_8$ & ${\rho^\prime,\epsilon^\prime}, v_x^\prime, v_y^\prime,v_z^\prime,B_x^\prime,B_y^\prime,{B_z^\prime}$ & $L_7$ & 7 & 8\\
  $c_f\leq v_z$        & $L_7,L_3,L_5,{L_1},{L_2},L_6,L_4,L_8$ & ${\rho^\prime,\epsilon^\prime}, v_x^\prime, v_y^\prime,v_z^\prime,B_x^\prime,B_y^\prime,{B_z^\prime}$ & - & 8 & 8 \\
  \enddata
\end{deluxetable}

Table~\ref{tab:incoming} lists all the possibilities for inward and outward propagating modes as $v_z$ is varied at a single location on a $z_\text{min}$ boundary.
When the bulk velocity $v_z$ is directed outward (in the -$\zhat{}$ direction) and has magnitude greater than the fast mode speed ($v_z<-c_f$) then every eigenvalue is negative, and all the modes propagate outward from the volume of interest, as indicated in the first row of the Table.
In this case, whatever dynamics are happening inside the volume completely determine the behavior of the plasma in the boundary, while anything happening outside the boundary cannot influence the boundary or interior in any way.
Any independent estimate for the evolution of the boundary data cannot be matched unless it is already consistent with the outgoing information.

If the magnitude of the outflowing bulk velocity is decreased to less than the fast speed but still greater than the Alfv\'en speed ($-c_f<v_z<-c_a$) then the forward propagating fast mode $L_8$ has a positive eigenvalue $\lambda_8$, and is therefore inward propagating relative to the $z_\text{min}$ boundary, as indicated in the second row of Table~\ref{tab:incoming}.
From Equation \ref{eq:l8z} and \ref{eq:lz-columns}, we see that the $L_8$ characteristic derivative, constructed from the $\vect{l}_8$ eigenvector, couples together the derivatives of seven of the eight primitive variables: everything but the derivative of the normal component of the magnetic field, $\partial_zB_z\equiv B_z^\prime$.
As discussed previously and reemphasized here, the \emph{single} inward propagating fast mode with characteristic derivative $L_8$ will simultaneously modify seven of the eight primitive variables in the boundary, \emph{but not arbitrarily so.}
The seven variables move in lock-step with fixed relative amplitudes given by the components of the eighth right eigenvector $\vect{r}_8$, all scaled by the (scalar!) value of the eighth element of characteristic derivative, $L_8$.
The external universe dictates the value of that scalar.

At the next level of Table \ref{tab:incoming}, the magnitude of $v_z$ is greater than the slow speed and less than the Alfv\'en speed, but all three are still negative valued $(-c_a<v_z<-c_s)$.
Then, the forward-propagating Alfv\'en mode $L_4$ joins the forward propagating fast mode $L_8$ as a positive mode, which we identify as an incoming mode through the $z_\text{min}$ boundary.
As seen in Equations \eqref{eq:l4z} and \eqref{eq:lz-columns}, the Alfv\'en mode couples together the derivatives of four primitive variables, but still not the derivative of the normal magnetic field $B_z$.
As indicated in the third column of Table~\ref{tab:incoming}, there is no new dependence on the primitive variable derivatives compared to the previous row.
However, because there are now two inward propagating modes, the external universe can modify the boundary in new ways through the joint action of both incoming modes.
Thus, the space of allowed changes to the boundary caused by the external universe has increased, while the control the internal plasma exerts on the boundary has concomitantly decreased.
Stepping the bulk velocity across each of the eigenvalue speeds in turn, this process continues until the bulk velocity is inward directed at a magnitude exceeding the local fast mode velocity $c_f$, in which case all characteristic modes are inward propagating.
In that case, the external universe completely dictates the evolution of the plasma in the boundary, and (consistent with that) the internal dynamics of the plasma have no effect on the boundary whatsoever.

As a final note about propagation directions, the decomposition into positive/negative and outward/inward propagating information happens on a point-by-point and time-by-time basis throughout the plasma. 
In particular, in our example of a constant-$z$ plane $\vect{x}_b$, the number of inward propagating modes is some function $f(x,y,t)$ in that plane.

Returning to the expanded characteristic form of the MHD equations given at \eqref{eq:mhd-chars}, it is interesting to note that each component of the momentum equation depends only on the difference between forward-- and reverse--propagating pairs of characteristics in each family (e.g., the difference between fast modes $L_7-L_8$). 
In contrast, the continuity, energy, and induction equations only involve summations between the pairs in each family (e.g., $L_7+L_8$).
This same property holds when the transverse terms in $\vect{C}$ are fully expanded in characteristic form, that is, when expanding out the $x$ and $y$ terms in Equation \eqref{eq:mhd-char-full} as we did the $z$ terms in \eqref{eq:mhd-chars}.
This pairing of sums and differences within each family of characteristics allows their combinations to describe both the propagation of disturbances within a plasma as well as the spatial structure of the plasma itself.

A simple example illustrates this point best.
Hydrostatic equilibrium (where $\partial_tv_z=0$) can be interpreted as a carefully balanced set of propagating disturbances that sum to support large scale gradients in the density and energy (i.e., in the energy and continuity equations) while simultaneously canceling to zero out any forces, i.e., the terms in the momentum equation.
From the perspective of the characteristics, the stratification is due to two counter-propagating modes whose net transfer of momentum is precisely balanced by gravity to produce a static structure.
This behavior is analogous to a zero-frequency standing wave.

Finally, as mentioned above, the characteristic derivative $L_1$ is a somewhat special case.
Its role in the present formulation is to preserve the value of $\nabla\cdot\vect{B}$ at each spatial location.  
We included the empirical lack of magnetic monopoles ($\nabla\cdot\vect{B}=0$) implicitly in the induction equation \eqref{eq:induction}.
Thus, if a problem's initial condition is solenoidal then $L_1$ ensures that the solution remains solenoidal over time.
\par

\section{Characteristic boundary conditions: overview and notation}\label{sec:characteristic-bounds}
Setting boundary conditions using the characteristics amounts to setting the value of each characteristic derivative $L_\sigma$ associated with an incoming mode $\sigma$.
\citet{Thompson:1990} created a catalog of standard hydrodynamic BCs written in terms of characteristics (no--slip walls, hard walls, uniform inflow, non--reflecting, constant pressure, etc.) while \citet{Poinsot:1992} extended the method to include nonhyperbolic terms, e.g., viscous terms in the Navier--Stokes equations.
Typically, these boundary conditions are derived by first assuming a physical constraint (such as constant pressure), calculating the values for the boundary--normal spatial derivatives of primitive variables implied by that constraint, and then solving for the incoming characteristic derivatives in terms of those normal derivatives.
The examples in \citet{Thompson:1990} essentially amount to applying the transformation at \eqref{eq:l2uprime} in situations where an easily calculable constraint can be provided, but the situation becomes much more complicated for our use case of data-driven MHD.

Numerous other authors have used the characteristics to formulate boundary conditions that are approximately transparent to waves, so-called ``nonreflecting boundary conditions.'' 
\citet{Hedstrom:1979} appears to be the first to have done this for a general system of nonlinear hyperbolic equations (which includes MHD) by setting the amplitudes of incoming characteristic derivatives to zero (e.g, setting $L_8=0$ when the system is at level 1 of Table~\ref{tab:incoming}); \citet{Grappin:2000}, \citet{Landi:2005}, \citet{Grappin:2008}, or \citet{Gudiksen:2011} provide more recent examples in the context of heliophysics.
These (perhaps misnamed) nonreflecting boundary conditions are only truly nonreflecting for simple waves, but as \citet{Hedstrom:1979} already pointed out, slightly more complex situations such as a weak shock or the interaction of multiple simple waves begin to present a problem.  
Those cases change the external system in ways that should eventually produce reflections.
However, a fundamental assumption of the nonreflecting boundary conditions is that the external system is unchanged by the information flowing out of the boundary: it acts as a passive information reservoir that does not directly influence the behavior inside the simulation.

Of course, this is exactly the \emph{opposite} behavior we want for data driving, where the external system explicitly forces the dynamics of the internal system, often on a rapid timescale.
In fact, when applied to photosphere-to-corona simulations whose purpose is to accurately model processes like the emergence of active regions, a wave picture is fundamentally inadequate.
We need to model the structural change of the magnetic field, including changes in the topology, bulk material flows, and so on.
Since the region just inside the boundary is also evolving, the boundary layer experiences a strong forcing from both sides.
For data driving, we therefore need to be able to isolate and decompose the evolution of the observed boundary into the portion that is due to internal disturbances propagating outward and the portion due to external disturbances propagating inward.
This requires very clearly identifying the types of information flowing in all directions through a given point in space.

Some additional notation is necessary to explicitly describe the flow of information throughout the system.
We start again with the MHD equations written in characteristic form and focusing on the \zhat{} direction,
\begin{equation*}
\partial_t\mhdstate + \underbrace{\smat\cdot\charderiv}_{\Matrix{A}_z\cdot\partial_z\mhdstate} + \vect{C} =\vect{0}, \tag{\ref{eq:mhdbc} revisited}
\end{equation*}
where we have also included the equivalent expression in terms of the primitive variables, for reference.
The decomposition defined at Equation \ref{eq:diagonalization} can be used to split the eigensystem into two subspaces that represent the positive $(+)$ and negative $(-)$ propagation of information relative to the diagonalization direction \zhat{}.
The split can be represented in multiple equivalent ways, all based on separating the positive and negative eigenvalues in the diagonal \evalmat{} matrix:
\begin{equation}
  \evalmat = \evalmat_+ + \evalmat_-\quad \text{with}\quad \evalmat_\pm = \frac{1}{2}(\evalmat \pm \abs{\evalmat}).\label{eq:lambdapm}
\end{equation}
The $\evalmat_\pm$ are two $8\times 8$ diagonal matrices with only the positive or negative eigenvalues at appropriate locations, and zeros elsewhere.
The ``$+$'' matrix selects modes for which $\lambda_i>0$ and ``$-$'' matrix selects modes for which $\lambda_i<0$.
Characteristics for which $\lambda_i=0$ have zero amplitude, propagate no information, and therefore require no special treatment in this formalism; see, e.g., Equations \eqref{eq:li} and \eqref{eq:mhd-chars}.

From Equation \eqref{eq:lambdapm} we can also define the positive and negative projection matrices, or equivalently, the projection operators
\begin{equation}
    \pmatpm = \evalmat_\pm\evalmat^{-1}\label{eq:Ppm},
\end{equation}
where $\evalmat^{-1}$ is again defined using the pseudoinverse.
The $\pmatp$ and $\pmatm$ operators project any object in MHD statespace onto the positive and negative orthogonal subspaces, respectively.
Like all orthogonal projection operators, these matrices have the property that repeated projection has no effect,
\begin{gather}
    \pmatp\pmatp = \pmatp,\quad \pmatm\pmatm = \pmatm,\label{eq:repeatprojections}
    \intertext{while orthogonality ensures that} 
    \pmatp\pmatm = \pmatm\pmatp = \Matrix{0}.
\end{gather}
Thus defined, the projection operators can be used as needed.

Equation \eqref{eq:Ppm} can be inverted to read
\begin{equation}
    \evalmat_\pm = \evalmat\pmatpm = \pmatpm\evalmat.
\end{equation}
Using these projection operators, the original coefficient matrices $\amat$ of the MHD Equations \eqref{eq:mhd-matrix} can be directionally split into the positive and negative coefficient matrices that govern the flow of information in the grid-positive and grid-negative directions:
\begin{gather}
    \amat = \amatp + \amatm = \smat(\evalmat_+ + \evalmat_-)\smatinv\label{eq:Apm}.
\end{gather}
Equivalently, the left and right eigenvector matrices can be split into positive and negative portions:
\begin{gather}
    \smat = \smat_+ + \smat_- = \smat\pmatp + \smat\pmatm\label{eq:s-posneg}\\
    \smatinv = \smat_+^{-1} + \smat_-^{-1} = \pmatp\smatinv + \pmatm\smatinv\label{eq:sinv-posneg},
\end{gather}
as well as the characteristic vectors:
\begin{gather}
    \charderiv_{z,\pm} = \pmat_{z,\pm}\charderiv_z=\pmat_{z,\pm}\smatinv_z\amat_z\partial_z\mhdstate{} \label{eq:l-posneg},
\end{gather}
where we have explicitly labeled a normalization direction ($z$) in this last equation.
Similar expressions for all of the above equations apply in the $x$ and $y$ directions, as well.

The eigensystem describes a basis for information propagation, and therefore, once a subspace has been selected for any element of the eigensystem it automatically applies to all elements of the system by the associative property (this follows from repeated projections via \eqref{eq:repeatprojections}).
That is to say, $\smat\charderiv_\pm = \smat_\pm\charderiv = \smat_\pm \charderiv_\pm = \pmatpm\smat\charderiv$.

The orthogonality of the positive and negative subspaces means that the elements of either system do not interact, for example, $\smat_\mp\charderiv_\pm = \vect{0}$.
What this means in terms of the MHD equations is that the spatial derivatives calculated inside the characteristic vectors (the $\partial_z U_\zeta$ terms in \eqref{eq:li-index}) are all taken in the upwind direction: the positive propagating modes have characteristic derivatives defined from below while negative propagating modes have characteristic derivatives defined from above\footnote{More generally, for a right handed coordinate system the positive (negative) propagating modes have derivatives defined from the left (right) in the standard sense along each of the coordinate axes, respectively.}.
For smooth systems these derivatives will be equal, but the characteristics intrinsically describe discontinuous solutions to the MHD equations, as well.

As described just after Equation \eqref{eq:aux-var}, we use subscripts to represent the subpaces associated with inward propagating (\incoming{}) and outward propagating (\outgoing{}) information through a boundary.
For instance, as indicated in Figure \ref{fig:axes-propagation}, at a $z_\text{min}$-boundary we identify inward with the positive subspace and outward with the negative subspace, $\smati = \smat_+$ and $\smato=\smat_-$, respectively.  
In general, the $+$ and $-$ notation applies anywhere in space, while we reserve the $\incoming$ and $\outgoing$ notation for a specified boundary.

As an example of how the projection onto subspaces works in practice, suppose a location on the $z_\text{min}$ boundary has an outward $(v_z<0)$ bulk velocity with amplitude less than the slow-mode speed ($-c_s \leq v_z \leq 0$). 
Then, Table~\ref{tab:incoming} shows that $L_6,\ L_4,$ and $L_8$ correspond to inward propagating modes, meaning that rows 6, 4, and 8 of the matrix $\smatinv_+ = \smatinvi$ are given as in Equation \eqref{eq:szm1}; the remaining rows contain zeros, because the projection $\pmat_+\smatinv$ gives zero for all those elements.
Similarly, $L_7,\ L_3,\ L_5,\ L_1,$ and $L_2$ correspond to outward propagating modes, the corresponding rows in $\smatinvo$ are given as in Equation \eqref{eq:szm1}, and the remaining three rows contain only zeros.
With the above notation in hand we are well equipped to describe the flow of information anywhere within the system, including any boundary.

\section{Data-driven boundary conditions: continuous case}\label{sec:ddbcs}
The data-driven boundary condition (\ddbc) problem can be stated concisely as follows: How is an MHD system evolved \emph{consistently} from $t$ to $t+\dtdata$ given the MHD state vector everywhere in the system at the \emph{initial} time $t$, i.e., $\mhdstate\left(t,\vect{x}\right)$, and the MHD state vector at the boundary(ies) \xb{} at the \emph{new} time $t+\dtdata$, i.e., $\mhdstate\left(t+\dtdata,\xb\right)$? 
Traditional boundary conditions set values and/or derivatives of the MHD primitive variables on the boundary, i.e., Dirichlet or Neumann boundary conditions. 
These standard boundary conditions effectively determine the characteristic derivatives $\charderiv$ from the values in the domain and boundary at time $t$. 
The boundary state evolves purely in terms of values determined\footnote{Here we have ignored the details and stability of the numerical time advance.  For example, predictor/corrector or implicit schemes \emph{appear} to formally use information from future times, e.g., $\mhdstate(\xb,t+\dtdata/2)$, to advance the MHD state, but in fact these estimates of intermediate MHD states are calculated deterministically from the values of $\mhdstate(\xb,t)$.} at time $t$
\begin{equation}
 \partial_t \mhdstate(\xb,t)=-\smat\cdot\charderiv(\xb,t) -\vect{C}(\xb,t)\Longrightarrow\mhdstate(\xb,t+\dtdata).\label{eq:Ut}
\end{equation}
The philosophy here is that $\mhdstate(\vect{x},t)$, $\smat$, $\charderiv(\xb,t)$, and $\vect{C}(\xb,t)$ are known and used to calculate the final unknown boundary state $\mhdstate(\xb,t+\dtdata)$.

 From the perspective of the method of characteristics, these standard boundary conditions determine the incoming characteristics purely from the evolution of the interior of the domain.
 For example, homogeneous Neumann boundary conditions could be used to determine the values of the primitive variables outside the simulation in terms of primitive variables inside the simulation by requiring that $\partial \vect{v}/\partial z|_{\xb}=0$.
 Traditional boundary conditions therefore correspond to an assumption (often \textit{ad hoc}) about the evolution of the external universe in response to the evolution of the MHD system.\par

In contrast to traditional boundary conditions described above, for \ddbc{}s we do not want to determine the incoming characteristics $\charderiv_\incoming(\xb,t)$ from the evolution of the interior of the domain because we want the external universe to inject \emph{independent information} into the MHD system through the boundary.
Instead we want to choose the incoming characteristics at the boundary such that the superposition of the incoming and outgoing characteristics leads to evolution of the boundary from $\mhdstate\left(t,\xb\right)$ towards the prescribed target state $\mhdstate\left(t+\dtdata,\xb\right)$:
\begin{equation}
    \smat\cdot\charderiv(t,\xb)=-\partial_t \mhdstate(t,\xb) -\vect{C}(t\xb)\Longrightarrow\mhdstate(t+\dtdata,\xb).\label{eq:L}
\end{equation}
The philosophy here is that $\partial_t\mhdstate(t,\xb)$, and through it the final state of the boundary $\mhdstate(t+\dtdata,\xb)$, and $\vect{C}(t,\xb)$ are known and used to calculate the best estimate of the incoming characteristic derivatives $\charderiv(t,\xb)$ to evolve the boundary state \emph{towards} $\mhdstate(t+\dtdata,\xb)$.
Put another way, the time evolution of the boundary $\partial_t \mhdstate(t,\xb)$ is known independently of~(\ref{eq:Ut}) and (conveniently) the MHD equations as represented in Equation \eqref{eq:mhdbc} in terms of the characteristic derivatives $\charderiv$ in Equation \eqref{eq:li-index} give no preference to either the temporal or the spatial derivatives, which means the unknown spatial derivatives can be solved for.
The result is that, by carefully and correctly choosing the spatial derivatives one can evolve the system to match the estimated boundary evolution.
We again stress that, by using the characteristic derivatives rather than the primitive variables directly, we are restricting ourselves to the subset of boundary evolutions allowed by MHD.
This will result in exact matching of the estimated boundary evolution provided the estimated evolution is already self-consistent with the MHD equations, but will only \emph{approximately} match when the estimated evolution contains errors.

The projection operator formalism introduced in the preceding \S\ref{sec:characteristic-bounds} can be applied to make these arguments mathematically rigorous. 
Equation \eqref{eq:Ut} can be rearranged to determine the spatial derivatives (via characteristic derivatives $\charderiv$) necessary to achieve the time derivative $\partial_t \mhdstate(t,\xb)$ which will produce the (possibly approximate) transformation $\mhdstate\left(t,\xb\right)\rightarrow\mhdstate\left(t+\dtdata,\xb\right)$.
To achieve this we substitute the decomposition \eqref{eq:s-posneg} into \eqref{eq:Ut}, multiply through by $\smatinv$, and use the fact that $\smat^{-1}\smat_{\incoming,\outgoing} = \pmatio$ to get
\begin{gather}
  \smatinv\partial_t\mhdstate + \pmati\charderiv + \pmato\charderiv + \smatinv\vect{C} = \vect{0},
\end{gather}
where each term in the expression is evaluated at the boundary \xb.
Multiplying through by \pmati{} the above equation decomposes completely for the incoming system
\begin{gather}
  \charderivi = -\smatinvi\bigl(\partial_t\mhdstate + \vect{C}\bigr)\label{eq:linc};
  \intertext{for completeness we could multiply by \pmato{} to do the same for the outgoing system}
  \charderivo = -\smatinvo\bigl(\partial_t\mhdstate + \vect{C}\bigr)\label{eq:lout},
\end{gather}
but the latter has no explicit dependence on anything outside of the domain and therefore it cannot be used to inject independent information into the domain.
Of course, Equation \eqref{eq:linc} is only rigorously valid when our estimate for $\partial_t \mhdstate(t,\xb)$ is completely consistent with the evolution of the MHD system.
This is unlikely when $\mhdstate\left(t,\xb\right)$ corresponds to observations.
We discuss this point in the next section, which focuses on the numerical integration in the discrete case.

We can also transform \eqref{eq:linc} into the space of spatial derivatives of primitive variables by substituting in for $\charderivzi = \evalmat\pmati\smatinv\partial_z\mhdstate$ and then solving for those derivatives, after which we find (written in several equivalent forms)
\begin{subequations}
    \begin{align}
        \partial_z\mhdstate & = -(\smat\evalmat\pmati\smatinv)^{-1}\bigl(\partial_t\mhdstate+\vect{C}\bigr)\\
        & = -\smat\evalmati^{-1}\smatinv\bigl(\partial_t\mhdstate+\vect{C}\bigr)\\
        & = -\amati^{-1}\bigl(\partial_t\mhdstate+\vect{C}\bigr)\label{eq:dzu-incoming}.
    \end{align}
\end{subequations}
Formally, the continuous problem is now solved: if we know the time derivative $\partial_t\mhdstate(t,\xb)$ and the transverse and inhomogeneous terms $\vect{C}(t,\xb)$ then we can directly solve for the (external, one-sided) spatial derivatives $\partial_z\mhdstate(t,\xb)$.
What's more, we can do so either in terms of the characteristic derivatives of inward propagating modes through Equation \eqref{eq:linc} or directly in terms of spatial derivatives of the primitive variables via Equation \eqref{eq:dzu-incoming}.
In the latter case, one cannot arbitrarily set derivatives of primitive variables but is instead restricted to the subspace of spatial derivatives that are linear combinations of the characteristic derivatives of inward propagating modes.
This property is enforced by the inverse of the positive subspace matrix, $\amati^{-1} = \smat\pmati\evalmat^{-1}\smatinv$.
On the other hand, if a given primitive variable is not a function of any incoming characteristic then its exterior derivative is unconstrained: it has no influence on the dynamics at or within the boundary and can therefore be set arbitrarily.
The most extreme example of this is when material leaves a boundary faster than the fast mode speed, so that all the characteristics are outward propagating.
In that case, it doesn't matter how the variables are set outside the simulation because there is no incoming information.
This is a common situation, for instance, at the outer boundary of solar wind simulations \citep{Grappin:2000}.

To complete the description of a characteristics-based boundary condition, the incoming characteristic derivatives are calculated using Equation \eqref{eq:linc} (or equivalently for many simulations, the values for simulation ghost cells are calculated using \eqref{eq:dzu-incoming}) while the outgoing characteristic derivatives $L_{\outgoing{},\sigma}$ (given in Table \ref{tab:incoming}) are still calculated using equations \eqref{eq:li}.
Then, all the $L_\sigma$ are substituted into the MHD equations \eqref{eq:Ut} and the boundary values are advanced in time by standard time--integration methods.
Because the MHD state vector on the boundary $\mhdstate(t,\xb)$ varies as a function of space, different locations on the boundary will simultaneously sit at different rows of Table~\ref{tab:incoming} and therefore have different numbers of incoming and outgoing characteristics.
In the same way, the number of incoming characteristics at a given location will vary as a function of time as the system evolves.  

Several points are worth emphasizing at this juncture:
\begin{itemize}
  {\item The MHD equations give no preference to either spatial or temporal derivatives.  If some subset of those derivatives are known, the others can be solved for.}
  {\item  There is no one-to-one mapping between characteristics and primitive variables, so arbitrary constraints on the primitive variables could lead to a consistent, under-, or over-determined system of equations when applied to MHD.}
  {\item Boundary conditions on the primitive variables that are also consistent with the MHD equations are necessarily expressible in terms of incoming characteristics only.}
  {\item Possibly the only guaranteed safe way to apply arbitrary boundary conditions is via the method of characteristics.}
\end{itemize}
The first point means we can use an estimate for the temporal evolution of the primitive variables on a boundary---for instance from a time series of observations---to find a set of incoming characteristic derivatives that are maximally consistent with the observed evolution at the boundary; this amounts to exchanging known temporal derivatives for unknown spatial derivatives in the MHD equations.
However, the second point cautions that the spatial derivatives of multiple primitive variables are simultaneously constrained by a single incoming characteristic, and the derivative of a single primitive variable simultaneously contributes to multiple characteristics: the derivatives are coupled together.
While the primitive variables themselves make no reference to the amount of independent information that can be self-consistently supplied to the system, this is exactly what the characteristics provide.
The number of incoming characteristics for a given boundary state $\mhdstate(t,\xb)$ is equivalent to the amount of independent information that can be supplied to the system.
The incoming eigvenvectors then give the precise, coupled forms that incoming information can take for an MHD system.
This leads to our third point, that any self-consistent boundary condition must be expressible in terms of the incoming characteristics.
Conversely, if a boundary condition is not expressible in terms of only incoming characteristics then the system being described is either under- or over- determined.
Finally, based on the above chain of logic, we posit that the characteristics likely provide the only generic method to guarantee that an arbitrary boundary condition will satisfy the MHD equations.

\section{Data-driven boundary conditions: finite volume discretization, time-integration, and data-optimization}\label{sec:fvddbc}
\begin{figure}
    \centering
    \includegraphics[width=0.9\textwidth]{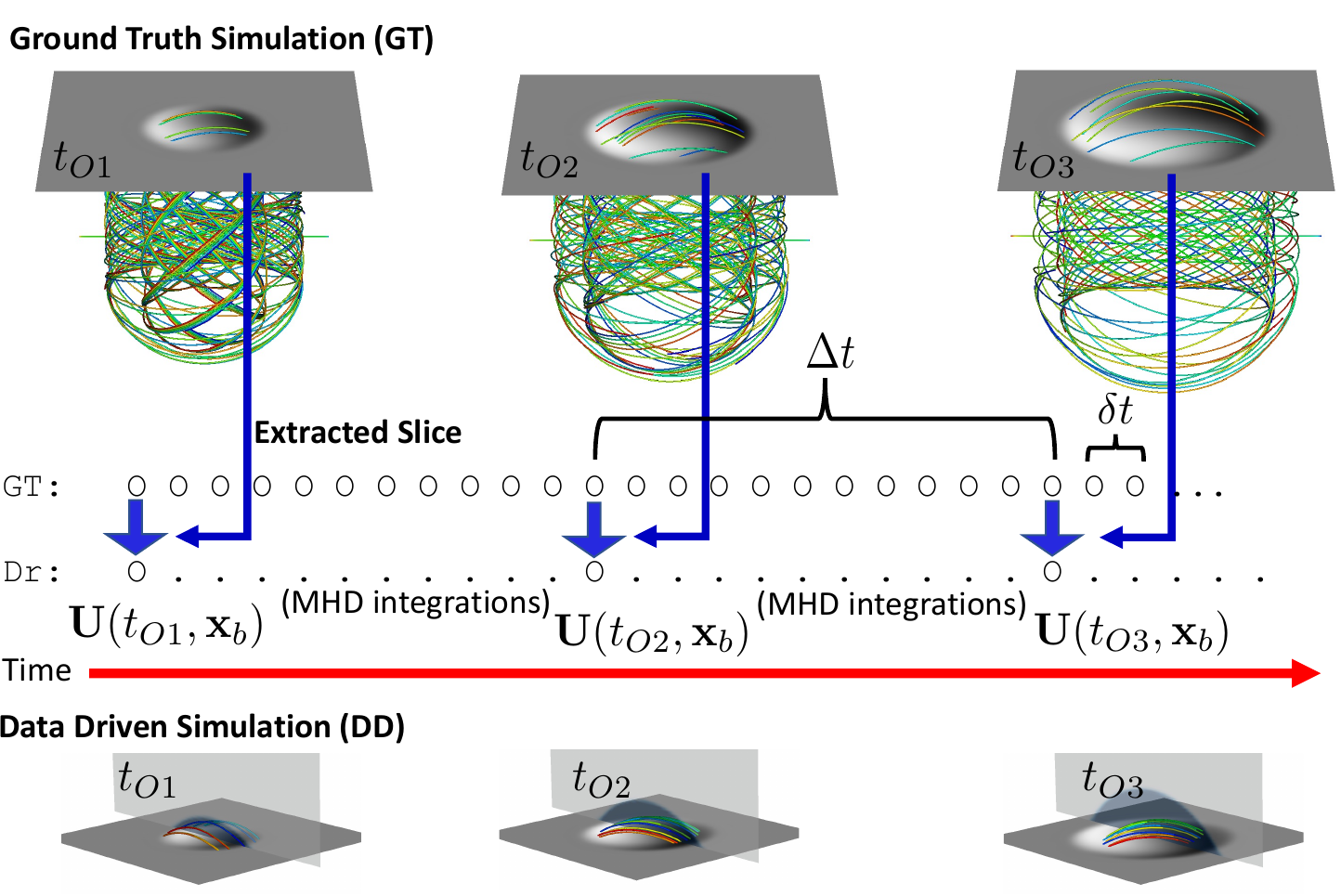}
    \caption{Schematic diagram of the data driving algorithm.  The upper row shows several times from a \emph{Ground Truth (GT)} simulation and highlights the vertical magnetic field in the layer that will be used for data driving (gray scale). Time flows left to right.  The time between each state O in the \emph{GT} simulation is \dtsim{} while the time between the extracted data series used for driving ($t_{O1}, t_{O2}, \ldots$) is \dtdata{}.  Blue arrows represent the intermittent injection of data extracted (or ``observed'') from the \emph{GT} simulation and used in the \emph{Driving} simulation (Dr)  The bottom row shows several times from the resulting \emph{Data-Driven} simulation from a side-on 3D perspective which shows vertical velocities in the vertical cut-plane and the magnetic field in the driving plane.  Field lines highlight the 3D magnetic structure for each simulation.  See text for the full relation between the \emph{GT}, \emph{Driving}, and \emph{Data-Driven} simulations.}
    \label{fig:schematic}
\end{figure}

Thus far we have described the problem of data-driven simulations with only minimal reference to the specifics of either the observations or the simulations.
Those details must now be brought to the forefront, but we keep the discussion as general as possible so that the technique may later be applied to diverse observational data sets or numerical schemes.
Our goal in the present investigation is to validate our data driving approach, and therefore an independently run ``ground truth'' simulation with a larger domain will serve as a stand-in for observations used to drive the boundary.
Figure~\ref{fig:schematic} provides a cartoon-overview of our data driving algorithm and its validation using the independent Ground Truth simulation, \emph{GT}.

In our numerical formulation, data driving results from performing a special time integration of the MHD equations in a single layer of driving cells.
In this section we describe the details of that time integration, where the driving layer is taken to be the $k=0$ cell index in the $z$-direction.
The integration extends temporally between one observation of the boundary at, say, time $\tstart$ and the next observation at time $\tend$, as indicated in Figure~\ref{fig:schematic}.
It is thus the numerical implementation of our data-driven boundary condition, \ddbc{}, which solves for the set of incoming characteristic derivatives at each space-time point on the driven boundary, $\charderivzi(t,\xb)$.
At the heart of this task is an optimization approach that finds the values of incoming characteristic derivatives that achieve the best possible consistency between the observations and the MHD equations.
Those incoming characteristic derivatives then determine the optimal evolution of primitive variables in the $k=0$ driving layer given the limited set of temporal observations of the MHD variables at that layer, i.e., the intermittently available observations (or the intermittently extracted driving layer from the \emph{GT} simulation). 
The use of intermittent data is depicted as blue arrows in Figure \ref{fig:schematic}.
As will become apparent, the boundary condition is effectively applied at the $k=-\frac{1}{2}$ interface, so the $k=-1$ cell is a ghost cell that does not directly come into play in our data driving algorithm; it is, however, used to couple separate simulations together.

In more detail, our implementation of data driving uses two simulations that partially overlap in space: the \emph{Driving} simulation and the \emph{Data-Driven} simulation.
The \emph{Driving} simulation spans a narrow band in height and is used exclusively to update the driving layer.
The \emph{Data-Driven} simulation spans the entire coronal volume above the driving layer and is numerically coupled to the \emph{Driving} simulation.
In addition to these, the present work also includes a third simulation, the \emph{GT} simulation, that is a stand-in for the observations.
This latter simulation does not exist when applying our method to the real Sun.

Figure~\ref{fig:schematic} sketches the relation between the Ground Truth simulation (top row), the time-integration of the MHD equations in the $k=0$ driving layer in the \emph{Driving} simulation (middle region), and the resulting \emph{Data-Driven} simulation (bottom row).
Full details of the simulations are given in \S\ref{sec:testcase}.
For now, the important points are that: (1) the driving layer sits at cell center index $k=0$; (2) the logically separate \emph{Driving} simulation (represented by the $\fnc{Dr: O\ldots\ldots O}$ row of the Figure), spanning $-1\leq k \leq 2$ in the $z$-direction, is used to numerically update in the $k=0$ driving layer ($k=-1$ is a ghost-cell for this simulation); (3) the \emph{Data-Driven} simulation (bottom portion of the Figure), spanning cells $1\leq k \leq N_z$ in the $z$ direction with numerical lower boundary at $k=\frac{1}{2}$, represents the simulated photosphere-to-corona MHD system above the driving layer; and (4) the \emph{Ground Truth (GT)} simulation (top portion of the Figure) spans the much larger range $k_\text{min, GT}<k<N_z$ with $k_\text{min, GT}\ll-1$, from which the driving layer at $k=0$ is occasionally extracted and supplied to the \emph{Driving} simulation.
A more realistic \emph{GT} simulation would span from the convection zone to the corona, similar to what was used in \citet{Toriumi:2020}.

\begin{figure}
    \centering
    \includegraphics[width=0.5\textwidth]{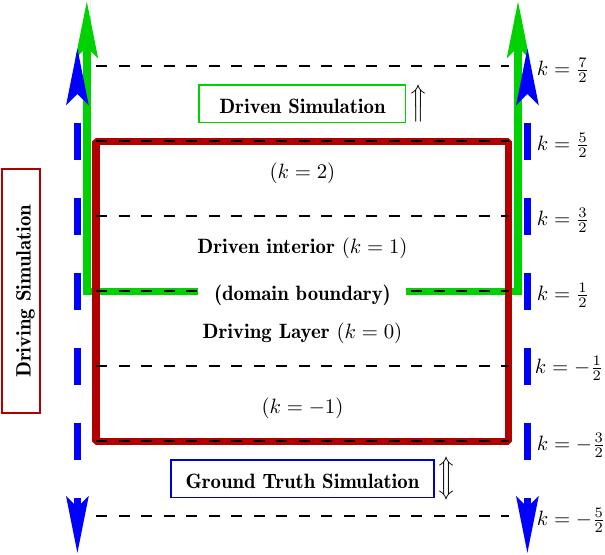}
    \caption{Relationship between the vertical grids ($z$ direction, indexed by $k$) in the three overlapping simulations.  The Ground Truth simulation (\emph{GT} ; blue) extends over a large domain.  The Driven Simulation (green) extends over the upper portion of the \emph{GT} domain.  The \emph{Driving} simulation (red) extends for four full computational cells that span the boundary of the Driven simulation.  Cell faces are shown as horizontal dashed lines.  Cell center locations are given integer indices and cell interfaces are given half-integer indices.  The $k=0$ cell-center layer is the driving layer and $k=\frac{1}{2}$ is the interface boundary of the Driven domain. The desired incoming characteristics are solved for at the $k=-\frac{1}{2}$ interface, $\charderiv_{z,\incoming}(t,i,j,k=-\frac{1}{2})$.
    }
    \label{fig:grid}
\end{figure}

Figure~\ref{fig:grid} describes the vertical overlap of numerical grids in the vicinity of the driving layer.
All three simulations use a common set of indices, i.e., a layer $k$ refers to the same height in all three simulations.
We use integer valued indices $(i,j,k)$ for cell centered locations and half-integer values for cell faces (e.g., the left and right $x$-faces $i\pm\frac{1}{2}$, front and back $y$-faces $j\pm\frac{1}{2}$, and top and bottom $z$-faces $k\pm\frac{1}{2}$).
In the figure, blue regions represent the extent of the \emph{GT} simulation, red the \emph{Driving} simulation, and green the \emph{Data-Driven} simulation.
Horizontal black dashed lines show the cell interface locations and are labeled by their half-integer indices in the $z$ direction.

Our numerical scheme to implement \ddbc{}s is ultimately cast in finite-volume form, but this specific choice is not required for characteristic based boundary conditions in general or data-driven boundary conditions in particular\footnote{Indeed, we originally implemented a finite-difference scheme, which worked well in some instances, but not others.}.
We call the characteristic-based finite-volume code \charc{}, and our implementation of data-driven boundary conditions in this code \charc-\ddbc{}.

In brief, a small-domain \emph{Driving} MHD simulation (represented by the dots in the \fnc{Dr} line in Figure~\ref{fig:schematic} and by the red region in \figref{fig:grid}) runs currently with and coupled to the main \emph{Data-Driven} MHD simulation (green region in Figure~/\ref{fig:schematic}).
The \emph{Driving} simulation is where the \ddbc{} is actually implemented.
Appropriate averaging of spatially overlapping cells couples the two simulations together.
In a real-world scenario these are the only two simulations that would run, and observational data would be ingested into cell $k=0$ of the \emph{Driving} simulation.
For testing purposes, a third MHD simulation, (\emph{GT}; blue), is used to extract synthetic observations that serve as input to the \emph{Driving} simulation.
Direct comparison between the \emph{GT} simulation and the \emph{Data-Driven} simulation in the cells $k\geq 0$ is used to assess how accurately the \ddbc{} performs. 
In practical application to data-driven simulations of the Sun, the extracted layer from the \emph{GT} will be replaced with a time series of plasma parameters derived from observations of the solar photosphere.
The main benefit of the logical separation into three simulations is that it keeps the key elements of our data driving method distinct from a specific numerical code and makes the method adaptable to a wide variety of numerical techniques.

In addition to the spatial dimensions, there are two time sequences to consider.
First, there is the time step between one observation and the next observation, \dtdata.
These are the only times at which boundary data (i.e., the driving layer) is known, $\mhdstate(t,\xb)\ldots\mhdstate(t+\dtdata,\xb)$.
In Figure~\ref{fig:schematic} these are represented, for example, by $t_{O1},\ t_{O2},\ldots$ with $\dtdata = t_{O2}-t_{O1}$.
Second, there is the simulation time step $\dtsim$, given by the simulation's Courant condition, which is typically much shorter than the cadence of boundary observations.  
For example, the standard HMI data series \texttt{hmi.ME\_720s\_fd10} provides vector magnetic field data at a cadence of 720 seconds, and the less frequently available high-cadence HMI data series provides data at 130 seconds, while a typical MHD code may have a Courant limited timestep of 1 second or less, depending on the simulation parameters.
The upper portion of \figref{fig:schematic} shows several times from the \emph{GT} simulation, the middle portion represents the occasional extraction (at cadence $\dtdata$) of the $k=0$ layer from the \emph{GT} simulation and its use in the driving simulation, and the lower portion shows the result of using the driving layer to produce the driven simulation in the reduced spatial domain.

In general there will be \totsteps{} simulation steps in the time \dtdata{} between one observation and the next. 
The driving simulation performs the optimization procedure described below $\totsteps$ times to determine the incoming characteristic derivatives at each time and integrates the MHD equations using the calculated derivatives.
This process will drive the \emph{Data-Driven} simulation, in a self--consistent way, between the two observation times.
Those $\totsteps$ steps comprise one observation cycle, and we repeat this process for as many observational cycles as exist.

In the following, we only need to describe a single observational cycle, which has an observation at the start of the cycle at time \tstart{} and an observation at the end of the cycle at time \tend{}.
The total duration of the cycle between two observations is $\dtdata=\tend-\tstart$.
The $\eta^\text{th}$ simulation step from one observed time to the next we designate by \tspj{}, where $0\leq\eta\leq\totsteps-1$, so that \tcycleend{}=\tend{}.
A single application of the optimization method is used to update the driving layer at $k=0$ between \tspj{} and \tspjpone{}, i.e., for the Courant-limited simulation time step $\dtsim = \tspjpone-\tspj$.
The optimization method is repeated until the \emph{Data-Driven} simulation has been numerically integrated from \tstart{} to \tend{}.
Note that neither \dtsim{} nor $\totsteps$ is known in advance, but this is unimportant.\par

We designate the observed state in the driving layer with a hat, $\observedstate$, so a single observational cycle is bounded by $\observedstate(\tstart,\xb)$ and $\observedstate(\tend,\xb)$.
The \emph{Data-Driven} simulation's boundary state may not exactly match any of the observed states, so that in general $\mhdstate(\tstart,\xb) \ne \observedstate(\tstart,\xb)$.
We wish to drive the current state $\mhdstate(\tstart,\xb)$ toward $\observedstate(\tend,\xb)$ as best we can in a ``hound chasing the hare'' fashion.

The observations in the driving layer allow us to estimate the time derivative we are attempting to reproduce by the boundary driving, i.e., this is the \emph{independent estimate} discussed in \S\ref{sec:ddbcs} in connection with Equation \eqref{eq:L}.
We designate the target time derivative by an over-bar, \targettimederiv{}, where the over-dot denotes a generic time derivative.
The simplest estimate is a linear interpolation over the two observed states,
\begin{equation}
  \dUdtest\approx\dUdtest{}_\text{simple}= \frac{\UBtendobs-\UBtstartobs}{\dtdata},
\end{equation}
so that the estimated time derivative is constant over the cycle.
A somewhat better approximation is also a linear estimate but uses the current value of the state variable (at step $\ts$) instead of the value at the beginning of the observation cycle:
\begin{equation}
  \dUdtest\approx\dUdtest{}_\text{current time}=\frac{\UBtendobs-\Uj}{\tend - \tspj}.\label{eq:targettimederiv}
\end{equation}
This has the effect of constantly nudging the current solution in the best direction to achieve the target final state \UBtendobs.
Higher order estimates could be explored, but we will take $\targettimederiv{} = \dUdtest{}_\text{current time}$ from here on.

Regardless of how it is estimated, \targettimederiv{} provides a numerical target value for the boundary state vector at the next simulation time, $t = \tspj+\dtsim$, for instance using a forward Euler update
\begin{equation}
  \targetstate(\tspj+\dtsim,\xb) = \UBtime{\tspj} +\dtsim\targettimederiv(\tspj,\xb).\label{eq:targetboundarystate}
\end{equation}
Just as for the time derivative, we could also use a higher order integration method to define a higher order estimate for the target boundary state.
But again, in this investigation we simply use Equation~\eqref{eq:targetboundarystate} to define the target boundary state.

\begin{figure}
    \centering
    \includegraphics[width=0.7\textwidth]{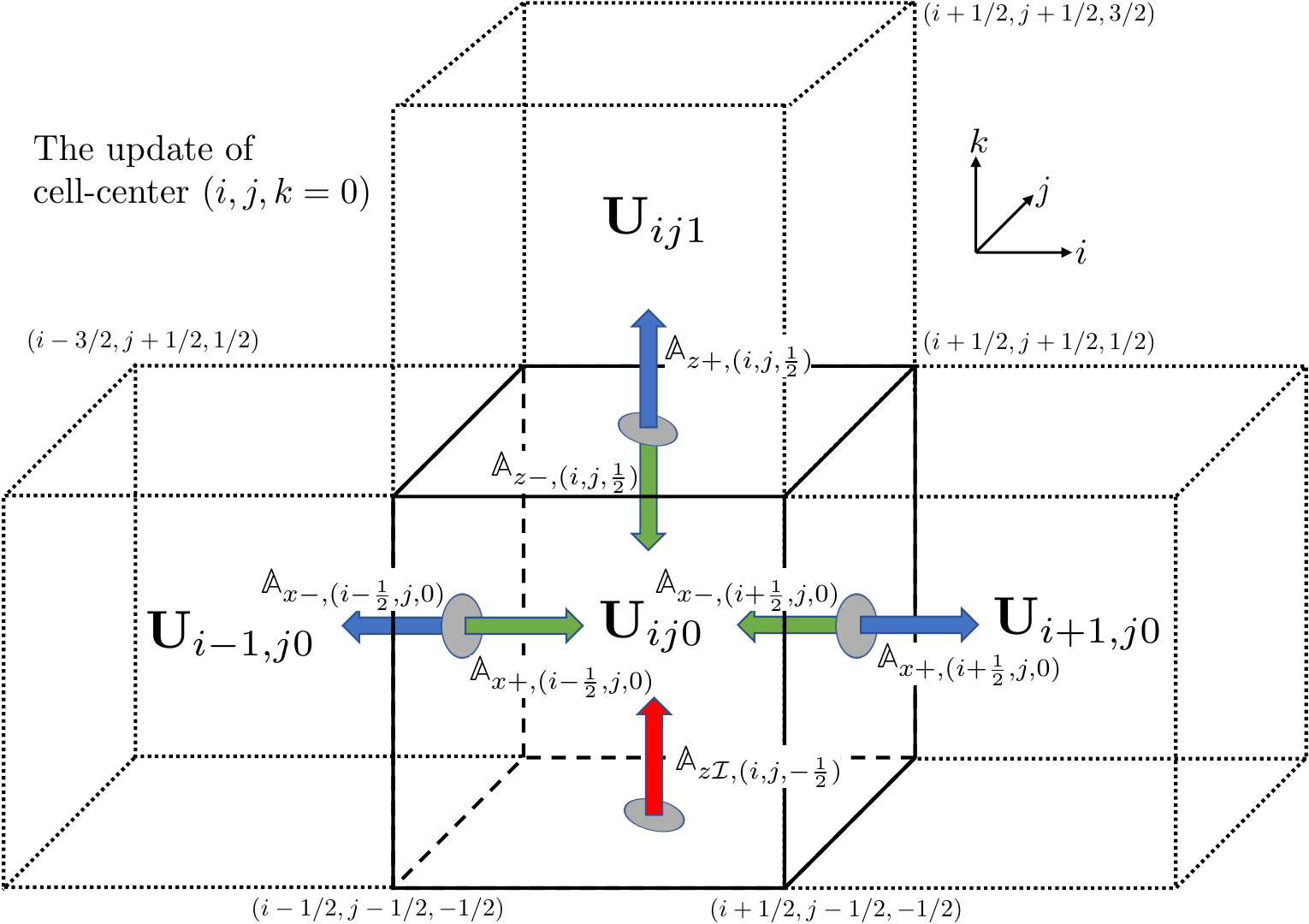}
    \caption{Key elements of finite volume scheme to update cell-centered primitive variables $\mhdstate_{ij0}$ in the driving layer $k=0$.  The directionally split coefficient matrices, $\amat_\pm$, are depicted at each interface (the y-faces are omitted for clarity).  Integer and half-integer indices refer to cell center and cell face locations, respectively, and several sets of vertex indices are included as examples.  Matrices associated with green arrows represent information coming into to the $ij0$ cell and those associated with blue arrows represent information leaving that cell.  The matrix at the lower face ($k=-\frac{1}{2}$) is given a red arrow to indicate that it needs to be calculated using independent information; i.e., the incoming information represented by $\amat_{\incoming,(i,j,-\frac{1}{2})}$ is the boundary condition.  The $k=-1$ cell is not shown as it is not used explicitly in the current numerical scheme (see discussion at the end of \S\ref{sec:fvddbc}).}
    \label{fig:fv_schematic}
\end{figure}

With the target state and target time derivative defined, we turn to the details of integrating the cells in the driving layer forward in time in a way that is consistent with both the estimated target states and the directional flow of information given by the characteristics.
To achieve this, we apply the characteristic eigendecomposition and projection formalism introduced in \S\ref{sec:characteristic-mhd} and \S\ref{sec:characteristic-bounds} to each face of a computational cell and thereby develop a finite volume scheme based on the characteristics, \charc{}.
This essentially treats each cell face as a boundary and we cyclically apply our formalism at each face in order to update the cell center values of the primitive variables.
By itself, this produces a fully functional 3D MHD code that works anywhere in a simulation interior.
Data driving results from applying the formalism developed in \S\ref{sec:ddbcs} to the one cell face through which new information enters the system: the lower $z$ face in the driving layer.
That new information is, of course, the target time derivative \targettimederiv{} just described (or, equivalently, the target state \targetstate).

\figref{fig:fv_schematic} depicts the \charc{} numerical scheme applied to a computational cell in driving layer $k=0$.
The MHD variables in \charc{} are cell averaged quantities that live at cell centers, $\mhdstate_{ijk} = \mhdstate(\vect{x}_{i,j,k}) = \mhdstate(x_i, y_j, z_k)$.
The highlighted cell in the figure is an element of the driving layer at $k=0$, so $\mhdstate_{ij0} = \mhdstate(\xb) = \mhdstate(x_i, y_j, z_b)$.
The time update to a given cell uses information from the six surrounding cell centers, as well as the average state at each cell interface (the $y$ direction is omitted, for clarity).
As mentioned above, the cell faces are identified by half-integer index labels, e.g., the right-side face of cell $(i,j,k)$ is labeled $(i+\frac{1}{2},j,k)$.
The objects in the Figure provide several more examples; for example, the right-back-top vertex of the highlighted $ij0$ cell is labeled $(i+\frac{1}{2},j+\frac{1}{2},\frac{1}{2})$.

The flow of information across each cell interface is defined in terms of the eigendecomposition described in \S\ref{sec:characteristic-bounds} that splits the system into positive and negative subspaces at each interface in the face-normal (\nhat) direction.
Each eigendecomposition is calculated using the \emph{interface} MHD state, defined as the average of the two surrounding cell centered states:
\begin{equation}
  \mhdstate_{n\pm\frac{1}{2}} = \frac{1}{2}(\mhdstate_n + \mhdstate_{n\pm1}),\qquad n\in\{i,j,k\}\label{eq:interfacestate}.
\end{equation}
For example, at the $i+\frac{1}{2}$ face, whose normal direction is \xhat{}, the eigendecomposition is
\begin{align}
  \Matrix{A}_{x,i+\frac{1}{2}} = \smat_{x,i+\frac{1}{2}} \evalmat_{x,i+\frac{1}{2}} \smatinv_{x,i+\frac{1}{2}}
  = \Matrix{A}_{x+,i+\frac{1}{2}} + \Matrix{A}_{x-,i+\frac{1}{2}},
\end{align}
where all versions of the $\amat_x$, $\smat_x$, and $\evalmat_x$ matrices are constructed using the $x$-interface-averaged MHD state $\mhdstate{}_{i+\frac{1}{2}, j,k}$.
The second equality shows the coefficient matrix $\amat_x=\amat_{x+}+\amat_{x-}$ split by projecting onto the positive and negative subspaces as in Equation \eqref{eq:Apm} (recall from the discussion following Equation \eqref{eq:l-posneg} that the projection of any of $\amat$, $\evalmat$, $\smat$, or $\charderiv$ onto a subspace necessarily projects the other terms onto the same subspace in the same expression).

In Figure~\ref{fig:fv_schematic}, the split at each cell face is represented by different colored arrows (for clarity we have only shown the $x$ and $z$ faces, but the front and back faces of the cell in the $y$ direction are to be included, as well).
The colors distinguish objects in \emph{incoming} subspaces (green) from \emph{outgoing} subspaces (blue), from the perspective of the $ij0$ cell. 
Thus, at the right cell face $(i+1/2,j,0)$ the incoming subspace in the $x$ direction to cell $ij0$ is the negative subspace, $\amat_{x-,(i+1/2,j,0)}$, while at the left cell face $(i-1/2,j,0)$ the incoming subspace is the positive subspace, $\amat_{x+,(i-1/2,j,0)}$.
The lower face of a $k=0$ cell is a special case, because new information from outside the simulation must be supplied at that location, hence we label it red in the figure.

The characteristics are incorporated into the calculation by projecting the normal derivative of the MHD state vector onto the split eigensystem at the interface state, e.g., in the \nhat{} direction, 
\begin{align}
    \amat_n\partial_n\mhdstate{} & = (\amat_{n+}+\amat_{n-})\partial_n\mhdstate{}\\
    & = \smat_{n}(\pmat_{n+} + \pmat_{n-})\evalmat_n\smat_{n}^{-1}\partial_n\mhdstate\\
    & = \smat_n(\charderiv_{n+} + \charderiv_{n-})\\
    & = \smat_{n+}\charderiv_{n+} + \smat_{n-}\charderiv_{n-}.\label{eq:sl2a}
\end{align}
The normal derivatives of the primitive variables are centered on the cell interface,
\begin{equation}
  \bigl(\partial_n\mhdstate\bigr)_{l+\frac{1}{2}} = \frac{\mhdstate_{l+1} - \mhdstate_l}{\Delta x_l},
\end{equation}
and these derivatives are projected onto the positive and negative eigensystems of the interface.
Here, $\Delta x_l = x_{l+1}-x_l$ is the distance between cell center locations on either side of the cell face, and the index $l$ is the index associated with the normal direction \nhat{} (as it will be throughout the following discussion).
The positive subspace \amatp{} (and associated derivatives $\charderiv_+$) contributes to the time-update of the cell on the right side of an interface, while the negative subspace \amatm{} (and associated $\charderiv_-$) contribute to the update of the cell on the left side of the interface. 

Taking the \xhat{} direction as an example, cell $i$ is updated according to its incoming characteristics, which means the positive subspace from the interface at $i-\frac{1}{2}$ and the negative subspace from the interface at $i+\frac{1}{2}$.
Each of these terms are labeled at the appropriate faces in Figure~\ref{fig:fv_schematic}.
Analogous expressions hold in \yhat{} and \zhat{} so that the total time update to the cell-centered state variable at cell $(i,j,k)$ is the sum of incoming characteristics over all six cell faces, plus the inhomogeneous terms:
\begin{align}\label{eq:fvmhd}
\partial_t\mhdstate_{ijk} + \sum_{(n,l)\in (x,i),(y,j),(z,k)} \biggl(
\Matrix{A}_{n+,l-\frac{1}{2}}\bigl(\partial_n\mhdstate\bigr)_{l-\frac{1}{2}}
+\Matrix{A}_{n-,l+\frac{1}{2}}\bigl(\partial_n\mhdstate\bigr)_{l+\frac{1}{2}}
\biggr)+\inhomoterms_{ijk}=\vect{0},
\end{align}
where the derivative direction $n$ is taken in the direction normal to each face, paired with each combination of $l\pm\frac{1}{2}$, so that $x$ is paired with $i$, $y$ is paired with $j$, and $z$ is paired with $k$.
Equation \eqref{eq:fvmhd} is a finite volume version of the characteristic form of the discretized MHD equations.
For $k>0$ (away from the driving layer) every term can be calculated and so the time derivative can be calculated and numerically integrated---this is just a standard expression for the numerical update anywhere in a simulation interior.
Applied to the $k=0$ driving layer, the only unknown terms in \eqref{eq:fvmhd} are associated with the incoming subspace at the $k=-\frac{1}{2}$ face, that is, $\Matrix{A}_{z\incoming,-\frac{1}{2}}\bigl(\partial_z\mhdstate\bigr)_{-\frac{1}{2}}$ where we now make the notational switch $\amat_{z+}\rightarrow \amat_{z\incoming}$ to reinforce that the positive subspace at face $k=-\frac{1}{2}$ is incoming.
This incoming matrix is marked by the red arrow in Figure~\ref{fig:fv_schematic}.

To make progress we apply the finite volume update to the $k=0$ cells, separate out the unknown terms from the summation, and solve for the time derivative to write
\begin{align}\label{eq:fv_tderiv}
    \partial_t\mhdstate_{ij0} = - \Matrix{A}_{z\incoming,ij,-\frac{1}{2}}(\partial_z\mhdstate)_{ij,-\frac{1}{2}}
    - \sum_{n,l\pm}\Matrix{A}_{n,l\pm}(\partial_n\mhdstate)_{l\mp} - \inhomoterms_{ij0}
    ,\quad  \substack{n\in (x,y,z)\\ l\pm\in (i\pm\frac{1}{2},j\pm\frac{1}{2},k+\frac{1}{2})}
\end{align}
where the $l\pm$ notation indicates that summation is taken over, and derivatives are evaluated at, all cell faces except the $k=-\frac{1}{2}$ face.
Next we define a residual vector \residual{} as the difference between the finite volume calculation of the time derivative and our observationally estimated time derivative from \eqref{eq:targettimederiv}:
\begin{align}
\residual &= \overbrace{\partial_t\mhdstate}^\text{calculation} - \overbrace{\targettimederiv}^\text{estimate} \\
& = -
\lefteqn{\overbrace{\phantom{
\Matrix{A}_{z\incoming,ij,-\frac{1}{2}}(\partial_z\mhdstate)_{ij,-\frac{1}{2}} -\sum_{n,l\pm}\Matrix{A}_{n,l\pm}(\partial_n\mhdstate)_{l\mp}-\inhomoterms_{ij0}  }}^\text{calculation}}
\underbrace{\Matrix{A}_{z\incoming,ij,-\frac{1}{2}}(\partial_z\mhdstate)_{ij,-\frac{1}{2}} }_\text{unknown} -
\underbrace{\sum_{n,l\pm}\Matrix{A}_{n,l\pm}(\partial_n\mhdstate)_{l\mp}-\inhomoterms_{ij0} -
\overbrace{\targettimederiv}^\text{estimate}}_\text{known},\quad \substack{n\in (x,y,z)\\ l\pm\in (i\pm\frac{1}{2},j\pm\frac{1}{2},k+\frac{1}{2})}.\label{eq:residual}
\end{align}
Each term is labeled according to whether it is a known or unknown quantity, and by  whether it is part of the finite volume calculation or the observational estimate.
Our goal is to find the unknown quantities, that is, the $k=-1$ cell centered state $\mhdstate_{-1}$ and the $k=-\frac{1}{2}$ incoming subspace $\amat_{\incoming, -\frac{1}{2}}$, that together minimize \residual.
Note that the first spatial derivative term on the right hand side of the equation implicitly includes the unknown external state via $(\partial_z\mhdstate)_{ij,-\frac{1}{2}} \approx\frac{1}{\Delta z}(\mhdstate_{ij0} - \mhdstate_{ij,-1})$.
Therefore, the problem is generally nonlinear because the explicit $\mhdstate_{-1}$ term from the derivative multiplies the coefficient matrix $\Matrix{A}_{\incoming,-\frac{1}{2}}$, which itself depends on $\mhdstate_{-1}$ though the $k=-\frac{1}{2}$ interface state as defined in \eqref{eq:interfacestate}. 

An optimal solution can be found by minimizing $\residual^2$ via a nonlinear optimization algorithm such as Gauss-Newton, Levenberg-Marquardt, conjugate gradient, etc.
In that case, the optimal result is the state $\mhdstate_{-1}$ that combines with $\mhdstate_0$ to simultaneously generate $\Matrix{A}_{\incoming,-\frac{1}{2}}$ and produce $\partial_t\mhdstate_{0} \rightarrow \targettimederiv$.
Instead, let us seek a solution for the $\residual=0$ case by naively taking the interface state as a given, inserting the finite difference version of the vertical derivative at the $k=-\frac{1}{2}$ face, and formally solving for the $\mhdstate_{-1}$ by matrix inversion:
\begin{gather}\label{eq:FV_formalsoln}
    \mhdstate_{ij,-1} = \mhdstate_{ij0} + \Delta z\biggl[ \Matrix{A}_{z\incoming,ij,-\frac{1}{2}}\biggr]^{-1}\biggl(\targettimederiv_{ij0} + \sum_{n,l\pm}\Matrix{A}_{n,l\pm}(\partial_n\mhdstate)_{l\pm}+\inhomoterms_{ij0}\biggr), \quad \substack{n\in (x,y,z)\\ l\pm\in (i\pm\frac{1}{2},j\pm\frac{1}{2},k+\frac{1}{2})}.
\end{gather}
The incoming coefficient matrix is invertible (see \eqref{eq:dzu-incoming}), so provided it is known \eqref{eq:FV_formalsoln} can be used to numerically calculate the state $\mhdstate_{ij,-1}$.

Equation \eqref{eq:FV_formalsoln} can in principle be used to find an iterative solution though the following process.
Starting with an initial guess for $\Matrix{A}_{\incoming,ij,-\frac{1}{2}}$, use \eqref{eq:FV_formalsoln} to solve for $\mhdstate{}_{ij,-1}$, use that state and $\mhdstate{}_{ij0}$ to calculate a new interface state, recalculate the eigendecomposition $\Matrix{A}_{\incoming,ij,-\frac{1}{2}}$, and then iterate.
It can be shown that this iterative state is stable, so that the correct solution reproduces itself on iteration.
However, convergence to the correct state is not guaranteed.

We tested the iterative approach above by using the positive subspace of the eigensystem at the $k=0$ cell center location in place of $\Matrix{A}_{\incoming,ij,-\frac{1}{2}}$ in \eqref{eq:FV_formalsoln} and then solving for the $k=-1$ state.
This simplification did work well for a variety of tests of our data-driven boundary condition described in \S\ref{sec:results}.
A major drawback, however, was that it could not be adapted to give different weights to estimates of different primitive variables, and could not be extended to deal with missing data; i.e., it was not a true optimization method.

A true optimization method, and the one we focus on for the remainder of this paper, was implemented using a Singular Value Decomposition (SVD) approach.
It still only provides a solution to the linear problem by approximating the incoming subspace at $k=-\frac{1}{2}$ using the values of primitive variables at $k=0$; that is, we will set $\smat_{z\incoming,-\frac{1}{2}} = \smat_{z\incoming,0}$ (or, equivalently, $\amat_{z\incoming,-\frac{1}{2}} = \amat_{z\incoming,0}$; see \eqref{eq:sl2a} or \eqref{eq:primitive2char}) and not consider additional updates to the $k=-\frac{1}{2}$ interface state.
This approximation can be relaxed by solving the nonlinear optimization problem in future work.

To derive the SVD solution, the ``unknown'' term in \eqref{eq:residual} is transformed from the space of primitive variable derivatives to the space of characteristic derivatives via
\begin{align}\label{eq:primitive2char}
    \amat_{z\incoming}\cdot\partial_z\mhdstate = \smat_z\evalmat_{z\incoming}\smatinv_z\cdot\partial_z\mhdstate = \smatzi\cdot\charderiv_{z\incoming},
\end{align}
while the ``known'' term is collected into a single vector $\svdrhs$:
\begin{equation}
    \svdrhs =-\sum_{n,l\pm}\Matrix{A}_{n,l\pm}(\partial_n\mhdstate)_{l\mp} - \inhomoterms_{ij0}-\targettimederiv,\quad  \substack{n\in (x,y,z)\\ l\pm\in (i\pm\frac{1}{2},j\pm\frac{1}{2},k+\frac{1}{2})}.
\end{equation}
Cast in the form of a linear programming problem, we hold $\smat_{z\incoming,-\frac{1}{2}}$ fixed and seek the least-squares solution for $\charderivzi$
\begin{equation}\label{eq:leastsquaresproblem}
    \smatzi\cdot\charderivzi = \svdrhs{},
\end{equation}
where all terms are evaluated at the $k=-\frac{1}{2}$ interface.
The solution minimizes the residual vector \residual{}:
\begin{equation}\label{eq:minnorm-residual}
    \residual = \min_{\charderivz}\Biggl(\smatzi\cdot\charderivzi - \svdrhs{}\Biggr).
\end{equation}

Applying SVD to solve this system of equations\footnote{As implemented in the code, we use the LAPACK \fnc{dgelsd}() routine to solve the SVD problem using the divide and conquer algorithm.  See documentation at \url{https://netlib.org/}.} determines the minimum-norm vector of incoming characteristic derivatives $\charderiv^*_{z\incoming}$ that results in the solution
\begin{equation}\label{eq:minnorm-solution}
    \minnormsoln = \smatzi\charderiv^*_{z\incoming},
\end{equation}
where the $*$ indicates the minimizing solution to the problem described by \eqref{eq:leastsquaresproblem}; all terms are again evaluated at the $k=-\frac{1}{2}$ interface.
We emphasize that $\minnormsoln$ may not equal $\svdrhs$ in general, but will if an exact solution exists.
Substituting the minimum-norm solution $\minnormsoln$ for the term $\Matrix{A}_{z\incoming,ij,-\frac{1}{2}}(\partial_z\mhdstate)_{ij,-\frac{1}{2}}$ into the finite-volume form of the time derivative given at Equation \eqref{eq:fv_tderiv}, we find
\begin{equation}
    \partial_t\mhdstate_{ij0} = -\minnormsoln - \sum_{n,l\pm}\Matrix{A}_{n,l\pm}(\partial_n\mhdstate)_{l\mp} - \inhomoterms_{ij0}
    ,\quad  \substack{n\in (x,y,z)\\ l\pm\in (i\pm\frac{1}{2},j\pm\frac{1}{2},k+\frac{1}{2})}.
\end{equation}

The interpretation of the SVD solution is that it is the physically admissible time derivative on the boundary, $\partial_t\mhdstate_0$, that most closely matches the target time derivative \targettimederiv{}.
To see this, formally decompose the minimum norm vector as
\begin{equation}
    \minnormsoln = - \sum_{n,l\pm}\Matrix{A}_{n,l\pm}(\partial_n\mhdstate)_{l\mp} - \inhomoterms_{ij0}-\dot{\mhdstate}^*.\label{eq:tderivminnorm}
\end{equation}
The time derivative is then
\begin{align}
\partial_t\mhdstate_{ij0} & = -\minnormsoln - \sum_{n,l\pm}\Matrix{A}_{n,l\pm}(\partial_n\mhdstate)_{l\mp} - \inhomoterms_{ij0}\\
    & = \Bigl(+\sum_{n,l\pm}\Matrix{A}_{n,l\pm}(\partial_n\mhdstate)_{l\mp} + \inhomoterms_{ij0}+\dot{\mhdstate}^*\Bigr) - \sum_{n,l\pm}\Matrix{A}_{n,l\pm}(\partial_n\mhdstate)_{l\mp} - \inhomoterms_{ij0}\\
    & = \dot{\mhdstate}^* = \svdrhs{}-\minnormsoln+\targettimederiv.\label{eq:timederivediff}
\end{align}
In essence, the SVD method produces a physically-admissible correction factor $(\svdrhs{}-\minnormsoln)$ to the requested target time derivative \targettimederiv{}.
The correction factor is nothing but the residual vector, $\residual=\svdrhs{}-\minnormsoln$. 
If $\residual=\vect{0}$ then $\minnormsoln = \svdrhs{}$ and the target update can be exactly matched using a linear combination of incoming characteristic derivatives; i.e., the target lies in the image of the matrix $\smatzi$.
On the other hand, any non-zero part of the residual vector lies in the null-space of the matrix $\smatzi$, which is the span of the outgoing characteristic modes that comprise the matrix $\smat_{z\outgoing}$.

Unlike the simple inversion method described following Equation \eqref{eq:FV_formalsoln}, the SVD method is a true optimization method.
In particular, different weights can be applied to each of the primitive variables to normalize their respective orders of magnitude, or additionally, to reflect the level of certainty with which the target value of each variable is known.
A typical situation in observational solar physics is that the magnetic fields and velocities are inferred with more certainty than the density or temperature.
SDO/HMI, for instance, is a highly optimized instrument for determining the vector magnetic field and Doppler velocities in certain regimes, has continuum data that can be used to infer the temperature, but gives little information on the density or pressure.
This could be handled by providing an estimate for $\rho$ and $\epsilon$ as constant in time, but weighting those variables much less than the target magnetic fields.
The SVD solution then matches the better-known variables and finds a self-consistent solution for the less well known fields, $\rho$ and $\epsilon$ in the current example.
In this way, the full information content of the incoming characteristics is harnessed to find a physically valid solution to the data driving problem.
We discuss the case of simply holding $\rho$ and $\epsilon$ uniform in space and constant in time on the boundary (as was done in, e.g., the data-driven simulations compared in \citet{Toriumi:2020}), in \S\ref{sec:typical-bounds}.

Uncertainties in the variables are handled by applying a weighting transformation to the SVD problem, which simply involves left multiplying Equation \eqref{eq:leastsquaresproblem} by a diagonal weighting matrix $\Matrix{W}$.
This transforms $\smatzi\mapsto\widetilde{\smatzi}=\Matrix{W}\smatzi$ and $\svdrhs{}\mapsto\widetilde{\svdrhs{}}=\Matrix{W}\svdrhs{}$.
SVD then produces the minimum-norm characteristic derivative vector $\charderivzi^*$. 
It can still be used directly in the numerical integration defined above because the form of \eqref{eq:leastsquaresproblem} is the same for both the weighted and unweighted cases; the minimum-norm characteristic derivative vector for the weighted system is, of course, different than that for the unweighted system.
In \S\ref{sec:results} we present results for the case with no weighting matrix, then demonstrate use of a weighting matrix in \S\ref{sec:typical-bounds}.
In both cases, $\minnormsoln$ is calculated by left-multiplying $\charderivzi^*$ by the \emph{unweighted} incoming subspace matrix $\smatzi$, as written at Equation \eqref{eq:minnorm-solution}.

Once the minimum norm solution $\charderivzi^*$ is known, the derivative of primitive variables between $k=0$ and $k=-1$, $(\partial_z\mhdstate{})_{-\frac{1}{2}}$, can be solved for by multiplying Equation \eqref{eq:primitive2char} from the left by $\amat_{z\incoming,-\frac{1}{2}}^{-1}$.
Looking at Table~\ref{tab:incoming}, the presence of any incoming characteristic modes involves at least 7 primitive variables, while the eighth ($B_z$) can always be defined using the solenoidal condition for magnetic fields.  
The derivative of the state vector is therefore always well defined for any number of incoming characteristic derivatives, or is unnecessary if there are no incoming characteristic derivatives.
After the spatial derivative is known the state vector at $k=-1$, $\mhdstate_{-1}$, can be set accordingly.
In practice, time updating the $k=0$ level in the \emph{Driving} simulation only requires the product $\amat_{z\incoming}\partial_z\mhdstate = \smatzi\charderivzi^*$, so the values of primitive variables at $k=-1$ are not needed.
They are still used, however, as part of the coupling between the \emph{Driving} and the \emph{Data-Driven} simulation.

The best fit solution for the time derivative, \eqref{eq:tderivminnorm}, is then time-integrated by any appropriate method to produce the time-updated estimate to the driving layer.
Again keeping things simple, we apply a first order forward-Euler scheme, for which the solution is
\begin{equation}
  \mhdstate(\tspj + \dtsim,\xb) = \mhdstate(\tspj, \xb) - \dtsim\biggl[\smatzi\charderivzi^* + \sum_{n,l\pm}\Matrix{A}_{n,l\pm}(\partial_n\mhdstate)_{l\mp} + \inhomoterms_{ij0}\biggr]
    ,\quad  \substack{n\in (x,y,z)\\ l\pm\in (i\pm\frac{1}{2},j\pm\frac{1}{2},k+\frac{1}{2})}.
\end{equation}
This equation defines the time-advance of the driving layer in the \emph{Driving} simulation \charc{}.
Stability is not an issue because new information is continually injected into the \emph{Driving} simulation at each time step.
Recall that this solution is still only approximately correct because we have used the directionally split subspace defined at $k=0$ in place of the subspace at $k=-\frac{1}{2}$. 
As discussed following Equation \eqref{eq:FV_formalsoln}, a nonlinear optimization method would determine a simultaneous solution for both the incoming subspace and the spatial derivative across the interface state, i.e., by finding $\amati\partial_z\mhdstate=\smati\charderivzi$ evaluated at $k=-\frac{1}{2}$.
This would generate a fully self-consistent solution, to which the current linear solution is a first approximation.
While we have explored using a nonlinear optimization method, thus far the solutions are not fully stable, and we therefore retain the linear optimization for now.

Above the boundary layer (in cells $k\geq1$), the regular numerical update to the \emph{Data-Driven} simulation is carried out for time step $\ts$ using whatever numerical scheme is implemented in that simulation.
At the end of the $\ts$ time step, the time-updated solution for the driving layer ($k=0$) and the ghost layer ($k=-1$) are copied from the \emph{Driving} simulation to the \emph{Data-Driven} simulation and the time-updated solution from the interior of the \emph{Data-Driven} simulation is copied into the \emph{Driving} simulation (cells $k=1,2$).
In this way the two simulations are kept coupled together and the \emph{Driving} simulation is ready to calculate the next step.
A new target time derivative is defined for the driving layer using Equation \eqref{eq:targettimederiv}, and the process repeats until the next observed state $\observedstate(\tend,\xb)$ is reached and new observational data needs to be ingested.

Summarizing all the steps in the algorithm, using the SVD based optimization method:
\begin{enumerate}
    {\item At the beginning of a timestep $\ts$ (which updates the simulation from \tspj{} to \tspjpone{}), the MHD state is known at all cell center locations $k\geq 0$, and is consistent between the \emph{Driving} simulation and the \emph{Data-Driven} simulation.\label{alg:start}}
    {\item An observationally derived target time derivative for the primitive variables in the $k=0$ driving layer is calculated using Equation \eqref{eq:targettimederiv}.  Equivalently, a target boundary state may be specified via \eqref{eq:targetboundarystate}.}
    {\item The interface states in the driving layer are calculated using \eqref{eq:interfacestate} at the five faces of each cell where all information is known (everywhere but $k= -\frac{1}{2}$).}
    {\item The eigensystem at each interface location is calculated through Equation \eqref{eq:diagonalization} and using the left and right eigenmatrices given in Appendices \ref{sec:Az}, \ref{sec:Ax}, and \ref{sec:Ay}. All quantities are calculated using the interface values of the primitive variables.}
    {\item The eigensystem of each interface state is split into the grid-positive and grid-negative subspaces using the projection operators \eqref{eq:Ppm} (also calculated using the interface states),  via \eqref{eq:Apm}.}
    {\item An optimization method is used to find the minimum-norm solution of \eqref{eq:residual} for incoming characteristic derivatives through the lower face, $\charderiv_{z\incoming}^*$ at $k=-\frac{1}{2}$.  This solution for the characteristic derivative vector implies an exterior state at $k=-1$ that minimizes the difference between the target time derivative and the numerical time derivative, i.e., the state $\mhdstate_{-1}(\tspj)$, while also producing the physically admissible time updated state at $k=0$, i.e., $\mhdstate_0(\tspjpone)$; these two states are stored in temporary variables.\label{alg:tupdate}}
    \item{The primitive variables in the \emph{Driving} simulation at level $k=-1$ are changed to the values just calculated ($\mhdstate_{-1}(\tspj)$), as this is the state of the exterior plasma that will produce the desired time update.}
    {\item The \emph{Data-Driven} simulation performs its standard numerical time integration of the MHD equations from time \tspj{} to \tspjpone{} for cells with $k\geq 1$.  When needed, the values of primitive variables at levels $k=0$ and $k=-1$ (or spatial derivatives defined in terms of them) are pulled from the \emph{Driving} simulation to be used by the \emph{Data-Driven} simulation during this step (these values may be spatially interpolated by the \emph{Data-Driven} simulation to other grid locations, as needed).}
    {\item The time-updated values in the $k=0$ layer calculated in Step~\ref{alg:tupdate}, $\mhdstate_{0}(\tspjpone)$, are applied in the \emph{Driving} simulation and copied to the \emph{Data-Driven} simulation.}
    {\item The time-updated values in layers $k=1,2$ are copied from the \emph{Data-Driven} simulation to the \emph{Driving} simulation.}
    {\item The full time integration is now complete for both the \emph{Driving} simulation and the \emph{Data-Driven} simulation, and their values are consistent with each other for $k\geq 0$ (the $k=-1$ state can only be calculated in Step~\ref{alg:tupdate}). \label{alg:end}}
    {\item Repeat Steps~\ref{alg:start}-\ref{alg:end} until the simulation reaches time $\tend$.}
\end{enumerate}

The major simplification in the current implementation described above comes in Step 6, where we find the linear SVD solution to the minimization problem as given at \eqref{eq:minnorm-solution}. 
For this, we approximated the incoming subspace at the cell interface $k=-\frac{1}{2}$ by the incoming subspace at $k=0$; that is, we set $\amat_{z\incoming,-\frac{1}{2}} = \amat_{z\incoming,0}$ (or the equivalent in the eigendecomposed system).
The assumption is that the cell centered state and the interface MHD states are similar enough, and the resulting eigendecompositions are similar enough, that this will produce an approximately correct answer, especially since the solution is constantly ``nudged'' in the right direction from timestep to timestep.
However, this does mean that the external state $\mhdstate_{-1}$, interface state $\mhdstate_{-\frac{1}{2}}$, and boundary state $\mhdstate_{0}$, and the associated matrices and characteristic derivatives projected onto the incoming subspace at the interface ($\smat_{z\incoming,-\frac{1}{2}}$, $\amat_{z\incoming,-\frac{1}{2}}$, and $\charderiv_{z\incoming,-\frac{1}{2}}$), are not exactly consistent.
A fully nonlinear optimization approach would self-consistently determine $\smat_{z\incoming,-\frac{1}{2}}$ and $\charderiv_{z\incoming,-\frac{1}{2}}$, and hence $\mhdstate_{-1}$ via $\smat_{z\incoming,-\frac{1}{2}}\charderiv_{z\incoming,-\frac{1}{2}}=\amat_{z\incoming,-\frac{1}{2}}(\partial_z\mhdstate)_{-\frac{1}{2}}$, but we have not yet fully implemented and tested the nonlinear approaches.

Finally, we note that any optimization approach to determining the optimal data-driven boundary condition can be extended to include other constraints.  
For instance, when applied to solar data and including gravity, it may make sense to preserve, say, the mean pressure on the boundary, the total mass flux through the boundary, or to match the energy or helicity flux through the boundary, as estimated by some other method.
This extendable feature makes the optimization method a robust choice for constructing physically motivated data-driven boundary conditions.    

\section{Test implementation using \charc{}-\ddbc{} to data drive \lare{}}\label{sec:testcase}
\begin{figure}
    \centering
    \includegraphics[width=0.82\textwidth]{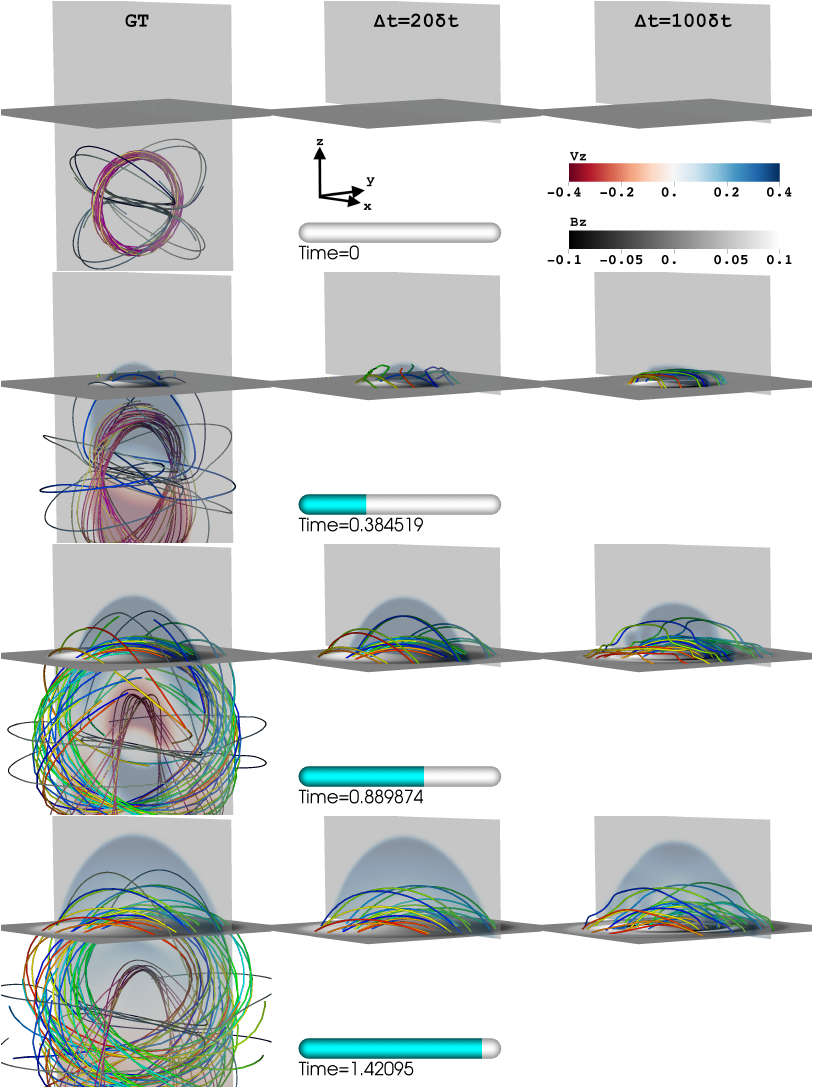}
    \caption{Perspective comparison of the \emph{GT} (left), $\dtdata{}=20\dtsim{}$ \emph{Data-Driven} (middle) and $\dtdata=100\dtsim{}$ \emph{Data-Driven} (right) simulations at four times throughout the simulation (top to bottom: 0, 125, 250, and 375\dtsim, or times $t=0,\ 0.38,\ 0.89,\ 1.42$).  The \emph{Data-Driven} models are driven with data extracted from the \emph{GT} simulation at $z=1.09$.  The vertical cut shows vertical velocities in the $y=0$ plane, and the horizontal cut shows vertical magnetic field in the $z=1.14$ plane (the first non-driven cell in the \emph{Data-Driven} simulations).  Seed points of field lines are initiated both below the driving layer (\emph{GT} simulation only) and above it (all simulations; see text in Appendix \ref{sec:initialconditions} following \eqref{eq:smoothing}).  An animation of this Figure is included in the online material.  The animation shows the expansion of the spheromak through the driving layer for the \emph{GT} simulation and the reproduction of those field lines in the two \emph{Data-Driven} simulations.  The low-cadence, $\dtdata = 100\dtsim$ simulation has notably deformed field lines compared to the other two simulations.}
    \label{fig:sideview}
\end{figure}

To demonstrate the generality of this approach we have developed a characteristic-based data-driven boundary condition (\ddbc{}) add-on module called \charc{}. 
This module is a separate, finite-volume MHD code where the time update is calculated in terms of the MHD characteristics. 
Importantly, the  \charc{} module is not specific to any particular meshing, gridding, or numerical time-advance. 
The interfaces to the \charc{} module were written so that it can be coupled to, and drive, any MHD code.
Furthermore, \charc{} can be used as a kind of numerical glue to couple different plasma codes to an MHD code, e.g., \charc{} could couple a kinetic plasma code to an MHD code. 
Analogously, \charc{} could even be used to drive an MHD simulation in a layer in the interior of an MHD domain. \par

As proof of concept, the characteristic based \charc{} module was coupled to the \lare{} MHD code which utilizes a Lagrangian remap method to advance the MHD equations \citep{Arber:2001}. 
We emphasize that the numerical algorithm and gridding used to advance the MHD equations is different in \charc{} and \lare{}.  
\charc{} runs concurrent with \lare{} and overlaps the two \lare{} ghost cells and first two interior cells along the \lare{} lower boundary, that is, \charc{} is the \emph{Driving} simulation and \lare{} is the \emph{Data-Driven} simulation introduced at the beginning of \S\ref{sec:fvddbc}. 
The vertical layout of the grids are shown in Figure~\ref{fig:grid}, in which \charc{} is the red simulation and \lare{} is the green simulation.

All the primitive variables in the characteristic code are stored at cell centers and represent cell averaged quantities, while the \lare{} variables are represented on a staggered grid. 
The \charc{} simulation is updated in-step with the normal \lare{} simulation, i.e., at the typical \lare{} Courant step.
The full bookkeeping is described in the algorithm summary at the end of \S\ref{sec:fvddbc}, but the cartoon version is that at the end of each numerical time step, the \charc{} variables that are located inside the \lare{} simulation boundaries are overwritten using values interpolated from the \lare{} simulation values, while the \lare{} ghost-cell values are overwritten using interpolated \charc{} values.
In this way, the two simulations are kept consistent with each other while still using the characteristic decomposition of MHD to define the boundary condition.
The reason for taking this hybrid approach is that \lare{} is fundamentally a Lagrangian code, not a characteristic-based code, so the characteristic decomposition is awkward to formulate purely in terms of the \lare{} numerics\footnote{The awkwardness arises for a variety of reasons that all stem from fundamental differences in the numerical schemes.  \lare{} treats variables on a staggered grid with a predictor-corrector update that incorporates mass and energy conservation by associating specific derivatives with specific control volumes.  Calculating all necessary quantities at the intermediate steps in \lare{} in terms of the characteristics would appear to require evaluating the characteristic derivatives projected onto MHD states at different locations and times and then combining them in a numerically stable way.  It is unclear how to do that.  The far more simple case of a incorporating the characteristics into a finite difference scheme failed for similar reasons, as described in Appendix~\ref{sec:transverse}.}.
Therefore, our tests below and those described in \citetalias{Kee:2023} for nonreflecting boundary conditions demonstrate that our method can be used to robustly couple together two MHD simulations with disparate numerical schemes or varieties of approximations. 
\par

\begin{figure}
    \centering
    \includegraphics[width=0.7\textwidth]{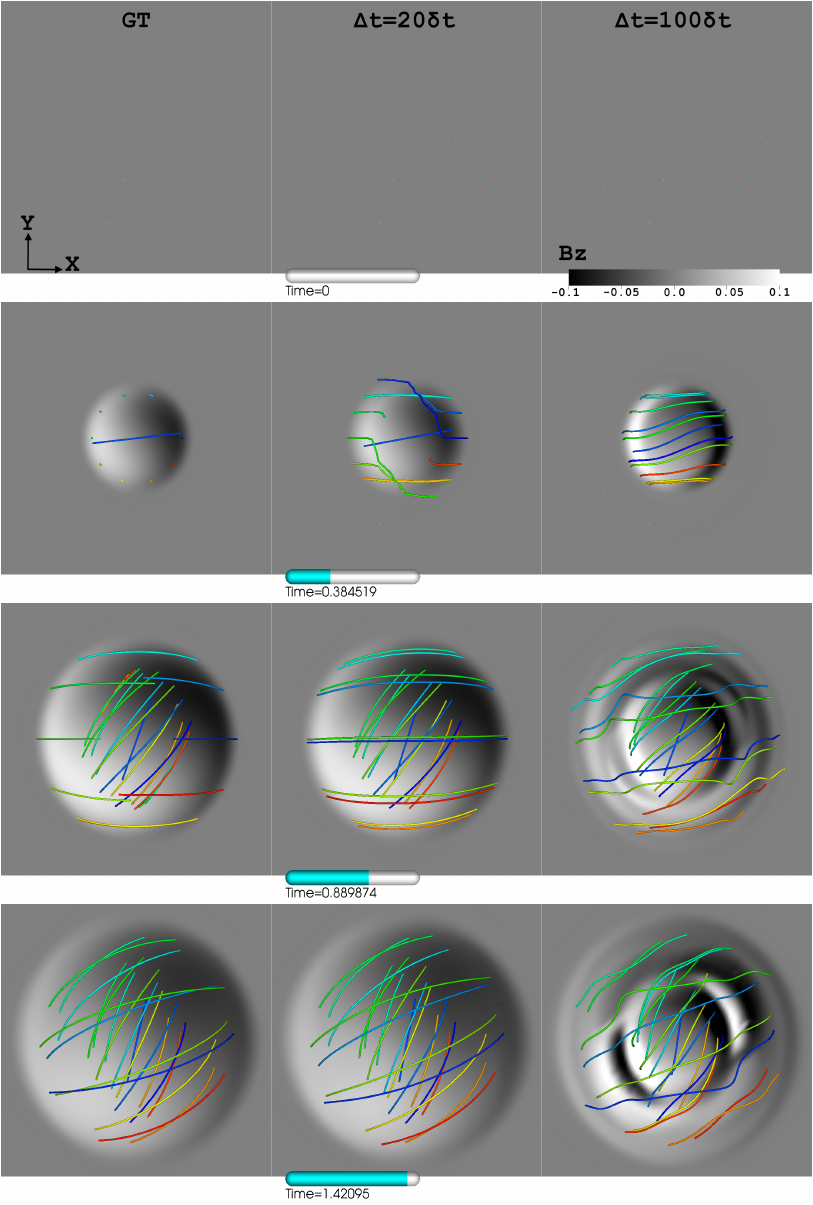}
    \caption{Same as Figure \ref{fig:sideview} but from a top-down perspective, showing \emph{GT} (left), $\dtdata{}=20\dtsim{}$ \emph{Data-Driven} (middle) and $\dtdata=100\dtsim{}$ \emph{Data-Driven} (right) simulations at four times throughout the spheromak expansion (top to bottom at Courant steps 0, 125, 250, and 375). The vertical magnetic field is shown in grayscale in the $z=1.14$ plane.  Seed points of the field lines are initiated at the $z=1.25$ plane (see text in Appendix \ref{sec:initialconditions} following \eqref{eq:smoothing}). An animation of this Figure is included in the online material.  The animation shows the expansion of the field lines through the driving layer.  The vertical magnetic field in the driving layer starts at zero. Subsequently the bipolar magnetic field appears as an expanding yin-yang pattern that rotates as the spheromak enters the driven volume.  Both the field lines and the vertical magnetic field in the driving layer match very well between the \emph{GT} and higher-cadence $\dtdata=20\dtsim$ \emph{Data-Driven} simulations, but the lower cadence simulation, with $\dtdata=100\dtsim$, is severely distorted.}
    \label{fig:topview}
\end{figure}

The data driving method developed in \S\ref{sec:fvddbc} was validated using a three-step process analogous to that described in \citet{Leake:2017}. 
First, a uniformly meshed ground truth (\emph{GT}) simulation was developed and run in \lare{}.
The initial condition was an initially unbalanced spheromak that expands into a field-free homogeneous plasma.
The numerical domain in the \emph{GT} simulation is larger than the total expansion of the spheromak in the portion we will eventually use for data driving.
The left hand columns of \figref{fig:sideview} and \figref{fig:topview} show side-on and top-down perspectives of the \emph{GT} simulation, respectively, in a cut-out domain that is horizontally the same size as the \emph{Data-Driven} simulations, to ease comparison.

Second, we extracted the eight primitive variables from a constant layer $z=z_b$ in the \emph{GT} simulation that lies just outside and above the initial spheromak. 
This layer is denoted by the dark plane in the left column of \figref{fig:sideview}. 
The values of variables on the staggered \lare{} grid are appropriately averaged to cell-centered locations and used for the extraction.
The gray scale images in left columns of \figref{fig:sideview} and \figref{fig:topview} show the vertical magnetic field $(B_z)$ in the extraction plane.
The unbalanced spheromak expands through this plane and, as the time-sequence in each figure and online animations show, the evolution of the magnetic field at and above the plane resembles many features of an emerging bipolar active region, including separation of the opposite magnetic polarities and rotation of a sigmoidally shaped polarity inversion line at the center of emergence.

Finally, we set up a suite of driven simulations in a reduced numerical domain that extends from the $z_b$ plane selected in Step 2 to the top edge of the \emph{GT} simulation.
Data from the extracted plane are used to test the data driving method.
That is, time series of primitive variables from the extracted plane are the synthetic observations that will be ingested by the driving layer at $k=0$, as described in Section~\S\ref{sec:fvddbc}.
This setup allows for a variety of tests. 
In the following investigation we perform two types of validation tests.  
The first focuses on varying the input cadence of the ``observations,'' similar in spirit to the tests performed by \citet{Leake:2017}, except in our data driving scheme the boundary is constrained to evolve consistently with MHD and the interior of the simulation.
The middle and right columns of \figref{fig:sideview} and \figref{fig:topview} compare two examples that data drive between progressively further spaced snapshots extracted from the \emph{GT} simulation, i.e., we vary the driving cadence \dtdata.
In a second type of test we do not provide our algorithm any information on the time evolution of either the density or internal energy density in the driving plane.
These results are discussed in detail in \S\ref{sec:results}. 
Appendix \ref{sec:initialconditions} provides the full analytic description of the spheromak and the numerical details of each simulation.
\par

In slightly more detail, the initial condition in the \emph{GT} simulation produces a blast wave with an embedded, expanding spheromak.
As described in Appendix \ref{sec:initialconditions}, the spheromak includes a shielding current sheet that magnetically isolates it from its homogeneous, field-free surroundings; when we speak of the ``spheromak'' we mean that closed magnetic volume of plasma.
The blast wave extends beyond the edge of the (magnetic) spheromak and hence the outer edge of the blast wave is a purely hydrodynamic shock that propagates to and through the edges of the \emph{GT} domain (absent numerical boundaries it would propagate to infinity).
The expansion of the blast wave is initially nonlinear, becomes increasingly linear after $t\approx 1.5$, and is overall a bit more rapid and extensive in the spheromak's rotationally symmetric $y$ and $z$ directions compared to the $x$ direction.
By $t=3.0$, when we stopped the \emph{GT} simulation, the blast wave had reached a distance of about $3.3$ in the $y$ and $z$ directions and $3.2$ in the $x$ direction.
At long times the blast wave would be roughly spherically symmetric, and represents mechanically radiated energy lost from the system.

The hydromagnetic spheromak undergoes a much more contained expansion, as defined by the expansion of the magnetic field.
Similar to the blast wave, the spheromak's initial expansion is nonlinear and becomes roughly linear around $t=1.0$, by which time it has expanded to a radius of $1.5$ in the $x$ direction and $1.7$ in the $y$ and $z$ directions. 
By $t=3.0$ it has expanded to $1.74$ in the $x$ direction and $2.0$ in the $y$ and $z$ directions.
We stop our analysis before any disturbance reaches the periodic boundaries into the \emph{Data-Driven} domain, above $z_b$ (which, again, has reduced size in $x$ and $y$ compared to the \emph{GT} domain).
Our simulations therefore fall into category (iii) of typical boundary conditions for simulated plasmas from the Introduction \S\ref{sec:introduction}.
\citetalias{Kee:2023} explores how the structure of a passively advected spheromak are modified by interaction with several standard nonreflecting boundary conditions.

In order to demonstrate the capabilities of our method and explore its limitations, we have performed two different types of data driving tests.
The first batch of tests is a series of driven simulations using progressively lower cadence subsets \dtdata{} of the full cadence boundary data extracted from the \emph{GT} simulation at every Courant-step \dtsim{}.
 \figref{fig:schematic} provides a schematic overview of the process.
At each Courant step in the \emph{Data-Driven} simulation, a target boundary state is defined and boundary data is temporally integrated from the current state to the state that is optimally close to the target state while still fulfilling the constraints imposed by MHD, as described in \S\ref{sec:fvddbc}.
At one extreme, the extracted boundary data is provided at every Courant step from the \emph{GT} simulation to the \emph{Driving} simulation so that \dtdata=\dtsim.
This best case scenario is only limited by the accuracy of the coupling between the two different numerical schemes (i.e., the characteristic code \charc{} and Lagrangian-remap code \lare{}).
At the other extreme, the extracted boundary state is only provided at one out of every hundred of the ground truth simulation time steps.
This simulation uses driving data that severely undersamples the dynamic evolution of the expanding spheromak (see \S\ref{sec:physical-courant}).
In total, driving was tested using $\dtdata = 1, 5, 20, 50,\text{ and }100$ timesteps \dtsim{} pulled from the \emph{GT} simulation. 
Taken together, this suite of simulations demonstrates the transition from an accurate best case reproduction of the \emph{GT} by the \emph{Data-Driven} simulation to a poor reproduction of the \emph{GT} once a dynamic sampling threshold is crossed.
This result echoes the findings of \citet{Leake:2017}.
However in contrast to \cite{Leake:2017} the boundary of each simulation in this suite remains consistent with MHD: our characteristic method does not allow setting the primitive variables at the boundary in an under- or over-determined way.

In the second batch of tests we apply our SVD-based \ddbc{} approach to the case where information on the thermodynamic variables is missing.
Here the ``target'' states for both $\rho$ and $\epsilon$ are set to their respective initial values and small relative weights are given to those variables when setting up the SVD optimization problem.  
This allows the \charc-\ddbc{} approach to find a better approximation to those variables.
The results are compared to the case in which the known solution for all primitive variables are provided to the SVD minimization and to the results of more typical boundary conditions found throughout the literature. 
These latter BCs use pure linear-interpolation of the velocity and magnetic fields between observed boundary states, and hold the density and internal energy density uniform and constant at their initial values.

\section{Results}\label{sec:results}

\figref{fig:sideview} shows a perspective overview of the ground truth simulation in the left column and two versions of \emph{Data-Driven} simulations with \dtdata=20 and 100\dtsim{} as the cadence between boundary observations in the middle and right columns, respectively.
\figref{fig:topview} displays the same simulations and times as \figref{fig:sideview} but from a top-down perspective.
Time increases downward along each column, and animations of both figures are available in the online material.
The vertical cut in the $y=0$ plane of \figref{fig:sideview} shows the vertical velocity $v_z$ in a red-blue color scale (the velocity cut-plane is omitted in \figref{fig:topview}).
The horizontal cut through the $z=1.14$ plane in each figure shows vertical magnetic field $B_z$ in grayscale.
This plane lies just outside of the initial spheromak, as seen in the upper left frame of \figref{fig:sideview}, and is the first non-driven cell in the driven domain, i.e., one cell above the driving layer at $z_b=1.109$.
The tubes in Figures \ref{fig:sideview} and \ref{fig:topview} trace out magnetic field lines, whose seed points are described in Appendix \ref{sec:initialconditions} just before \eqref{eq:pressureperturbation}.
The locations of the seed points are constant in time and the field lines are identified by random colors (which are also constant in time for a given seed point).

The first column in both figures show that the combined spheromak and blast wave from the \emph{GT} simulation expands through the driving layer and produces a magnetic configuration similar to a sheared arcade above the driving surface.
This process is similar to how flux emerges on the Sun where the moment of emergence is preceded by a sheath current that forms the boundary between the emerging active region and ambient corona. 
The flux then expands into the ambient corona.
This is especially easy to see towards the end of the simulation in the bottom row of \figref{fig:topview}, where the field lines originating further from the polarity inversion line (where $B_z=0$) have different orientations when they cross the inversion line compared to field lines originating closer to the inversion line.  
Comparison between the left and middle columns of each figure shows that running a \emph{Data-Driven} simulation with a driving cadence of $\dtdata = 20\dtsim$ produces fields and flows in the driven simulation that are very close to the \emph{GT} simulation (though some slight differences are apparent).
However, when the driving frequency decreases to just one out of every hundred snapshots ($\dtdata = 100\dtsim$), as shown in the right hand column of both figures, the differences relative to the ground truth simulation are more pronounced.

The top-down perspective in \figref{fig:topview} makes it very easy to see that the difference between the \emph{GT} and $20\dtsim{}$ \emph{Data-Driven} case (left and middle panels) are quite subtle while for the $100\dtsim{}$ driven case (right panel) the difference, even qualitatively, is drastic.
In the latter case, the distribution of magnetic flux near the driven boundary developed a remarkably different pattern compared to the \emph{GT} case as a result of the data-driven algorithm; correspondingly distorted field lines were produced higher up in the \emph{Data Driven} simulation volume.
The large disparity between the true and driven solutions, even near the driving layer, is caused by the combination of a temporal sampling of the driving layer that under-resolves the dynamic evolution of the system and our algorithm's prioritization of the physics of the MHD equations over temporally-interpolated boundary evolution.

Put simply, the discrepancy with the \emph{GT} simulation at the largest driving cadence of $\dtdata=100\dtsim{}$ demonstrates that the linear interpolation between two driving states on the lower boundary is inconsistent with the MHD equations for that cadence.
Instead, when a boundary is so sparsely observed in time that a linear interpolation between two states is not a valid solution to the MHD equations, our data driving method will instead produce an optimal solution that \emph{is} consistent with the MHD equations.
We discuss the issue of a maximum allowed driving cadence further in \S\ref{sec:physical-courant}.
If one wishes to data drive a simulation under those conditions, either additional independent information or a different set of assumptions must be provided to give a better estimate of the boundary evolution.

Note that any mismatch between the requested and produced boundary states serves as a check on the validity of a \emph{Data-Driven} simulation regardless of the presence of ground truth data.
This check between requested versus produced boundary states in the driving layer is an inherent feature of our data driving method and, critically, is present in the scenario where the data are derived from actual observations.

\begin{figure}
    \centering
    \includegraphics[width=\textwidth]{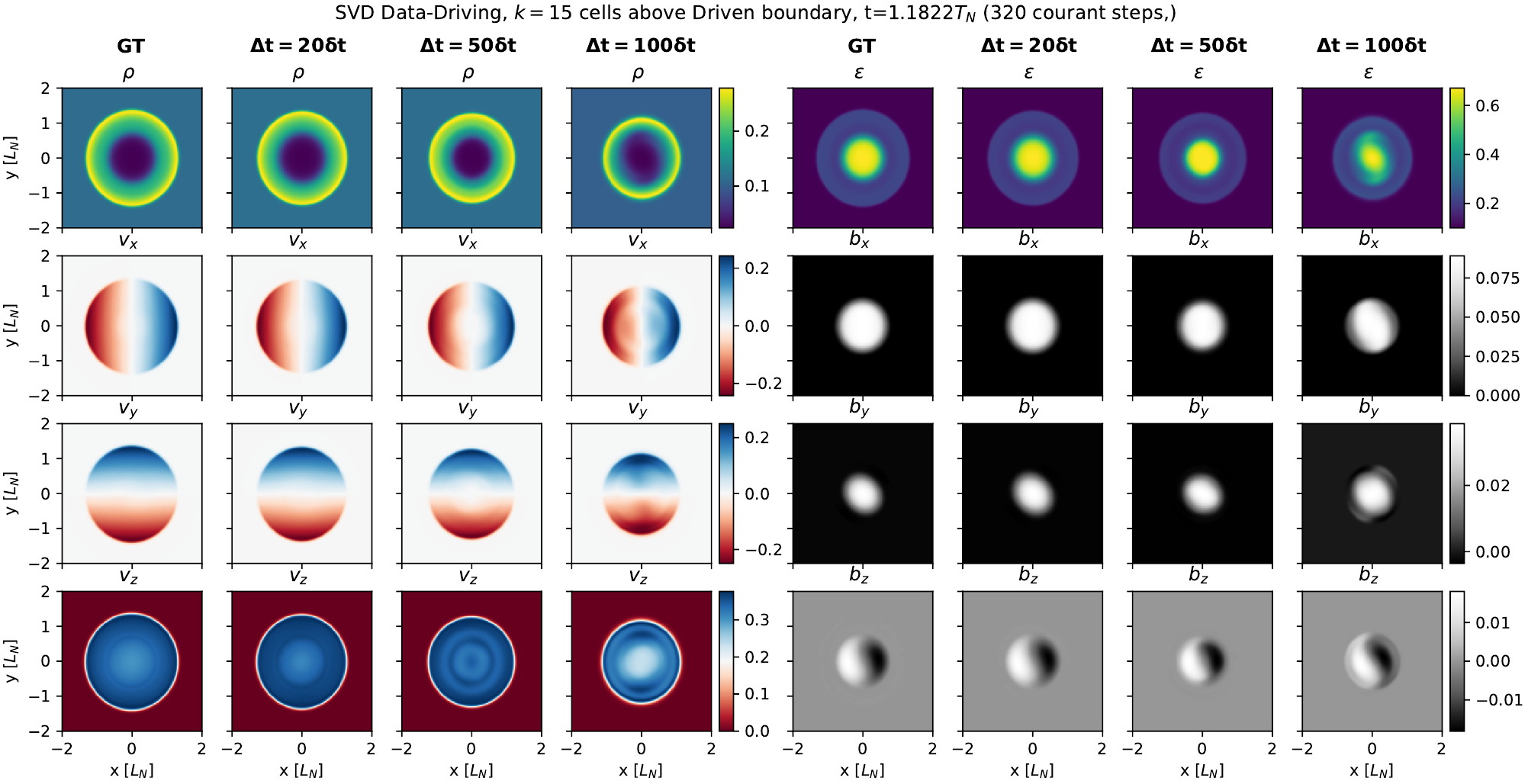}
    \caption{Comparison of all primitive variables in the interior of the \emph{Data-Driven} simulation at a horizontal slice at $k=15$ cells above the driven boundary $(z\approx1.59)$, between the \emph{GT} simulation (columns 1, 5) and simulations driven every 20\dtsim{} (columns 2,6), 50\dtsim{} (columns 3,7), and 100\dtsim{} (columns 4,8).  The horizontal axis is $x$, the vertical axis is $y$, and data corresponds to Courant step $320$, or $t=1.18$, .  The color scale for each primitive variable is common for all simulations.  The top row compares the density in the left four columns and internal energy density in the right four columns.  The next three rows compare components of the velocity field (left) in a Red-Blue color map and magnetic field (right) in a Grayscale color map.}
    \label{fig:GT_DD_compare_all}
\end{figure}

\figref{fig:GT_DD_compare_all} compares the values of primitive variables directly between four sets of simulations.
Each panel is a horizontal slice extracted from 15 cells above the driving layer ($z = 1.59$) at a single time midway through the spheromak expansion ($t=1.1822$ or 320 Courant steps).
The panels are labeled by primitive variable and simulation, and similar variables are further distinguished by different color scales: the \fnc{python} standard \emph{viridis} for thermodynamic variables, red-blue for velocities, and grayscale for magnetic fields.

The first row shows slices through density ($\rho$, left 4 columns) and internal energy density ($\epsilon$, right 4), and the next three rows show components of the velocity field ($v_x,v_y,v_z$, left columns) and magnetic field ($B_x,B_y,B_z$, right).  
Columns 1 and 5 are for the \emph{GT} simulation, 2 and 6 for the $20\dtsim{}$ driving case, 3 and 7 for $50\dtsim{}$ driving, and 4 and 8 for the $100\dtsim{}$ driving simulation.
The color bar for a given primitive variable is shared across all simulations.
Here we see that all of the primitive variables are similar between the \emph{GT} and \emph{Data-Driven} simulations up to the $50\dtsim{}$ driving cadence case, with decreasing fidelity with increased cadence.
In contrast, we once again see that the $\dtdata=100\dtsim{}$ simulation differs substantially from the \emph{GT} case in all primitive variables: it has a  qualitatively different appearance.

\begin{figure}
    \centering
    \includegraphics[width=0.6\textwidth]{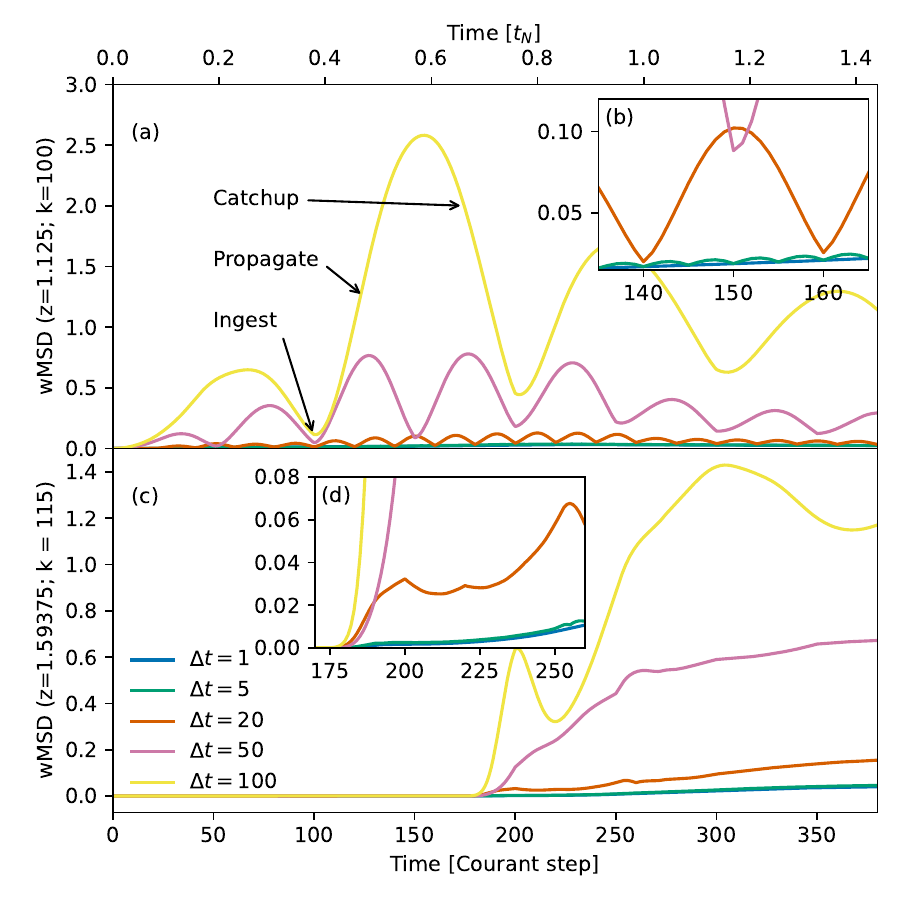}
    \caption{Weighted mean-squared-difference (\wmsd{}) as a function of time using different driving cadences, with $\dtdata= 1, 5, 20, 50,$ and $100\dtsim$ for the blue, green, brown, violet, and yellow lines, respectively.  (a) \wmsd{} calculated in the driving boundary layer, i.e., cell $k = 100\ (k=0)$ in the \emph{GT} (\emph{Data-Driven}) simulation.  (b) Zoom-in near the middle of the simulation.  (c) Same as (a) but 15 cells higher into the simulation (i.e., in the plane shown in \figref{fig:GT_DD_compare_all}).  (d) Zoom-in shortly after the spheromak edge reaches the $k=15$ layer in the \emph{Data-Driven} simulation.}
    \label{fig:wMSE}
\end{figure}

To demonstrate the effect of varying the driving cadence in a more quantitative form we calculated the weighted mean-squared difference between each of the \emph{Data-Driven} simulations and the \emph{GT} simulation.
For a driven state $\mhdstate{}$ and ground truth state $\gtstate{}$ we define the weighted mean squared difference as
\begin{equation}\label{eq:wMSE}
    \wmsd(\mhdstate,\vect{G};t) = \bigl\langle(\mathbf{U}-\mathbf{G})\mathbb{K}^{-1}(\mathbf{U}-\mathbf{G})^T\bigr\rangle,
\end{equation}
where \mhdstate{} and \gtstate{} are both functions of time.
The weighting matrix $\mathbb{K}^{-1}$ (not a function of time) is defined as the inverse of the covariance matrix $\mathbb{K}$ between each pair of primitive variables in the \emph{GT} data calculated at a single reference time:
\begin{equation}
    \mathbb{K}_{\sigma\zeta} = \bigl\langle (G_\sigma - \langle G_\sigma\rangle)(G_\zeta - \langle G_\zeta\rangle) \bigr\rangle.
\end{equation}
$\mathbb{K}$ is a measure of the width of each variable's inherent distribution and the correlations between different variables.
It is therefore a suitable normalization coefficient between the variables. 
We calculate $\mathbb{K}$ at the reference time $t_\eta=280$ Courant steps, midway through the spheromak expansion, and use that single covariance matrix in the calculation of \wmsd{} at all times.

In the above equations, the expectation value of a quantity $q$ is defined 
\begin{equation}
    \langle q \rangle = \frac{1}{N_q}\sum_{i\in \text{domain}} q_i    
\end{equation}
where the domain can be any subset of the simulation, and has $N_q$ cells.
In the following, we calculate expectation values separately in each constant-$z$ plane, so that $N_q = N_x\times N_y$.

The weighting matrix $\mathbb{K}^{-1}$ serves as a unit of measure of the difference between variables in the \emph{GT} and \emph{Data-Driven} simulations.
Consider the most simple scenario in which all variables are uncorrelated and normally distributed.
In that case the covariance matrix $\mathbb{K}$ is diagonal and each element is the variance of the associated variable (i.e., the square of its standard deviation).
Now consider the argument of the \wmsd{} function applied to a single cell, in several situations:
if one of the variables in the \emph{Data-Driven} simulation differs from its \emph{GT} value by one standard deviation of that variable's inherent distribution, and all other variables exactly match their \emph{GT} values, then applied to that single cell $\wmsd{} = 1$; if every variable differs from its \emph{GT} value by one standard deviation of its inherent distribution, then $\wmsd{}=8$; and if each variable is off by $1/\sqrt{8}$ of the standard deviation of its distribution in the \emph{GT} simulation then the $\wmsd{}$ will be 1.
This discussion makes clear that, when considering the full 8-variable MHD state vector, a \wmsd{} value of 1 represents very little difference between the \emph{GT} and \emph{Data-Driven} simulations.
While correlations between the variables will make this simple interpretation of \wmsd{} more complex, we have found that the metric provides a decent measure of the relative similarity between two simulations.

\figref{fig:wMSE} shows the result of calculating \wmsd{} at each simulation time and in two different simulation layers: the driving layer $k=0$ in Panel (a) and $k=15$ cells into the \emph{Data-Driven} simulation in Panel (c).
The second layer corresponds to roughly half the maximum expansion of the spheromak in the vertical direction.
The expectation value in Equation \eqref{eq:wMSE} is calculated over all $N_x\times N_y$ cells in each layer\footnote{We have also experimented with calculating the expectation value over the subset of cells that have either $\abs{B}>10^{-10}$ or $\abs{\rho-\rho_0}/\rho_0 > 10^{-5}$ in either the \emph{Data-Driven} or \emph{GT} simulation.  The intent was to better penalize either over- or under-expansion of the sphereomak relative to the \emph{GT} simulation.  Consistently defining the ``outer edge'' of the sphereomak proved problematic.  The end result is that, while calculating expectation values over all cells in a given layer reduces the relative value of \wmsd{} for layers further from the boundary, it produces an easier to understand metric of comparison across many simulations.}.
The inset Panels (b) and (d) show zoomed-in regions of each parent plot.
Blue, green, brown, purple, and yellow lines correspond to driving cadences of $\dtdata=$ 1, 5, 20, 50, and 100 Courant steps $\dtsim$, respectively.

The driving cadence for each of the \emph{Data-Driven} simulations is clearly visible in \figref{fig:wMSE}(a) as the periodic rise and fall of each mean difference line.
The minima of each curve coincide simply because the longer driving cadence simulations ingest data at integer multiplies of most of the shorter cadence simulations.
An exception to this is seen by comparing \wmsd{} curves for the $\dtdata = 20\dtsim$ and $\dtdata=50\dtsim$. 
As the inset Panel (b) shows, the former (brown curve) has a difference maxima at $t=150$, in the middle of its driving cadence, while the latter has a minima, right at the moment when it ingests fresh data.

The linear interpolation of the state vector between observation times means that, each time new data is provided to the driving algorithm, that information is immediately incorporated into every location across the lower boundary (albeit with a low amplitude).
For longer driving cadences, and in the context of this expanding spheromak simulation, this results in instantaneous transverse phase velocities across the driving layer, encompassing the full spatial extent of what will become newly emerged field.
This is, of course, unphysical behavior, because the true evolution in this case is a gradual expansion outward from the central region.
Instead, the instantaneously provided data then propagates away from the driven boundary everywhere it was introduced, increasing the \wmsd{} metric in the \emph{Data-Driven} simulations. 
The differences only decrease once the true, gradual expansion in the GT simulation catches up with the prematurely injected data in the \emph{Data-Driven} simulations, at which point the metric begins to fall.
In all cadence regimes, our optimization based SVD solution to determine boundary conditions in terms of the characteristic derivatives does find a \emph{physically plausible} evolution of the boundary.
For shorter driving cadences, it reproduces the true, GT solution; for longer cadences, the solution can differ substantially.

These three phases  of the driving process are labeled in Panel (a) of Figure~\ref{fig:wMSE}\----as ``Ingest'', ``Propagate'', ``Catchup''\----for the extreme case of driving at $\dtdata=100\dtsim$, where some new, and rather inconsistent, data is introduced at the ``Ingest'' time.
Note that, through the definition of \targettimederiv, this includes information at $t+\dtdata$.
The knowledge that new information was just introduced everywhere at the lower boundary slowly diffuses through this system during ``Propgate'' phase, and the errors increase during this time as the optimization method struggles to find a physically possible way in which such information could have been introduced.
Once the midpoint of the driving cadence is reached, the information has propagated through enough of the system that the algorithm is able to find a physically allowed dynamical pathway towards the next observation point; that is, the driven system begins to ``Catch-up'' to the ground truth system.
At the end of the cycle, a new batch of data is introduced, and the process repeats.

Simulations with the most frequently ingested data naturally have the lowest error.
Note that driving at the ground truth's Courant step, $\dtdata= 1\dtsim$ (blue line), does not result in $\wmsd=0$ because \lare{} and \charc{} use two different numerical schemes.
Coupling the two codes together essentially introduces some slight numerical diffusion due to the interpolation of primitive variables to different grid locations.
As a result, the \wmsd{} for the $\dtdata=1$ case represents a best-case scenario for data driving, with as faithful a reproduction as is possible to achieve in the present case, and has $\wmsd\lesssim 0.01$ throughout the simulation.
If we restrict the area over which \wmsd{} is calculated to only regions into which the spheromak has expanded, then the maximum difference for the $\dtdata=1$ case is $\wmsd{}\approx0.33$.
Very roughly, this means that, on average, the MHD state vector in a random cell in the \emph{Data-Driven} simulation will differ from the $\emph{GT}$ values in that same cell by approximately $\sqrt{0.33/8}\approx0.2$ of a standard deviation of each variable's inherent distribution.
Of course, some variables may adhere much more closely to their \emph{GT} counterparts than others.

From the inset panel \figref{fig:wMSE}(b) it is clear that both the $\dtdata=$ 5 and 20\dtsim{} simulations are nearly able to match the quality of the $\dtdata=1$ case in the boundary layer at each of their respective injection times (green and brown curves).
That is to say, in between the driving times at which the solution in the boundary is known, the \emph{Data-Driven} simulation becomes progressively worse up to the midpoint of the driving cadence and then progressively better.
Increasing the time between driving data leads to correspondingly greater excursions away from the true evolution of the driving boundary, but for the cases $\dtdata = 1,5,20\dtsim{}$, our data driving algorithm is essentially able to match the observed boundary states. 
For the $\dtdata=50\dtsim{}$ case, however, our MHD-constrained data driving approach is no longer able to match the driving boundary data, as seen when the minima of the purple curve at Courant step $150$ is nearly the same as the maxima of the brown curve: the best 50\dtsim{} time step has the same error as the worst 20\dtsim{} time step.
The $\dtdata=100$ case (yellow line) is correspondingly worse.

Similar properties hold in the interiors of the \emph{Data-Driven} simulations, as seen in \figref{fig:wMSE}(c) and (d).
These panels show the \wmsd{} calculated in a plane 15 cells into the \emph{Data-Driven} simulation ($z\approx1.6$), about halfway between the driving layer and the greatest extent of the expanding spheromak in the $z$ direction.
The slower driving cadences for $\dtdata=50$ and $\dtdata=100$, and the nonphysical linearly interpolated target states between the starting and ending states, cause signals to reach this part of the volume earlier than they should, i.e., the purple and yellow difference curves deviate from zero prior to the blue, green, and brown curves.
The simulations with faster driving cadences have complex, quasi-periodic deviations from the \emph{GT} case as new information propagates through the system, as seen in the inset Panel (d), especially for the $\Delta t=20$ case.
The overall difference to the \emph{GT} simulation remains low in the spheromak interior in that case, however.
In contrast, simulations with slower driving cadences periodically develop much worse reproductions of the \emph{GT} simulation as new driving data is ingested and (improperly) linearly interpolated in time.

\subsection{Condition to determine the maximum allowed driving time}\label{sec:physical-courant}
\begin{figure}
    \centering \includegraphics[width=0.5\textwidth]{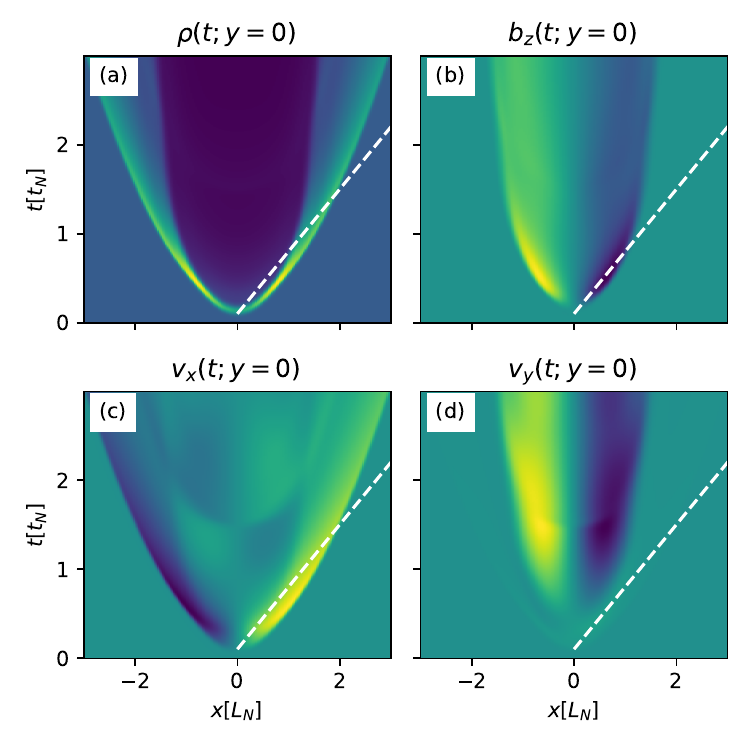}
    \caption{Time-distance plot showing the horizontal expansion of the spheromak along the $x$-direction along the line $(y=0, z_b=1.11)$ for four primitive variables: (a) $\rho$, (b) $B_z$, (c) $v_x$, and (d) $v_y$.  The approximate average expansion rate of $dx/dt\approx2/1.4 \left[ \lnorm/\tnorm\right]$ is indicated by the white dashed line.}
    \label{fig:expansionspeed}
\end{figure}

The above results indicate that there is a critical sampling cadence $\dtsample$ beyond which data driving will become increasingly incorrect.
This maximum time can be of thought of as an analogue of a Courant condition for a discretely sampled physical system: an object that moves by more than its own physical length scale between two observations cannot be unambiguously identified.

\citet{Leake:2017} estimated the critical sampling cadence in their 2D flux emergence simulation.
Following their example, we define the actual observation cadence $\tau$, apparent horizontal feature velocity $v_h$, and feature extent $L$ (the half-width-at-half-max of the feature).
An unambiguous association of a single spatial feature observed at two times requires $\tau<\dtsample\equiv L/v_h$.
Note that this physical sampling condition is independent of the simulation's numerical Courant condition, which is defined with reference to the numerical grid via $\dtsim = dx/max(|v|, c_f)$.

Figure \ref{fig:expansionspeed} shows the application of the critical sampling cadence to our simulation.
The four panels are time-distance plots extracted from the driving plane along the $x$-axis on the line $(y,z) = (0,z_b=1.109375)$ for four primitive variables---(a) $\rho(x,t)$, (b) $B_z(x,t)$, (c) $v_x(x,t)$, and (d) $v_y(x,t)$---and show the expansion of the spheromak through the driving plane.
We define the physical extent $L$ as the half-width-half-max of one of the lobes in $B_z$ at time $t=0.5$, which turns out to be $L\approx0.2$.
A by-eye estimate for the horizontal velocity is indicated by the dashed white line, which roughly corresponds to either the fastest expansion of the vertical magnetic field (Panels b or d) or to the average nonlinear expansion of the fast-mode shock traveling into the surrounding field-free plasma (Panels a or c).  
These velocities are approximately equal and have the value $v_h = dx/dt = 2/1.4\approx1.4$.
This gives a critical sampling rate for this feature of $\dtsample=L/v_h \approx0.2/1.4 = 0.14$.
In order to compare with the tests of various sampling rates earlier in this section, we note that the average Courant step across the simulations is $\langle\dtsim\rangle = 0.004$.
So, we should see the fidelity of the reconstruction diminish when driving at a cadence lower than $\dtsample/\langle\dtsim\rangle \approx 36$ (after using the actual numerical values instead of the one decimal values above).
This rough estimate is indeed consistent with the analysis of how \wmsd{} varies with driving cadence, reported in Figure \ref{fig:wMSE}.

Clearly, the estimate of a critical sampling time will depend on the type of feature used to perform the estimate. 
In general, it will be given by the smallest, fastest moving feature that is present in a system that can be observed for a given instrumental setup.
As a point of comparison, \citet{Leake:2017} found that most of their 2D flux emergence simulations produced features with horizontal lengths and velocities that would be accurately sampled by the $12\unit{min}$ cadence of HMI vector magnetograms, but not always. 
Their simulations with the fastest emergence violated this condition, and as they discussed in some detail, their data driving algorithm, which linearly interpolated the MHD state vector between input states, produced poor results in those cases.

\subsection{Comparison to typical data-driven approaches}\label{sec:typical-bounds}
By far the most common variety of data-driven boundary conditions for solar research currently found in the literature are those that linearly interpolate in time a set of observed magnetic field variables while holding the density and energy fixed to their initial (often uniform) values \citep{Bourdin:2013, Hayashi:2018, Warnecke:2019,Jiang:2020,Kaneko:2021, Jiang:2021,Guo:2021, Inoue:2023}.
This is because the most consistently available synoptic data set for solar observations over the last 13 years are provided by the SDO/HMI instrument, which is optimized for magnetic field and Doppler velocity measurements at moderate spatial resolution ($\approx 1^{\prime\prime}$) and temporal cadence (typically 12 minutes and 45 seconds for vector and line-of-sight magnetograms, respectively); other synoptic programs, such as SDO's predecessor SOHO/MDI or the ground-based NSO/GONG network, are similar but have lower temporal- and spatial-resolution.
The plasma density and temperature (and thus pressure or internal energy density) can be inferred by matching observed continuum intensity to a model solar atmosphere, however, one typically desires multiple spectral regions and lines from different elements to reduce uncertainties and biases \citep{Maltby:1986}.
Even the most sophisticated inversions that combine multiple spectral regions, high spectral resolution, sensitive polarimetry, and advanced model atmospheres can contain order of magnitude errors in some quantities and regions \citep{Borrero:2019, Borrero:2023}.
Beyond those issues, such data are not available from the typical synoptic data sets, and we are not aware of any data driving approaches that currently attempt to incorporate those derived quantities.

Of the data driving approaches cited above, some use the entire magnetic field vector while others use only the line-of-sight (or, approximately, radial) component, and this choice varies from group to group and even paper to paper within a single research group, depending on the scientific goals.
In addition to the magnetic field and Doppler velocity, transverse velocities can be derived by applying feature tracking algorithms to magnetograms or other data series \citep{Welsch:2004, Schuck:2008, Fisher:2015}, but this extra step is not used in all of the above references.
Instead, different authors again employ many different velocity boundary conditions, some linearly interpolating the results of feature tracking algorithms, some using symmetric boundary conditions, and some simply setting all the velocities to zero on the boundary, even though this choice is physically inconsistent with the induction equation.
The through-line of this discussion is that the simple approximation of holding $\rho$ and $T$ (or its equivalent) fixed while linearly interpolating all or part of the magnetic and velocity fields remains a popular data driving approach, even for groups that have previously used a version of characteristic boundary conditions (e.g., compare \citet{Jiang:2020} or \citet{Jiang:2021} to \citet{Jiang:2016}).

We therefore perform a new test to compare our \charc-\ddbc{} algorithm to these more typical approaches.
In the following, we refer to simulations that use these typical (interpolation-only) boundary conditions as the ``baseline'' simulations;  note that the baseline simulations do not use the characteristics at all.
We develop an apples-to-apples comparison between our characterstic approach and the baseline simulations by considering the case where no information about either the density or internal energy density is supplied to any driving simulation, i.e., these variables are estimated as being constant and uniform at their initial values.
All other primitive variables (velocity and magnetic field) are assumed to be perfectly known at the driving cadence.
As such, this new test still represents a greater degree of certainty in the MHD state vector at each location than one typically has from synoptic observations.
Specifically, the baseline data-driven simulations use the typical approach of linearly interpolating $\vect{B}$ and $\vect{v}$ in the $k=0$ layer between injection times, using a zero-gradient assumption for the $k=-\frac{1}{2}$ layer (that is, setting $\mhdstate_{k=-1} = \mhdstate_{k=0}$), and then modifying the magnetic field normal to the driving plane (\zhat{}) as needed to enforce $\nabla\cdot\vect{B}=0$.
The mass density and internal energy density are held constant and uniform at their initial values in both the $k=0$ and $k=-1$ cells.
As described, this procedure roughly mimics typical approaches found in the literature cited above and related works by many of the same authors.

\begin{figure}
    \centering
    \includegraphics[width=0.6\textwidth]{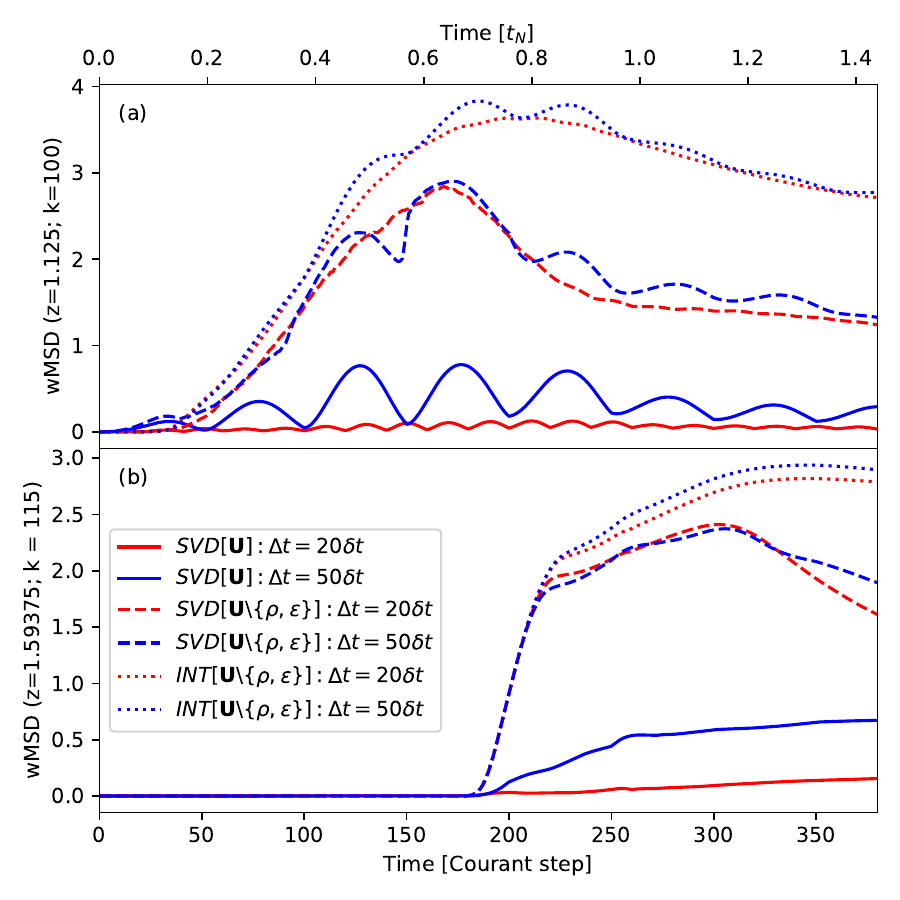}
    \caption{Comparison of \wmsd{} for our data driving algorithm versus linear ``baseline'' interpolation algorithms.  For each line type, the red (blue) line shows driving at $\dtdata=20(50)\dtsim$.  The solid curves reproduce the SVD based \charc-\ddbc{} results from Figure \ref{fig:wMSE}, for reference, the dashed curves show the difference metric for SVD \charc-\ddbc{} when no information is known about either the density $\rho$ or internal energy density $\epsilon$, and the dotted curves show the metric when $\vect{B}$ and $\vect{v}$ are linearly interpolated while $\rho$ and $\epsilon$ are held fixed to their initial values.  Panel (a) gives \wmsd{} calculated in the driving boundary layer, i.e., the $k = 100\ (k=0)$ cell in the \emph{GT} (\emph{Data-Driven}) simulation.  (b) Same as (a) but 15 cells higher into the simulation, i.e., the layer from \figref{fig:GT_DD_compare_all}.}
    \label{fig:typical-ddbcwMSE}
\end{figure}

We then compare those baseline simulations to our data driving approach by running another set of the SVD-based \charc{}-\ddbc{} simulations using the same boundary data as provided to the baseline simulations (i.e., with constant and uniform $\rho$ and $\epsilon$ on the boundary). 
This new set of \charc-\ddbc{} simulations incorporates the weighting matrix for the SVD problem, as described following Equation~\eqref{eq:timederivediff}. 
The diagonal weighting matrix $\Matrix{W}$ has values for $\rho$ and $\epsilon$ of $1\times10^{-6}$, while the weights for magnetic and velocity variables have values of $1.0$.
After applying the weighting transformation the \charc-\ddbc{} algorithm runs as described in \S\ref{sec:fvddbc}.

The results of both sets of simulations are summarized in \figref{fig:typical-ddbcwMSE}, which compares the difference metric \wmsd{} for the baseline data driving methods that use only linear interpolation (dotted lines, labeled $\text{\textit{INT}}[ \mhdstate\setminus\{\rho,\epsilon\}]$) and our approach based on the characteristics (dashed lines, labeled $\text{\textit{SVD}}[\mhdstate\setminus\{\rho,\epsilon\}]$); the notation $\setminus\{\rho,\epsilon\}$ emphasizes that information about the evolution of density and internal energy density are not provided to either the baseline or our characteristic boundary conditions.
For reference, the solid lines reproduce curves from Figure \ref{fig:wMSE}, i.e., they are the results of providing the full MHD state vector to the \charc{}-\ddbc{} algorithm at every driving time step.
Shown in Panel (a) are results for the driving layer and in Panel (b) for the interior of the driven region, i.e., $k=15$ cells above the driving layer, as in Figure \ref{fig:typical-ddbcwMSE}.
Red (blue) curves correspond to driving at a cadence of $\dtdata=20(50)\dtsim$.

Comparing the dashed and dotted curves clearly shows that our \charc{}-\ddbc{} algorithm performs better than the baseline approach to data driving, given identical input.
By the end of the simulation, the mean difference to the \emph{GT} simulation is reduced by more than a factor of two for our approach over the baseline case.
Comparison to the solid curves shows the effect of withholding any information about the evolution of the thermodynamic variables.
The large jump in \wmsd{} from the case where the correct evolution of mass density and internal energy density are given to the case where they are not (solid curves to broken curves) is due, in large part, to a reduced expansion of the spheromak and a decrease in amplitude of the blast wave relative to the \emph{GT} simulation.
These systematic errors are greater when the mass and internal energy density are held fixed along the lower boundary (dotted lines), hence the corresponding larger values in \wmsd{} of the baseline simulations compared to our characteristic-based (dashed).
This demonstrates the ability of our approach, based on the characteristic form of MHD, to fill in missing information about the evolution of state vector at the driving boundary (upper panel), leading to improved results in the simulation interior (lower panel).

In the current implementation of our approach, the weighted SVD algorithm tends to find solutions for which the mass and internal energy densities evolve in tandem, whereas the true solution has a low plasma density and high internal energy density central region (see Figure~\ref{fig:GT_DD_compare_all}, top row).
This behavior likely stems from the fact that SVD produces a minimum norm solution.
Additionally, as currently set up, the SVD problem attempts to find a solution that will drive the density and internal energy density back to their initial values.
The result is that the weighted version of \charc-\ddbc{} produces low density and low energy in the spheromak core, and so the \wmsd{} metric is dominated by an inaccurate (but still physically realizable!) evolution of the internal energy density.
This is a primary reason why the \wmsd{} curves in Panel (a) are only improved by a factor of 2 over the base baseline approach that uses linear interpolation for $\vect{B}$ and $\vect{v}$.
Improvements to our method are expected to more closely match the case where the full MHD state vector is known on the boundary at the driving times (solid lines), and these will be explored in future work.

As mentioned above, there exist in the literature many variations on (non-characteristic based) data driven boundaries that linearly interpolate some subset of the primitive variables. 
In addition to the baseline simulation described above, we have tested several additional versions of these driven boundary conditions.
We omit the full results to save space, but do summarize a few important points here that help provide broader context for the comparison of these other versions of baseline (interpolation-only) driven boundaries to our \charc-\ddbc{} method.
First, we explored several variations of holding components of the velocity fixed at zero and linearly interpolating the other variables.
In all cases this generates similar results to the baseline linear interpolation case presented in \figref{fig:typical-ddbcwMSE} (dotted) but with correspondingly larger values of the difference metric.
Second, we ran a set of simulations that linearly interpolate all of the primitive variables in time using different driving cadences.
This approach produces \wmsd{}s that are close to those based on the characteristics, i.e., as presented in \figref{fig:wMSE}, and outperform it in some instances. 
This is to be expected: a simple linear interpolation is how the target update is defined in \charc{}-\ddbc{} to begin with at \eqref{eq:targettimederiv}, so at high cadence the two approaches should give nearly identical results.

This latter test demonstrates that a linear interpolation is sufficient to produce high fidelity results \emph{provided the dynamic evolution of the system is adequately sampled and accurate for every primitive variable.}
However, the utility of a simple linear interpolation decreases when information about the system's evolution is missing, either by low cadence observations or by variables with large observational bias and uncertainty, in which case the observed boundary evolution is unlikely to be consistent with the evolution of the simulation interior.
We would expect that a similar challenge would arise when there is inconsistency between well observed variables, for example, when the spectral lines that are used to determine temperature and magnetic field are sensitive to different heights in the solar atmosphere.
In that case, even though the evolution of each variable may be well determined, they may not relate to each other in a physically meaningful way, at least when used to directly define a boundary condition.
Indeed, \citet{Carlsson:1994} already encountered this issue in the relatively simple case of a piston driver at the lower boundary of a 1D radiative hydrodynamic simulation. 
The piston velocity was to be given by the observed Doppler velocity of an Fe I line, but the mismatch between the formation height of the line and their actual lower boundary caused significant errors in the resulting synthesized Fe I line from the simulation.
These arose, essentially, because they had not properly encoded the physics of wave propagation through the solar atmosphere.
While they were able to mitigate the problem in a clever way in that case, it well illustrates the challenges posed by even slight mismatches between the observations and simulations.
Such inherent inconsistencies are expected whenever primitive variables are inferred from observations. 

The above tests show that the baseline approach of holding the density and energy to fixed values at the boundary while linearly interpolating other variables over time can significantly alter the expansion of the spheromak into the driven simulation.
This is likely because the spheromak expansion is primarily a fast-mode type disturbance that requires gradients in both the density and energy.
Instead, when density and energy are held fixed, the first phase of expansion away from the boundary has to be driven purely by the magnetic field.
If we analyze the baseline simulation's behavior in terms of the characteristics then the linear interpolation boundary driving is most similar to one of the Alfv\'en modes (however, we again stress that there is no guarantee that the behavior of the driven cells, when simply set using linearly interpolated primitive variables, represents any allowed motion under MHD).
Just above the driven boundary---one cell into the driven simulation---the Alfv\'enic disturbance at the boundary must couple to a combination of fast- and slow-modes to drive further expansion away from the boundary.
This results in less expansion, weaker field strengths, and thus a smaller modeled spheromak away from the boundary compared to the \emph{GT} simulation.
In contrast, our approach based on the characteristics allows the mass and internal energy density to self-consistently evolve with the other variables and therefore mitigates these issues, but (so far) does not fully eliminate them.
We hope to address remaining issues in future work.

Finally, we note that the magnetic topology of all the driven simulations appear similar to the \emph{GT} simulation. 
However, we caution that the spheromak test case has a rather simple topology to begin with.
Our conclusion from this final suite of simulations is that using simple linear interpolations of primitive variables when holding some number of those variables fixed can produce inaccurate results.
The output may appear reasonable in the simulated domain for some of the variables, but it is unclear how one could tell that the recovered solution is far from the truth, especially if the linear driving at the boundary is inconsistent with the actual equations being solved in the rest of the simulation.
In contrast, our use of the characteristics to ensure an MHD-allowed evolution of the boundary, combined with an optimization approach to find a solution that matches the observations as closely as possible, produces a substantial improvement over the baseline approach.
Additionally and critically, for the characteristics, one can tell from the difference between the requested and produced boundary evolution when the requested evolution can no longer be accommodated by the MHD equations.
Finally, the optimization method can be extended to incorporate additional constraints, such as estimated relative helicity, mass flux, or Poynting flux, that could help to provide a better solution when limited, uncertain, or inconsistent observations are available.

\section{Discussion}\label{sec:discussion}
We have presented a self contained description of the theory, numerical implementation, and testing of a solution to the rather nuanced problem of data-driven boundary conditions for magnetohydrodynamic simulations.
The previous sections already contain a good deal of discussion, so here we summarize only at a high level and try to weave together some thoughts that are only possible after all of the foregoing exposition.

The main problem we are trying to solve is that, in order to study the dynamics of the solar atmosphere, we need an accurate representation of the corona, especially (\emph{but not solely!}) the magnetic field.
Direct inference of the complete 3D coronal state from observations, for instance using polarized spectral lines that form under coronal conditions, is at least difficult if not practically impossible on a routine basis with current technology.
This is despite very real and ongoing improvements in that field in terms of both instrumentation \citep{Tomczyk:2008,Tomczyk:2016, Rimmele:2020} and reconstruction techniques based on observations \citep{Kramar:2016, Casini:2017, Dima:2020, Li:2021, Schad:2023}. 
On the other hand, photospheric observations, even with low spectral resolution, allow routine inference of much of the MHD state vector at the solar surface, for instance using GONG \citep{Harvey:1996} or SDO/HMI \citep{Pesnell:2012} for magnetic fields and Doppler velocities, optical flow techniques for transverse velocities \citep{Schuck:2008}.  
Inferring thermodynamic quantities typically requires simultaneous observations of multiple spectral lines whose ratios are sensitive to density and temperature.
When such observations are not available, as is the case for synoptic solar observations, estimates based on empirically- and model-derived relations can be used \citep[see][and references in the Introduction of that work]{Jaeggli:2012}.
The time evolution of the photospheric state vector then provides a dynamic constraint to the plasma above the photosphere.
That fact raises the question of how best to incorporate the observational data into the boundary of an MHD simulation.

The main goals of this paper were developed to answer that question: first, to develop general boundary conditions that require all the primitive variables on the boundary to be set in a way that is consistent with the MHD equations, and second, to apply that formulation of boundary conditions to the data driving problem.
This is far from a trivial task because the dynamics of all the variables are coupled together in a limited number of physically admissible ways, a constraint that is very much at the heart of our approach to data driving.

Setting one primitive variable to some value makes implicit assumptions about how the system is evolving and thus implies specific values for a subset of the other variables.
As a simple example, suppose a boundary cell is updated in time by $B_x\rightarrow B_x+\delta B_x$, which is meant to represent a propagating Alfv\'en wave.
In order to be self-consistent, this requires that the velocity field should also updated by $v_x\rightarrow v_x-\delta B_x/\sqrt{\rho}$ (for propagation in the same direction as the background field $B_z$).
Setting the velocity to anything else will not accurately represent a propagating Alfv\'en wave, and may not represent \emph{any} physical way the plasma could move.
This freedom to incorrectly prescribe the boundary is permitted when the boundary conditions are set arbitrarily in terms of the primitive variables.

For more complicated scenarios (say, a multitude of overlapping waves or a dynamically emerging flux rope) it is much more difficult, if not impossible, to directly determine how the variables should be set in a self-consistent manner.
Depending on the state of the system, certain types of usual MHD dynamics may not even be allowed.
Again using a simple wave as an example, suppose we set the boundary updates for each primitive variable, $\mhdstate\rightarrow\mhdstate+\delta\mhdstate$, to represent an upward propagating slow mode wave, but did so in a location where the plasma bulk velocity is directed downward at greater than the local Alfv\'en speed (row 2 of Table \ref{tab:incoming}).
This is not a physically admissible situation, as the plasma cannot propagate information in that way.
Those boundary conditions are therefore nonsense! 
\emph{However, almost any MHD code will happily apply nonsensical boundary conditions anyway}, and this includes our own workhorse code, \lare{}.
It is unknown how that mismatch will affect the results of the simulation, and almost impossible to determine after the fact that nonsense information has been injected into the simulation.
In general, setting any of the primitive variables independently will not represent a physically admissible solution to the MHD equations in the boundary, and it is unclear what implications this may have for the ongoing simulation.
What is clear is that, because MHD is a hyperbolic system that fundamentally describes the propagation of information, bad information injected though the boundary in this way will eventually corrupt the dynamics of the entire system. 
\emph{Bad boundary conditions lead to bad solar simulations!}

Our solution to this problem is to describe the plasma in the boundary purely in terms of the ways it can actually move, which means implementing the boundary conditions purely in terms of the characteristics of the system, specifically the incoming characteristics, as described in \S\ref{sec:characteristic-bounds}.
This is wonderful because it forces all the primitive variables to be set in a self-consistent manner, i.e., we are prevented from over-specifying the system.
The self-consistency arises because the characteristic modes are, effectively, the allowed Lagrangian displacements of the MHD state variable, so updates to the primitive variables calculated in terms of them are solutions to the MHD equations.
Put another way, setting the boundary conditions in terms of the incoming characteristic derivatives implies that, in principle, one could build a real machine in the laboratory that physically implements those boundary conditions on the plasma.

However, this solution produces a fresh challenge, which is how to set the incoming characteristic derivatives to represent the desired physics.
Fortunately, the data driving problem in solar physics likely represents the best possible scenario for determining the desired physics because all the hard work has already been done: the Sun already did it, and we observed it happening!
But typically, we have only \emph{sparsely} observed the Sun, either temporally or in terms of the number of primitive variables that can be reliably inferred from the observations.
That fact lead us to the optimization approach described in \S\ref{sec:fvddbc}.

The next major task was to test all of the above developments, and for this we chose to recreate a ground-truth simulation of an expanding spheromak.
The dynamics of the \emph{GT} simulation mimic some typical behaviors of emerging active regions on the Sun: the magnetic field expands outward from the first emergence location; the polarity inversion line in the extracted plane, where the normal component of the field passes through zero, has a sigmoidal shape; and the polarity inversion line appears to rotate during emergence.
In this regard, the expanding spheromak is a nice test case because it does contain interesting and applicable dynamics while still allowing relatively simple analysis.  
Notably, we did not include any inhomogeneous terms in this validation, such as gravity or radiative transfer effects (although we did carry those terms through the entire calculation as the variable \inhomoterms{}).
Future investigations will include these effects, which are required for more realistic photosphere-to-corona simulations and when applying our method to actual observational data.

In this investigation we performed an initial validation by varying the input cadence of synthetic observations extracted from the \emph{GT} simulation. 
These tests are similar in spirit to those performed by \citet{Leake:2017}, except in our data driving scheme the boundary is constrained to evolve consistently with both MHD equations and the interior of the driven simulation.
Instead, \citet{Leake:2017} simply used linear interpolation in time of all of the primitive variables extracted from their \emph{GT} simulation to set the values in ghost cells of their driven simulation.
As they showed, if the dynamic evolution of the boundary is well approximated by a linear interpolation at each location then this approach can produce good results.
However, it was not always obvious when the linear interpolation approximation was no longer valid.
In their case, they could introduce strong and completely spurious currents into their simulation interior that dominate the system's dynamics, all while exactly matching the true solution at the driving times (this produced substantially more ``free magnetic energy'' in the driven system than existed in the true system).
In contrast, a major strength of our optimization approach is that it naturally makes known when the driving cadence is too infrequent: our optimization based boundary condition will still produce a \emph{valid} MHD solution for the estimated boundary evolution, but one that no longer matches the observed evolution of the boundary.
When present, this discrepancy amounts to a ``cry for help'' from the \emph{Data-Driven} simulation: it is being asked to conform to a dynamical trajectory disallowed by the physical laws that govern our universe.

Our initial test cases revealed the transition from good to poor reconstructions for driving cadences of $\dtdata\gtrsim 50\dtsim$, which was ultimately determined by length and velocity scales present in the system dynamics (\S\ref{sec:physical-courant}).
At that low sampling rate, the driving layer was so sparsely sampled in time that a linear interpolation between the two observed states was no longer close to a valid solution of the MHD equations.
Instead, our data driving method produced optimal solutions that are consistent with the MHD equations, in the sense that there could, in principle, be a series of external states that drive the internal plasma through the optimal evolution.

As a second validation test, we directly compared the result of our data driving method against a more common approach wherein $\rho$ and $\epsilon$ are held fixed to their initial values while the remaining primitive variables are linearly interpolated between the observed states---we called these ``baseline'' simulations.
Many variations of this approach are found throughout the literature \citep{Bourdin:2013, Galsgaard:2015, Leake:2017, Jiang:2020, Kaneko:2021, Guo:2021, Inoue:2023}.
In our tests, our characteristic based optimization method was able to produce better results than the baseline approach.
This test is the closest approximation to the type of data driving that can currently be achieved using the routine observations provided by SDO/HMI, which are the best suited for studying active region evolution.
Note that the linear interpolation tests performed in \citet{Leake:2017} took the extreme case where all eight primitive variables were assumed to be known.
We also performed such tests, and like \citet{Leake:2017}, found that linear interpolation can produce excellent results \emph{provided that the evolution of all the primitive variables in the driven boundary are already self-consistent with each other and the interior evolution of the simulation}.
If the evolution is not self-consistent, either because the observational cadence is too low and/or some variables are less well observed, then the fidelity of the reconstruction will decrease.

The main issue with simple interpolation is that the resulting boundary condition is over-determined: more primitive variables are independently and inconsistently set than allowed by the MHD equations, i.e., more than the number of incoming characteristics given in Table \ref{tab:incoming}.
This problem does not arise when using high cadence input that is already self-consistent, as, for instance, in the high cadence tests of \citet{Leake:2017}: even though those authors always set the values of all primitive variables, their driving data were pulled from a self-consistent MHD simulation.
At high cadence, they essentially provided \emph{redundant} information about the simultaneous evolution of multiple primitive variables.
The errors in their data driving tests only grew significant when the information they provided was no longer self-consistent due to interpolation times that exceeded the physically defined critical cadence \dtsample, as was the case for us.

This point is essentially a manifestation of Kirchoff's integral theorem for boundary conditions.
If both the value and derivative of a function are set on the boundary they are not generally consistent (for arbitrary values of both values and derivatives).
But if one sets both the values and derivatives \emph{self-consistently} then the system is not actually overdetermined because the specific combination of values and derivatives are just linear combinations of the correct characteristics.
The case of \citet{Leake:2017} is an example of this, where every variable and its derivative was set in the BC but they were all pulled self-consistently from an MHD simulation. 

Now consider how this issue of over-specifying the system applies to a case of real data driving, for example, using primitive variables derived from spectropolarimetric observations of the photosphere.
In the limit that all variables are well observed at high enough temporal and spatial resolution and have co-spatial origin, a simple interpolation between observed states will give a correct solution to the system's evolution. 
It will not be over-specified because the evolution of all the variables would be self-consistent and include mutually redundant information.
Real life seldom works that way, but instead includes noise, bias, and other inconsistences. 
Dealing with such issues is a major strength of our method based on the characteristics and employing an optimization approach that prioritizes the validity of the underlying equations.
For the case of insufficient temporal data, where interpolation between known states at distant times produces inconsistent estimates for MHD states at intermediate times, our characteristic method instead provides a plausible physical trajectory through state space between the two known states, a trajectory constrained by the MHD equations themselves.

Similarly, when some portion of the MHD state vector is either wholly unknown or estimated at lower precision than others, the optimization method should again provide an optimal trajectory in a least-squares sense between estimated states, constrained to be as consistent as possible with the observed states.
In such cases it may be necessary to resort to an \textit{ad hoc} assumption about how a variable evolves.
While this is not ideal, it can still be done in a way that prioritizes a self-consistent evolution of the entire MHD state vector by combining the characteristics with an optimization approach.
For instance, if we are missing information about the time evolution of the density, we could assume a model where it exponentially relaxes back to a background stratified value over some timescale $t_R:$ $\rho(t)\rightarrow \rho_0e^{-t/t_R}$.
The optimization procedure would then accommodate this prescription, but only to the extent that the evolution of $\rho$ was consistent with that of the other, better known variables.
Our application of the weighted SVD solution, described in \S\ref{sec:fvddbc} and tested in \S\ref{sec:typical-bounds}, demonstrated the validity of this approach; refinements to it are forthcoming.
Finally, by applying an optimization technique to this problem, additional constraints could also be imposed.
The most immediately useful constraints for solar physics applications are independent estimates for the energy, helicity, or mass fluxes through the boundary.
We will explore the capabilities of our method in these cases in future work.

Other groups have performed validation studies for data-driven MHD simulations that are similar to our own, most notably the test of several data driving methods in \citet{Toriumi:2020} and the follow-up study focused on the DARE code in \citet{Jiang:2020}.
In this instance, the DARE code did not use a characteristic method for the driven boundary condition, although these same authors have done so in the past \citep{Jiang:2016}.
These authors tested DARE's ability to recreate the coronal portion of a self-consistent, \textit{ab initio} ground truth simulation of a buoyant flux rope emerging through a gravitationally stratified atmosphere.
As stated in \S\ref{sec:introduction}, they produced a decent reconstruction of the magnetic field when using data extracted from a height with minimal Lorentz forces to drive the MHD simulation, i.e., data from the coronal base.
However, the technique struggled when driven using data from closer to the photosphere.
They pointed to the presence of strong Lorentz forces in the photosphere, which are approximately balanced by pressure gradients in the ground truth simulation but not in the data-driven simulation.
They state that these photospheric Lorentz forces are too large to be effectively handled by DARE, but we find this to be a puzzling argument: DARE is an MHD-based code and Lorentz forces are certainly within the scope of MHD, so DARE should be able to handle such a case.
However, we do note that their ground truth simulation, which includes gravitational stratification and an emerging flux rope, is a much more difficult challenge to reproduce than our pressure-gradient-driven expanding spheromak.
In a future work we will apply our method to the \citet{Toriumi:2020} test case, which will hopefully provide another data point for when and where data-driven MHD approaches succeed or fail.

A number of authors of coronal simulations do use a form of the characteristics to set at least some of their boundary conditions.
A long series of work originating with \citet{Nakagawa:1980} uses a technique they name Projected Normal Characteristics (PNC) to set boundary conditions \citep{Nakagawa:1980,Nakagawa:1987,Wu:1987,Hayashi:2005, Wu:2006,Jiang:2011, Yang:2012,Hayashi:2013,Feng:2015,Jiang:2016,Feng:2017,Hayashi:2018,Hayashi:2019,Hayashi:2022}, while recently a separate characteristic based method was developed by \citet{SarpYalim:2017} and used in \citet{Singh:2018} and \citet{SarpYalim:2020}.
The PNC method is used for multiple purposes, often as a form of nonreflecting boundary condition where the evolution of the boundary is calculated by setting the incoming characteristics to cancel the effects of transverse terms.  
In our notation, this is implemented as $\charderiv_{z\incoming} = -\smat_{z\incoming}^{-1}\vect{C}$, and again, we present an extended discussion of this and other nonreflecting boundary conditions in \citetalias{Kee:2023}.
For instance, in Appendix B of \citet{Hayashi:2018} they describe using the PNC approach for the side and top boundaries of a driven simulation, but use a non-characteristic based method for the bottom, driven boundary.

The PNC method has also been applied to the problem of data driving \citep[e.g., ][]{Hayashi:2005,Wu:2006,Yang:2012,Feng:2015, Jiang:2016}.
Most cite \citet{Wu:2006} as the originator of this approach, and that approach seems to differ from what we have described in this paper.
In particular, in the discussion in \citet{Wu:2006} following their Equation~(A9) they appear to rely on a one-to-one mapping between between each incoming characteristic and the value of a single primitive variable on the boundary.
This one-to-one mapping is also present in the work of \citet{SarpYalim:2017}, \citet{Singh:2018}, and \citet{SarpYalim:2020}; see, e.g., Equation (11) in the first of these, where the number of primitive variables to be set arbitrarily is assumed to be equal to the number of incoming characteristics.
This one-to-one mapping runs counter to our approach, where the values of multiple primitive variables are coupled together through the eigenstructure of each characteristic mode and all incoming characteristic derivatives are solved for simultaneously, as we discussed in Section \ref{sec:characteristic-mhd} in connection with Table \ref{tab:incoming}.

What is clear is that the application of the PNC method to data driving is quite challenging. 
The introduction of \citet{Jiang:2021} describes some of the issues these authors have encountered, and highlight two difficulties in particular: (1) dealing with situations where the number of variables they wish to set using observations does not precisely match the number of incoming characteristics; and (2) dealing with locations near a PIL, where the normal magnetic field passes through zero.
The approach we have described does not encounter issues in either of these cases (at least for the tests we have performed so far), which reinforces the notion that these authors' use of the characteristics to implement boundary conditions differs from our own.

A related use of the characteristics that has appeared in several previous solar simulations is to use a single incoming characteristic derivative to enforce a single specific boundary condition, for instance, constant mass flux \citep{Hayashi:2005,Lionello:2013, Feng:2015}. 
While this is valid approach in the regime in which it was applied (solar wind simulations with an inner boundary condition located outside the solar Alfv\'en surface, so that the wind is always super Alfv\'enic and directed into the simulation), these types of assumptions will likely conflict with observational data in the sub-Alfv\'enic corona at some location on the boundary.
This may partially explain the difficulties discussed in the introduction of \citet{Jiang:2021}.

In contrast to data driving, most previous formulations of boundary conditions in terms of the characteristics are mostly concerned with getting information \emph{out} through the simulation boundary rather than in.
The goal in these cases is to implement a nonreflecting boundary condition that leaves the ongoing simulation itself relatively unmolested \citep{Hedstrom:1979,Thompson:1990,Grappin:2000,Gudiksen:2011}.
Achieving this actually makes rather strong assumptions about the state of the external universe, but those assumptions are not always clearly specified.
In particular, we have used the characteristic formulation of general boundary conditions described in this paper to test several standard flavors of ``nonreflecting'' boundaries and found that, while they all work reasonably well for transient disturbances such as waves and shocks, they can fail spectacularly when attempting to advect larger scale structures out of the simulation (for instance, a spheromak).
For these results, we refer the reader to an extensive discussion of the topic in Paper II of this series \citep{Kee:2023}.

As the previous example highlights, the formalism we have developed has application beyond data-driven boundary conditions, and indeed, beyond boundary conditions themselves.
One application has already been demonstrated, which is the ability to couple together numerical codes that are based on different numerical schemes, as was done to couple the finite-volume wave-splitting \charc{} code to the Lagrangian-remap \lare{} code.
A closely related application is to use \charc{} to implement data driving in the middle of a simulation, which would represent an important step towards using multi-height observations of the Sun for data assimilation.
This idea could possibly be extended to couple together codes that implement different physics, for instance to create a bi-directionally coupled PIC and fluid reconnection simulation.
Coupling in terms of the characteristics would ensure that, at least at the level of the MHD approximation, information is correctly exchanged between the two systems.
Another application of the formalism we have developed is to the analysis of MHD systems in general and simulations in particular.
The characteristics seem under-utilized for this task, at least in the context of solar simulations.
For instance, one could use the characteristics to identify individual wave modes (in the infinitesimal limit) or types of information (in the general case) propagating through a simulation, study their interactions, and the conversion of wave energy from one type to another.
Again, subsequent investigations will explore these topics.

\section{Conclusions}\label{sec:conclusions}
We have developed a general framework for formulating boundary conditions in MHD simulations based on the method of characteristics and have applied it to the problem of photosphere-to-corona simulations driven by observations in the photospheric layer.
Our philosophy has been to prioritize consistency with the underlying physical equations of MHD over strict adherence to temporally interpolated observational data because, in practice, such data will contain a variety of random and systematic errors.
The result of our approach is a data-driven boundary layer that is a valid solution to the time-integrated MHD equations which matches the observed boundary evolution as closely as possible.
This holds to the extent that the numerically discretized equations match the true equations.

This investigation provides a detailed and practical guide for the implementation of our data-driven boundary conditions using the combination of a characteristic-based finite volume scheme, \charc{}, and an optimization approach to produce an optimal data-driven boundary condition, \charc{}-\ddbc{}.
The optimization approach has the effect of continually ``nudging'' the boundary in an MHD-allowed direction in state-space towards the best estimate of the state derived from observations.
Critically, our approach can directly handle missing, noisy, and/or systematically biased data.

We validated our approach by reproducing a ground-truth \lare{} simulation of an expanding spheromak.
This test problem was chosen because it is as simple as possible while still retaining many dynamical properties in common with emerging active regions on the Sun.
The results were satisfyingly unexciting.
When high cadence synthetic observations were provided to the \charc{}-\ddbc{}, the output matched the ground truth simulation at the expected level of accuracy, given the different numerical schemes of the two codes.
On the other hand, when the input cadence was lowered, so that a great deal of (physically incorrect) interpolation was used between observed boundary states, our approach lost the ability to accurately replicate the ground truth simulation. 
Further, the loss of fidelity occurred at a previously predicted critical cadence determined by physical properties of the system dynamics.

Our validation also included the case where we removed all information about the evolution of some variables, namely, the mass and internal energy densities.
This mirrors the practical application to SDO/HMI data, which will remain the most promising observational source for data driving for the foreseeable future.
In this case, setting up an SVD problem to find an optimized solution for incoming characteristic derivatives was able to produce a solution with improved fidelity compared to holding those variables constant in the lower boundary.
This test demonstrates that the characteristics can be used in practical applications to constrain variables about which one has little reliable observational information.

Our approach to boundary conditions, and their implementation in the \charc{} code, has uses beyond data-driven boundary conditions.
For one, the use of characteristics for analyzing simulation interiors seems underappreciated.
Because the characteristics are associated with the distinct types of information that propagate through a plasma, they could be used, for example, to track energy associated with different modes as they propagate through a system, and thus identify locations of mode conversion or dissipation.

As a second example, we have already demonstrated that MHD codes with different numerical schemes can be coupled together with \charc{}.
This ability could be taken further, for instance to join together PIC codes with MHD codes, or to incorporate multi-height observations of the solar atmosphere into a single driven simulation.

Third, as we have stated, \emph{any} valid boundary conditions can be represented in characteristic form.
In particular, arbitrary open boundary conditions can be represented in characteristic form.
Obviously, this has long been recognized \citep{Hedstrom:1979}, but mostly applied to cases where one wishes waves to leave a simulation with minimal back-reaction on the internal dynamics.
But for photosphere-to-corona simulations, we often want to study the eruptive phenomena of flares and coronal mass ejections that produce great changes both locally and in the far-field regions. 
It is unclear what boundary conditions for the side and top boundaries are best suited to that task.
We refer the reader to \citetalias{Kee:2023} for an initial exploration of that topic, in which we implement and compare two common versions of nonreflecting boundary conditions using \charc{}.
In this light, data-driven open boundary conditions are far easier to defend than other open boundary conditions, which essentially impose \textit{ad hoc} constraints at each location in space and time.  
In contrast, for data-driven boundary conditions, the data themselves guide what the spatially and temporally varying boundary condition must be in order to accommodate the ongoing simulation, the observations at the boundary, and the propagation of information governed by the MHD equations.

Finally, for data driving itself, the framework we have developed in this paper can be greatly extended.
The current tests are the initial proof-of-concept, but have already demonstrated the ability of combining the characteristics with an optimization approach to back out the evolution of primitive variables about which one nominally knows nothing.  
Future work will apply the method to actual observations of solar active regions, extend to a nonlinear optimization method, and include other physical constraints.

\acknowledgments
This work is supported by the National Solar Observatory, the Office of Naval Research, and the NASA HSR and LWS Programs.  
MGL acknowledges support from the Office of Naval Research. 
JEL, MGL, PWS, and LAT acknowledge support from NASA grant NNH16ZDA001N-LWS “Implementing and Evaluating a Vector-Magnetogram-Driven Magnetohydrodynamic Model of the Magnetic Field in the Low Solar Atmosphere.” MGL acknowledges support from NASA grants NNH17ZDA001N-LWS “Investigating Magnetic Flux Emergence with Modeling and Observations to Understand the Onset of Major Solar Eruptions,” NNH18ZDA001N-HSR “Investigating Magnetic Flux Rope Emergence as the Source of Flaring Activity in Delta-Spot Active Regions,” and NNH20ZDA001N-HSR “Investigating the Influence of
Coronal Magnetic Geometry on the Acceleration of the Solar Wind.”
PWS and JEL acknowledge support from the NASA Internal Science Funding Model (H-ISFM) program “Magnetic Energy Buildup and Explosive
Release in the Solar Atmosphere.”
This research has made use of NASA's Astrophysics Data System.

\software{
  Correctness of the MHD eigensystem was verified using \textsl{Mathematica} \citep{Mathematica}.
  The primary 3D MHD code, \lare{} v.2.10 \citep{Arber:2001}, is written in Fortran 2003 and parallelized with MPI.
  3D renderings were generated using the \fnc{VisIt} visualization software \citep{HPV:VisIt}.
  The majority of the numerical analysis was carried out using the \fnc{NumPy} package \citep{Harris:2020} and figures were prepared using the \fnc{Matplotlib} library \citep{Hunter:2007} in \fnc{python3}.
  The \fnc{dgelsd} algorithm to solve the SVD problem is included in LAPACK \citep{lapack}.
}

\bibliography{localstuff}
\bibliographystyle{aasjournal}

\appendix
\section{Left and Right Eigenvectors of \texorpdfstring{$\Matrix{A}_z$}{Az}}\label{sec:Az}
The right eigenvectors, which satisify $\Matrix{A}_z\cdot\vect{r}_\sigma=\lambda_\sigma\vect{r}_\sigma$, are as follows:
\begin{equation}
  \vect{r}_1=\left.\begin{aligned}\begin{pmatrix}
      0    \\
      0 \\
      0 \\
      0 \\
      0 \\
      0 \\
      0 \\
      1 
  \end{pmatrix}\end{aligned}\right., \vect{r}_2=\left.\begin{aligned}\begin{pmatrix}
      -\rho    \\
      \epsilon \\
      0 \\
      0 \\
      0 \\
      0 \\
      0 \\
      0 
  \end{pmatrix}\end{aligned}\right., \vect{r}_{3,4}^\pm=\left.\begin{aligned}\begin{pmatrix}
      0                         \\
      0                         \\
      \pm\beta_y                     \\
      \mp\beta_x                     \\
      0                         \\
      -\beta_y\sqrt{\rho}s_z \\ 
      \beta_x\sqrt{\rho}s_z \\ 
      0                         
  \end{pmatrix}\end{aligned}\right., \vect{r}_{5,6}^{\pm}=\left.\begin{aligned}\begin{pmatrix}
      \rho\alpha_s                         \\
      \frac{a^2}{\gamma}\alpha_s           \\
      \pm \beta_x\alpha_fc_fs_z  \\
      \pm \beta_y\alpha_fc_fs_z  \\
      \pm c_s\alpha_s                \\
      -\beta_x\sqrt{\rho}a\alpha_f \\
      -\beta_y\sqrt{\rho}a\alpha_f \\
      0                         
  \end{pmatrix}\end{aligned}\right.,\vect{r}_{7,8}^{\pm}=\left.\begin{aligned}\begin{pmatrix}
      \rho\alpha_f                         \\
      \frac{a^2}{\gamma}\alpha_f           \\
      \mp \beta_x\alpha_sc_ss_z  \\
      \mp \beta_y\alpha_sc_ss_z  \\
      \pm c_f\alpha_f                         \\
      \beta_x\sqrt{\rho}a\alpha_s \\
      \beta_y\sqrt{\rho}a\alpha_s \\
      0                         
  \end{pmatrix}\end{aligned}\right..
\end{equation}

Likewise, the left eigenvectors, which satisfy $\vect{l}_\sigma^T\cdot\Matrix{A}_z = \lambda_\sigma\vect{l}_\sigma^T$, are
\begin{equation}\label{eq:lz-columns}
  \vect{l}_1=\left.\begin{aligned}\begin{pmatrix}
      0    \\
      0 \\
      0 \\
      0 \\
      0 \\
      0 \\
      0 \\
      1 
  \end{pmatrix}\end{aligned}\right., \vect{l}_2=\left.\begin{aligned}\begin{pmatrix}
      \frac{1-\gamma}{\gamma\rho} \\
      \frac{1}{\gamma\epsilon} \\
      0 \\
      0 \\
      0 \\
      0 \\
      0 \\
      0 
  \end{pmatrix}\end{aligned}\right.,(\vect{l}_{3,4}^\pm)=\frac{1}{2}\left.\begin{aligned}\begin{pmatrix}
      0                         \\
      0                         \\
      \pm\beta_y                     \\
      \mp\beta_x                     \\
      0                         \\
      -\frac{\beta_ys_z}{\sqrt{\rho}} \\ 
      \frac{\beta_xs_z}{\sqrt{\rho}} \\ 
      0                          \\
  \end{pmatrix}\end{aligned}\right.,(\vect{l}_{5,6}^{\pm})=\frac{1}{2a^2}\left.\begin{aligned}\begin{pmatrix}
      \frac{\alpha_s a^2}{\gamma\rho}          \\
      \frac{\alpha_s a^2}{\gamma\epsilon}      \\
      \pm \beta_x\alpha_fc_fs_z  \\
      \pm \beta_y\alpha_fc_fs_z  \\
      \pm c_s\alpha_s                         \\
      -\beta_x\frac{a \alpha_f}{\sqrt{\rho}} \\
      -\beta_y\frac{a \alpha_f}{\sqrt{\rho}} \\
      0                         
  \end{pmatrix}\end{aligned}\right.,(\vect{l}_{7,8}^{\pm})=\frac{1}{2a^2}\left.\begin{aligned}\begin{pmatrix}
      \frac{\alpha_f a^2}{\gamma\rho}          \\
      \frac{\alpha_f a^2}{\gamma\epsilon}      \\
      \mp \beta_x\alpha_sc_ss_z  \\
      \mp \beta_y\alpha_sc_ss_z  \\
      \pm c_f\alpha_f                         \\
      \beta_x\frac{a \alpha_s}{\sqrt{\rho}} \\
      \beta_y\frac{a \alpha_s}{\sqrt{\rho}} \\
      0                         
  \end{pmatrix}\end{aligned}\right..
\end{equation}
In all these expressions, the first subscript refers to the lower sign, and the second to the upper sign.  
The eigenvectors are biorthonormal in the sense that $\vect{l}_\sigma^T\cdot\vect{r}_\zeta = \delta_{\sigma,\zeta}$.
Note that, because $\Matrix{A}_z$ is not orthogonal matrix, the left and right eigenvectors are elements of two distinct vector spaces.

The right eigenmatrix has columns given by the right eigenvectors, while the left eigenmatrix has rows given by the left eigenvectors:
\begin{equation}
  \rightzevecmat = \left.\begin{bmatrix}\begin{array}{cccccccc}
      \vert & \vert & \vert & \vert & \vert & \vert & \vert & \vert \\
      \vect{r}_1 & \vect{r}_2 & \vect{r}_3 & \vect{r}_4 & \vect{r}_5 & \vect{r}_6 & \vect{r}_7 & \vect{r}_8 \\
      \vert & \vert & \vert & \vert & \vert & \vert & \vert & \vert
      \end{array}\end{bmatrix}\right., \quad  \smatinv_z = \left.\begin{bmatrix}\begin{array}{ccc}
      \horzbar & \vect{l}_1^T & \horzbar \\
      \horzbar & \vect{l}_2^T & \horzbar \\
      \horzbar & \vect{l}_3^T & \horzbar \\
      \horzbar & \vect{l}_4^T & \horzbar \\
      \horzbar & \vect{l}_5^T & \horzbar \\
      \horzbar & \vect{l}_6^T & \horzbar \\
      \horzbar & \vect{l}_7^T & \horzbar \\
      \horzbar & \vect{l}_8^T & \horzbar
      \end{array}\end{bmatrix}\right..
\end{equation}
Fully written out, the matrices are

\begin{equation}\label{eq:sz}
  \rightzevecmat = \begin{bmatrix}
    0  & -\rho    & 0                        & 0                     & \rho\alpha_s                &  \rho\alpha_s                 & \rho\alpha_f                  &  \rho\alpha_f                 \\
    0  & \epsilon & 0                        & 0                     & \frac{a^2}{\gamma}\alpha_s  &  \frac{a^2}{\gamma}\alpha_s   & \frac{a^2}{\gamma}\alpha_f    &   \frac{a^2}{\gamma}\alpha_f   \\
    0  & 0        & -\beta_y                 &  +\beta_y             & - \beta_x\alpha_fc_fs_z      &  + \beta_x\alpha_fc_fs_z       & + \beta_x\alpha_sc_ss_z     &    - \beta_x\alpha_sc_ss_z      \\
    0  & 0        & +\beta_x                 &  -\beta_x             & - \beta_y\alpha_fc_fs_z      &  + \beta_y\alpha_fc_fs_z       & + \beta_y\alpha_sc_ss_z     &    - \beta_y\alpha_sc_ss_z      \\
    0  & 0        & 0                        & 0                     & - c_s\alpha_s               &  + c_s\alpha_s                 & - c_f\alpha_f               &   + c_f\alpha_f               \\
    0  & 0        & -\beta_y\sqrt{\rho}s_z   & -\beta_y\sqrt{\rho}s_z &-\beta_x\sqrt{\rho}a\alpha_f &  -\beta_x\sqrt{\rho}a\alpha_f  & \beta_x\sqrt{\rho}a\alpha_s   &   \beta_x\sqrt{\rho}a\alpha_s    \\
    0  & 0        & \beta_x\sqrt{\rho}s_z    & \beta_x\sqrt{\rho}s_z  &-\beta_y\sqrt{\rho}a\alpha_f &  -\beta_y\sqrt{\rho}a\alpha_f  & \beta_y\sqrt{\rho}a\alpha_s   &   \beta_y\sqrt{\rho}a\alpha_s    \\
    1  & 0        & 0                        & 0                     & 0                           &  0                            & 0                             &  0                          
  \end{bmatrix}
\end{equation}

and

\begin{equation}\label{eq:szm1}
  \leftzevecmat = \left.\begin{aligned}\begin{bmatrix}
 0 & 0 & 0 & 0 & 0 & 0 & 0 & 1 \\
 -\frac{\gamma -1}{\gamma  \rho } & \frac{1}{\gamma  \epsilon } & 0 & 0 & 0 & 0 & 0 & 0 \\
 0 & 0 & -\frac{\beta _y}{2} & \frac{\beta _x}{2} & 0 & -\frac{\beta _y s_z}{2 \sqrt{\rho }} & \frac{\beta _x s_z}{2 \sqrt{\rho }} & 0 \\
 0 & 0 & \frac{\beta _y}{2} & -\frac{\beta _x}{2} & 0 & -\frac{\beta _y s_z}{2 \sqrt{\rho }} & \frac{\beta _x s_z}{2 \sqrt{\rho }} & 0 \\
 \frac{\alpha _s}{2 \gamma  \rho } & \frac{\alpha _s}{2 \gamma  \epsilon } & -\frac{c_f \alpha _f \beta _x s_z}{2 a^2} & -\frac{c_f \alpha _f \beta _y s_z}{2 a^2} & -\frac{c_s \alpha _s}{2 a^2} & -\frac{\alpha _f \beta _x}{2 a
   \sqrt{\rho }} & -\frac{\alpha _f \beta _y}{2 a \sqrt{\rho }} & 0 \\
 \frac{\alpha _s}{2 \gamma  \rho } & \frac{\alpha _s}{2 \gamma  \epsilon } & \frac{c_f \alpha _f \beta _x s_z}{2 a^2} & \frac{c_f \alpha _f \beta _y s_z}{2 a^2} & \frac{c_s \alpha _s}{2 a^2} & -\frac{\alpha _f \beta _x}{2 a
   \sqrt{\rho }} & -\frac{\alpha _f \beta _y}{2 a \sqrt{\rho }} & 0 \\
 \frac{\alpha _f}{2 \gamma  \rho } & \frac{\alpha _f}{2 \gamma  \epsilon } & \frac{c_s \alpha _s \beta _x s_z}{2 a^2} & \frac{c_s \alpha _s \beta _y s_z}{2 a^2} & -\frac{c_f \alpha _f}{2 a^2} & \frac{\alpha _s \beta _x}{2 a
   \sqrt{\rho }} & \frac{\alpha _s \beta _y}{2 a \sqrt{\rho }} & 0 \\
 \frac{\alpha _f}{2 \gamma  \rho } & \frac{\alpha _f}{2 \gamma  \epsilon } & -\frac{c_s \alpha _s \beta _x s_z}{2 a^2} & -\frac{c_s \alpha _s \beta _y s_z}{2 a^2} & \frac{c_f \alpha _f}{2 a^2} & \frac{\alpha _s \beta _x}{2 a
   \sqrt{\rho }} & \frac{\alpha _s \beta _y}{2 a \sqrt{\rho }} & 0 \\
   \end{bmatrix}\end{aligned}\right..
\end{equation}

\section{Left and Right Eigenmatrices of \texorpdfstring{$\Matrix{A}_x$}{Ax}}\label{sec:Ax}
The coefficient matrix in the x-direction, $\Matrix{A}_x$, from Equation \eqref{eq:mhd-matrix} is given by
\begin{equation}
  \Matrix{A}_{x}=\left.\begin{aligned}\begin{bmatrix}
      v_x &   0 &   \rho             &   0 & 0 & 0          & 0        & 0 \\
      0   & v_x & (\gamma-1)\epsilon &   0 & 0 & 0          & 0        & 0 \\
      \frac{(\gamma-1)\epsilon}{\rho} & (\gamma-1) & v_x & 0 & 0 & 0 & B_y/\rho & B_z/\rho \\
      0 & 0 & 0 & v_x & 0 & 0 &-B_x/\rho & 0 \\
      0 & 0 & 0 & 0 & v_x & 0 & 0 & -B_x/\rho \\
      0 & 0 & 0 & 0 & 0 & v_x & 0 & 0\\
      0 & 0 & B_y & -B_x & 0 & 0  & v_x & 0\\
      0 & 0 & B_z & 0    & -B_x & 0 & 0 & v_x\\
  \end{bmatrix}\end{aligned}\right..
\end{equation}

The right and left eigenmatrices of $\Matrix{A}_x$ are then calculated in a similar fashion as for $\Matrix{A}_z$, except the correct projections of the Alfv\'en velocities onto the normal and perpendicular directions must be used in the definitions of the eigenvalues and vectors, i.e., $v_z\rightarrow v_x$, $c_a\rightarrow \abs{b_x}$ (including inside the definitions of $c_{s,f}$ and $\alpha_{s,f}$), $b_\perp^2\rightarrow b_y^2+b_z^2$, and the tangential ratios are now $\beta_y = b_y/b_\perp$ and $\beta_z = b_z/b_\perp$.
With those definitions, the right and left eigenmatrices in the x-direction are
\begin{equation}\label{eq:sx}
  \rightxevecmat = \left.\begin{aligned}\begin{bmatrix}
      0 & -\rho  & 0 & 0 & \rho  \alpha _s & \rho  \alpha _s & \rho  \alpha _f & \rho  \alpha _f \\
 0 & \epsilon  & 0 & 0 & \frac{a^2 \alpha _s}{\gamma } & \frac{a^2 \alpha _s}{\gamma } & \frac{a^2 \alpha_f}{\gamma } & \frac{a^2 \alpha _f}{\gamma } \\
 0 & 0 & 0 & 0 & -c_s \alpha _s & c_s \alpha _s & -c_f \alpha _f & c_f \alpha _f \\
 0 & 0 & -\beta _z & \beta _z & -s_x c_f \alpha _f \beta _y & s_x c_f \alpha _f \beta _y & s_x c_s \alpha _s \beta _y & -s_x c_s \alpha _s \beta _y \\
 0 & 0 & \beta _y & -\beta _y & -s_x c_f \alpha _f \beta _z & s_x c_f \alpha _f \beta _z & s_x c_s \alpha _s \beta _z & -s_x c_s \alpha _s \beta _z \\
 1 & 0 & 0 & 0 & 0 & 0 & 0 & 0 \\
 0 & 0 & -\sqrt{\rho } s_x \beta _z & -\sqrt{\rho } s_x \beta_z & -a \sqrt{\rho } \alpha _f \beta _y & -a \sqrt{\rho } \alpha _f \beta _y & a \sqrt{\rho } \alpha _s\beta _y & a \sqrt{\rho } \alpha _s \beta _y \\
 0 & 0 & \sqrt{\rho } s_x \beta _y & \sqrt{\rho } s_x \beta_y & -a \sqrt{\rho } \alpha _f \beta _z & -a \sqrt{\rho } \alpha _f \beta _z & a \sqrt{\rho } \alpha _s\beta _z & a \sqrt{\rho } \alpha _s \beta _z \\
  \end{bmatrix}\end{aligned}\right.
\end{equation}

and

\begin{equation}\label{eq:sxm1}
  \leftxevecmat = \left.\begin{aligned}\begin{bmatrix}
    0 & 0 & 0 & 0 & 0 & 1 & 0 & 0 \\
 -\frac{\gamma -1}{\gamma  \rho } & \frac{1}{\gamma  \epsilon } & 0 & 0 & 0 & 0 & 0 & 0 \\
 0 & 0 & 0 & -\frac{\beta _z}{2} & \frac{\beta _y}{2} & 0 & -\frac{s_x \beta _z}{2\sqrt{\rho }} & \frac{s_x \beta _y}{2 \sqrt{\rho }} \\
 0 & 0 & 0 & \frac{\beta _z}{2} & -\frac{\beta _y}{2} & 0 & -\frac{s_x \beta _z}{2\sqrt{\rho }} & \frac{s_x \beta _y}{2 \sqrt{\rho }} \\
 \frac{\alpha _s}{2 \gamma  \rho } & \frac{\alpha _s}{2 \gamma  \epsilon } & -\frac{c_s \alpha _s}{2 a^2} &-\frac{s_x c_f \alpha _f \beta _y}{2 a^2} & -\frac{s_x c_f\alpha _f \beta _z}{2 a^2} & 0 & -\frac{\alpha _f \beta _y}{2 a \sqrt{\rho }} & -\frac{\alpha _f \beta_z}{2 a \sqrt{\rho }} \\
 \frac{\alpha _s}{2 \gamma  \rho } & \frac{\alpha _s}{2 \gamma  \epsilon } & \frac{c_s \alpha _s}{2 a^2} &\frac{s_x c_f \alpha _f \beta _y}{2 a^2} & \frac{s_x c_f\alpha _f \beta _z}{2 a^2} & 0 & -\frac{\alpha _f \beta _y}{2 a \sqrt{\rho }} & -\frac{\alpha _f \beta_z}{2 a \sqrt{\rho }} \\
 \frac{\alpha _f}{2 \gamma  \rho } & \frac{\alpha _f}{2 \gamma  \epsilon } & -\frac{c_f \alpha _f}{2 a^2} &\frac{s_x c_s \alpha _s \beta _y}{2 a^2} & \frac{s_x c_s\alpha _s \beta _z}{2 a^2} & 0 & \frac{\alpha _s \beta _y}{2 a \sqrt{\rho }} & \frac{\alpha _s \beta_z}{2 a \sqrt{\rho }} \\
 \frac{\alpha _f}{2 \gamma  \rho } & \frac{\alpha _f}{2 \gamma  \epsilon } & \frac{c_f \alpha _f}{2 a^2} & -\frac{s_x c_s \alpha _s \beta _y}{2 a^2} & -\frac{s_x c_s\alpha _s \beta _z}{2 a^2} & 0 & \frac{\alpha _s \beta _y}{2 a \sqrt{\rho }} & \frac{\alpha _s \beta_z}{2 a \sqrt{\rho }} \\
  \end{bmatrix}\end{aligned}\right..
\end{equation}

The left and right eigenmatrices have been verified (using \textsl{Mathematica}) to be biorthonormal in the sense that $\leftxevecmat\rightxevecmat = \Matrix{I}$.

\section{Left and Right Eigenmatrices of \texorpdfstring{$\Matrix{A}_y$}{Ay}}\label{sec:Ay}
The coefficient matrix in the y-direction, $\Matrix{A}_y$, from Equation \eqref{eq:mhd-matrix} is given by

\begin{equation}
  \Matrix{A}_{y}=\left.\begin{aligned}\begin{bmatrix}
      v_y &   0 & 0 & \rho & 0 & 0 & 0 & 0 \\
      0   & v_y & 0 & (\gamma-1)\epsilon & 0 & 0 & 0 & 0 \\
      0   & 0   & v_y & 0 & 0 &-B_y/\rho & 0 & 0 \\
      \frac{(\gamma-1)\epsilon}{\rho} & (\gamma-1) & 0 & v_y & 0 & B_x/\rho & 0 & B_z/\rho \\
      0 & 0 & 0 & 0 & v_y & 0 & 0 & -B_y/\rho \\
      0 & 0 & -B_y & B_x & 0 & v_y & 0 & 0 \\
      0 & 0 &  0   & 0   & 0 & 0  & v_y & 0\\
      0 & 0 & 0 & B_z & -B_y & 0 & 0 & v_y\\
  \end{bmatrix}\end{aligned}\right..
\end{equation}

The right and left eigenmatrices of $\Matrix{A}_y$ are then calculated in a similar fashion as for $\Matrix{A}_z$, except the correct projections of the Alfv\'en velocities onto the normal and perpendicular directions must be used in the definitions of the eigenvalues and vectors, i.e., $v_z\rightarrow v_y$, $c_a\rightarrow \abs{b_y}$ (including inside the definitions of $c_{f,s}$ and $\alpha_{s,f}$), $b_\perp^2\rightarrow b_x^2+b_z^2$, and the tangential ratios are $\beta_x = b_x/b_\perp$ and $\beta_z = b_z/b_\perp$.
With those definitions, the right and left eigenmatrices are

\begin{equation}\label{eq:sy}
  \rightyevecmat = \left.\begin{aligned}\begin{bmatrix}
 0 & -\rho  & 0 & 0 & \rho  \alpha _s & \rho  \alpha _s & \rho  \alpha _f & \rho  \alpha _f \\
 0 & \epsilon  & 0 & 0 & \frac{a^2 \alpha _s}{\gamma } & \frac{a^2 \alpha _s}{\gamma } & \frac{a^2 \alpha
   _f}{\gamma } & \frac{a^2 \alpha _f}{\gamma } \\
 0 & 0 & \beta _z & -\beta _z & -s_y c_f \alpha _f \beta _x &
   s_y c_f \alpha _f \beta _x & s_y c_s \alpha _s \beta _x &
   -s_y c_s \alpha _s \beta _x \\
 0 & 0 & 0 & 0 & -c_s \alpha _s & c_s \alpha _s & -c_f \alpha _f & c_f \alpha _f \\
 0 & 0 & -\beta _x & \beta _x & -s_y c_f \alpha _f \beta _z &
   s_y c_f \alpha _f \beta _z & s_y c_s \alpha _s \beta _z &
   -s_y c_s \alpha _s \beta _z \\
 0 & 0 & \sqrt{\rho } s_y \beta _z & \sqrt{\rho } s_y \beta
   _z & -a \sqrt{\rho } \alpha _f \beta _x & -a \sqrt{\rho } \alpha _f \beta _x & a \sqrt{\rho } \alpha _s
   \beta _x & a \sqrt{\rho } \alpha _s \beta _x \\
 1 & 0 & 0 & 0 & 0 & 0 & 0 & 0 \\
 0 & 0 & -\sqrt{\rho } s_y \beta _x & -\sqrt{\rho } s_y \beta_x & -a \sqrt{\rho } \alpha _f \beta _z & -a \sqrt{\rho } \alpha _f \beta _z & a \sqrt{\rho } \alpha _s\beta _z & a \sqrt{\rho } \alpha _s \beta _z \\

  \end{bmatrix}\end{aligned}\right.,
\end{equation}

and

\begin{equation}\label{eq:sym1}
  \leftyevecmat = \left.\begin{aligned}\begin{bmatrix}
    0 & 0 & 0 & 0 & 0 & 0 & 1 & 0 \\
    -\frac{\gamma -1}{\gamma  \rho } & \frac{1}{\gamma  \epsilon } & 0 & 0 & 0 & 0 & 0 & 0 \\
    0 & 0 & \frac{\beta _z}{2} & 0 & -\frac{\beta _x}{2} & \frac{s_y \beta _z}{2
      \sqrt{\rho }} & 0 & -\frac{s_y \beta _x}{2 \sqrt{\rho }} \\
    0 & 0 & -\frac{\beta _z}{2} & 0 & \frac{\beta _x}{2} & \frac{s_y \beta _z}{2
      \sqrt{\rho }} & 0 & -\frac{`s_y \beta _x}{2 \sqrt{\rho }} \\
    \frac{\alpha _s}{2 \gamma  \rho } & \frac{\alpha _s}{2 \gamma  \epsilon } &
    -\frac{s_y c_f \alpha _f \beta _x}{2 a^2} & -\frac{c_s \alpha _s}{2 a^2} &
    -\frac{s_y c_f \alpha _f \beta _z}{2 a^2} & -\frac{\alpha _f \beta _x}{2 a
      \sqrt{\rho }} & 0 & -\frac{\alpha _f \beta _z}{2 a \sqrt{\rho }} \\
 \frac{\alpha _s}{2 \gamma  \rho } & \frac{\alpha _s}{2 \gamma  \epsilon } &
 \frac{s_y c_f \alpha _f \beta _x}{2 a^2} & \frac{c_s \alpha _s}{2 a^2} &
 \frac{s_y c_f \alpha _f \beta _z}{2 a^2} & -\frac{\alpha _f \beta _x}{2 a
   \sqrt{\rho }} & 0 & -\frac{\alpha _f \beta _z}{2 a \sqrt{\rho }} \\
 \frac{\alpha _f}{2 \gamma  \rho } & \frac{\alpha _f}{2 \gamma  \epsilon } &
 \frac{s_y c_s \alpha _s \beta _x}{2 a^2} & -\frac{c_f \alpha _f}{2 a^2} &
 \frac{s_y c_s \alpha _s \beta _z}{2 a^2} & \frac{\alpha _s \beta _x}{2 a
   \sqrt{\rho }} & 0 & \frac{\alpha _s \beta _z}{2 a \sqrt{\rho }} \\
 \frac{\alpha _f}{2 \gamma  \rho } & \frac{\alpha _f}{2 \gamma  \epsilon } &
 -\frac{s_y c_s \alpha _s \beta _x}{2 a^2} & \frac{c_f \alpha _f}{2 a^2} &
 -\frac{s_y c_s \alpha _s \beta _z}{2 a^2} & \frac{\alpha _s \beta _x}{2 a
   \sqrt{\rho }} & 0 & \frac{\alpha _s \beta _z}{2 a \sqrt{\rho }} \\
  \end{bmatrix}\end{aligned}\right..
\end{equation}

The left and right eigenmatrices have been verified (using \textsl{Mathematica}) to be biorthonormal in the sense that $\leftyevecmat\rightyevecmat = \Matrix{I}$.

\section{Transverse terms; finite volume versus finite difference approaches}\label{sec:transverse}
The vector (in state-space) of transverse-to-\zhat{} terms $\vect{C}$ defined in Equation \ref{eq:c} has elements corresponding to each primitive variable given by:
\begin{subequations}
  \begin{align}
    C_\rho & = v_x\partial_x\rho + \rho \partial_xv_x + v_y\partial_y\rho+\rho\partial_yv_y\\
    C_\epsilon & = v_x\partial_x\epsilon + (\gamma-1)\epsilon \partial_xv_x + v_y\partial_y\epsilon + (\gamma-1)\epsilon\partial_yv_y\\
    C_{v_x} & = (\gamma-1)\frac{\epsilon}{\rho}\partial_x\rho + (\gamma-1)\partial_x\epsilon + v_x\partial_xv_x + \rho^{-1}B_y\partial_xB_y \notag\\
    & \qquad\qquad\qquad\qquad + \rho^{-1}B_z\partial_xB_z + v_y\partial_yv_x - \rho^{-1}B_y\partial_yB_x\\
    C_{v_y} & = v_x\partial_xv_y -\rho^{-1}B_x\partial_xB_y + (\gamma-1)\frac{\epsilon}{\rho}\partial_y\rho + (\gamma-1)\partial_y\epsilon \notag \\
    & \qquad\qquad\qquad\qquad + v_y\partial_yv_y + \rho^{-1}B_x\partial_yB_x + \rho^{-1}B_z\partial_yB_z \\
    C_{v_z} & = v_x\partial_xv_z - \rho^{-1}B_x\partial_xB_z + v_y\partial_yv_z - \rho^{-1}B_y\partial_yB_z + g\\
    C_{B_x} & = v_x\partial_xB_x - B_y\partial_yv_x + B_x\partial_yv_y+v_y\partial_yB_x\\
    C_{B_y} & = B_y\partial_xv_x-B_x\partial_xv_y + v_x\partial_xB_y + v_y\partial_yB_y\\
    C_{B_z} & = B_z\partial_xv_x - B_x\partial_xv_z + v_x\partial_xB_z+B_z\partial_yv_y - B_y\partial_yv_z + v_y\partial_yB_z.
  \end{align}
\end{subequations}

In the boundary layer, the transverse derivatives are treated in the same way as derivatives in the normal direction, i.e., taking the finite volume approach and splitting the eigensystem into positive and negative subspaces at each interface.
For the present investigation we avoided introducing additional complications by using periodic side boundaries for all simulations and stopping our analysis before any disturbance reached those boundaries.

On the other hand, our first attempt at implementing characteristic boundary conditions used a finite difference approach for treating the transverse terms in the boundary layer.
This method proved to be a poor numerical representation of MHD and also numerically unstable, which precipitated our switch to the finite volume approach described in \S\ref{sec:fvddbc}.
We briefly sketch the finite difference method as a record of what did not work, which may be of use to other researchers.

Our unstable finite difference approximation to the transverse terms was nearly identical to the finite volume approach of \S\ref{sec:fvddbc}, except that the eigensystems were calculated at cell centers instead of cell interfaces.
The projected characteristic speeds (i.e., the eigenvalues of each of the transverse coefficient matrices $\Matrix{A}_x$ and $\Matrix{A}_y$) were then used as weights for upwind finite differences.
This approach is similar to the one described in the appendix of \citet{Jiang:2011}, but likely differs in the details.
First, we diagonalized the transverse (\xhat{} and \yhat{}) terms as at \eqref{eq:diagonalization}:
\begin{equation}
  \Matrix{A}_x = \rightxevecmat\xevalmat\leftxevecmat,\ \Matrix{A}_y = \rightyevecmat\yevalmat\leftyevecmat.
\end{equation}
The eigenvalues in \xevalmat{} and \yevalmat{} are the same as in Equation \eqref{eq:mhd-evals} after the replacements specified in Appendices \ref{sec:Ax} and \ref{sec:Ay}, e.g. for the \yhat{}-direction: $v_z\rightarrow v_y$, $c_{z,a}\rightarrow c_{y,a}(=\abs{b_y})$ (including inside the definitions of $c_{s,f}$ and $\alpha_{s,f}$), and $b_\perp^2\rightarrow b_x^2+b_z^2$; similar replacements are applied in $\xhat$.

The nonnegative and nonpositive coefficient matrices in each transverse direction were calculated according to
\begin{equation}
  \Matrix{A}_{y+} = \rightyevecmat\frac{\Matrix{\Lambda}_y+\abs{\Matrix{\Lambda}_y}}{2}\leftyevecmat,\quad  \Matrix{A}_{y-} = \rightyevecmat\frac{\Matrix{\Lambda}_y-\abs{\Matrix{\Lambda}_y}}{2}\leftyevecmat,
\end{equation}
where $\abs{\Matrix{\Lambda}} = \fnc{diag}[\abs{\lambda}]$ in each direction.  
As before, the positive (negative) matrix corresponds to grid-positive (negative) propagating disturbances.  
Decomposed this way, the transverse derivatives are
\begin{equation}
  \Matrix{A}_x\cdot\partial_x\vect{U}+\Matrix{A}_y\cdot\partial_y\vect{U} = (\Matrix{A}_{x+}+\Matrix{A}_{x-})\cdot\partial_x\vect{U}+(\Matrix{A}_{y+}+\Matrix{A}_{y-})\cdot\partial_y\vect{U}.
\end{equation}

To approximate this equation as a finite difference, the positive and negative eigenmatrices were evaluated at cell centers while the derivatives were taken in the upwind direction for each of the positive and negative subspaces.
Effectively, an eigenspeed-dependent set of cell face centered derivatives were back projected onto the cell center eigensystem to update the cell centered state variable.
Thus, we used a backward Euler finite difference approximation projected onto the positive matrix and a forward Euler finite difference approximation projected onto the negative matrix.
In the \xhat{} direction the approximation is
\begin{equation}
  \Matrix{A}_x\cdot\partial_x\vect{U} \approx \Matrix{A}_{x+,i}\cdot\frac{\vect{U}_{ijk}-\vect{U}_{i-1,j,k}}{x_i-x_{i-1}} 
  + \Matrix{A}_{x-,i}\cdot\frac{\vect{U}_{i+1,j,k}-\vect{U}_{i,j,k}}{x_{i+1}-x_{i}},
\end{equation}
and in \yhat{}
\begin{equation}
  \Matrix{A}_y\cdot\partial_y\vect{U} \approx 
  \Matrix{A}_{y+,j}\cdot\frac{\vect{U}_{ijk}-\vect{U}_{i,j-1,k}}{y_j-y_{j-1}} 
  + \Matrix{A}_{y-,j}\cdot\frac{\vect{U}_{i,j+1,k}-\vect{U}_{i,j,k}}{y_{j+1}-y_{j}}.
\end{equation}

Written out explicitly, the MHD equations were approximated in a finite difference sense as
\begin{align}
  \partial_t\mhdstate(x_i,y_j,z_k) & = -\smat_z\cdot\charderivz - \Matrix{A}_y\cdot\partial_y\vect{U}-\Matrix{A}_x\cdot\partial_x\vect{U} - \vect{D}\\
  & \approx -(\smat_z\charderivz)_{ijk} - \Bigl[\Matrix{A}_{x+,ijk}\cdot\frac{\vect{U}_{ijk}-\vect{U}_{i-1,j,k}}{x_i-x_{i-1}} + \Matrix{A}_{x-,{ijk}}\cdot\frac{\vect{U}_{i+1,j,k}-\vect{U}_{i,j,k}}{x_{i+1}-x_{i}} \Bigr]\notag\\
  & - \Bigl[\ \Matrix{A}_{y+,{ijk}}\cdot\frac{\vect{U}_{ijk}-\vect{U}_{i,j-1,k}}{y_j-y_{j-1}} + \Matrix{A}_{y-,{ijk}}\cdot\frac{\vect{U}_{i,j+1,k}-\vect{U}_{i,j,k}}{y_{j+1}-y_{j}}\Bigr] -\vect{D}_{ijk},
\end{align}
where the first term, $\smat_z\charderivz$, could use any representation of the characteristic derivatives in the normal direction.

As stated, we found that the approach for transverse terms sketched above was not numerically stable.
The reason is that, individually, the positive and negative subspace matrices $\amat_\pm$ contain terms in each entry, $A_{\pm,\sigma\zeta}$, that are not included in the MHD equations.
However, the combination $\amat = \amatp + \amatm$ has detailed cancellations between the nonphysical terms that, in sum, reproduce the MHD equations.
The detailed balance fails when different derivatives are projected onto the split subspaces of a single eigensystem, or, equivalently, when a single derivative is projected onto two different (split) eigensystems.
For example, the face-centered finite difference $\partial_x\mhdstate\rvert_{i+1/2} = \frac{1}{\Delta x}(\mhdstate_{i+1}-\mhdstate_i)$ is projected onto cell centered eigensystem $\amat_{-,i}$ to update $\mhdstate_{i}$, and onto eigensystem $\amat_{+,i+1}$ to update $\mhdstate_{i+1}$.
The combination of these projections do not produce consistent pairwise cancellations.
Taylor expanding the equations about a given location on the grid shows that the errors are of second order.

In contrast, for the finite volume method we ultimately implemented, a single centered derivative is always projected onto the pair of subspaces derived from the single eigensystem at the cell interface.
Even though those projections update the two different cells on each side of the interface, they do so in a numerically balanced way.
This is why the finite volume method works while the finite difference method fails in this case.

\section{Relation between characteristic derivatives and vertical derivatives of primitive variables}\label{sec:LtoU}
The normal derivative of the primitive variables can be expressed directly in terms of the characteristic derivatives in that direction.
In \zhat{} this is
\begin{gather}
  \partial_z\mhdstate = \Matrix{E}^{-1}\cdot\charderivz = \smat_z\zevalmat^{-1}\cdot\charderivz \tag{\ref{eq:l2uprime} revisited}
\end{gather}
where the inverse matrix is
\begin{gather}
  \Matrix{E}^{-1}=\left.\begin{aligned}\begin{bmatrix}
      0 & -\frac{\rho}{\lambda_2} & 0 & 0 & \frac{\rho\alpha_s}{\lambda_5} & \frac{\rho\alpha_s}{\lambda_6} & \frac{\rho\alpha_f}{\lambda_7} & \frac{\rho\alpha_f}{\lambda_8} \\
      0 & \frac{\epsilon}{\lambda_2} & 0 & 0 & \frac{(\gamma-1)\epsilon\alpha_s}{\lambda_5} & \frac{(\gamma-1)\epsilon\alpha_s}{\lambda_6} & \frac{(\gamma-1)\epsilon\alpha_f}{\lambda_7} & \frac{(\gamma-1)\epsilon\alpha_f}{\lambda_8} \\
      0 & 0 & -\frac{\beta_y}{\lambda_3} & \frac{\beta_y}{\lambda_4} & -\frac{c_f\alpha_f\beta_x}{\lambda_5 s_z} & \frac{c_f\alpha_f\beta_x}{\lambda_6 s_z} & \frac{c_s\alpha_s\beta_x}{\lambda_7 s_z} & -\frac{c_s\alpha_s\beta_x}{\lambda_8 s_z} \\
      0 & 0 & \frac{\beta_x}{\lambda_3} & -\frac{\beta_x}{\lambda_4} & -\frac{c_f\alpha_f\beta_y}{\lambda_5 s_z} & \frac{c_f\alpha_f\beta_y}{\lambda_6 s_z} & \frac{c_s\alpha_s\beta_y}{\lambda_7 s_z} & -\frac{c_s\alpha_s\beta_y}{\lambda_8 s_z}\\
      0 & 0 & 0 & 0 & -\frac{c_s\alpha_s}{\lambda_5} & \frac{c_s\alpha_s}{\lambda_6} & -\frac{c_f\alpha_f}{\lambda_7} & \frac{c_f\alpha_f}{\lambda_8} \\
      0 & 0 & -\frac{\sqrt{\rho}\beta_y}{\lambda_3s_z} & -\frac{\sqrt{\rho}\beta_y}{\lambda_4s_z} & -\frac{a\sqrt{\rho}\alpha_f\beta_x}{\lambda_5} & -\frac{a\sqrt{\rho}\alpha_f\beta_x}{\lambda_6} & \frac{a\sqrt{\rho}\alpha_s\beta_x}{\lambda_7} & \frac{a\sqrt{\rho}\alpha_s\beta_x}{\lambda_8}\\
      0 & 0 & \frac{\sqrt{\rho}\beta_x}{\lambda_3s_z} & \frac{\sqrt{\rho}\beta_x}{\lambda_4s_z} & -\frac{a\sqrt{\rho}\alpha_f\beta_y}{\lambda_5} & -\frac{a\sqrt{\rho}\alpha_f\beta_y}{\lambda_6} & \frac{a\sqrt{\rho}\alpha_s\beta_y}{\lambda_7} & \frac{a\sqrt{\rho}\alpha_s\beta_y}{\lambda_8}\\
      \frac{1}{\lambda_1} & 0 &   0  & 0    & 0 & 0 & 0 & 0
  \end{bmatrix}\end{aligned}\label{eq:Einverse}\right.
\end{gather}

Carrying through the matrix multiplication, the vertical derivatives of primitive variables are given explicitly as
\begin{subequations}\label{eq:uprimefroml}
  \begin{align}
    \partial_z\rho & = -\frac{\rho}{\lambda_2}L_2 + \frac{\rho\alpha_s}{\lambda_5}L_5 + \frac{\rho\alpha_s}{\lambda_6}L_6 + \frac{\rho\alpha_f}{\lambda_7} L_7+ \frac{\rho\alpha_f}{\lambda_8}L_8 \\
    \partial_z\epsilon & = \frac{\epsilon}{\lambda_2}L_2 + \frac{(\gamma-1)\epsilon\alpha_s}{\lambda_5}L_5 + \frac{(\gamma-1)\epsilon\alpha_s}{\lambda_6}L_6 + \frac{(\gamma-1)\epsilon\alpha_f}{\lambda_7} L_7 + \frac{(\gamma-1)\epsilon\alpha_f}{\lambda_8}L_8 \\
    \partial_zv_x & = -\frac{\beta_y}{\lambda_3}L_3 + \frac{\beta_y}{\lambda_4}L_4 - \frac{c_f\alpha_f\beta_x}{\lambda_5 s_z}L_5 + \frac{c_f\alpha_f\beta_x}{\lambda_6 s_z} L_6 + \frac{c_s\alpha_s\beta_x}{\lambda_7 s_z}L_7 -\frac{c_s\alpha_s\beta_x}{\lambda_8 s_z}L_8 \\
    \partial_zv_y & = \frac{\beta_x}{\lambda_3} L_3 -\frac{\beta_x}{\lambda_4} L_4 -\frac{c_f\alpha_f\beta_y}{\lambda_5 s_z}L_5 + \frac{c_f\alpha_f\beta_y}{\lambda_6 s_z}L_6 + \frac{c_s\alpha_s\beta_y}{\lambda_7 s_z}L_7 -\frac{c_s\alpha_s\beta_y}{\lambda_8 s_z}L_8\\
    \partial_zv_z & =-\frac{c_s\alpha_s}{\lambda_5}L_5 + \frac{c_s\alpha_s}{\lambda_6}L_6 -\frac{c_f\alpha_f}{\lambda_7}L_7 + \frac{c_f\alpha_f}{\lambda_8}L_8 \\
    \partial_zB_x & = -\frac{\sqrt{\rho}\beta_y}{\lambda_3s_z}L_3 -\frac{\sqrt{\rho}\beta_y}{\lambda_4s_z}L_4 -\frac{a\sqrt{\rho}\alpha_f\beta_x}{\lambda_5} L_5 -\frac{a\sqrt{\rho}\alpha_f\beta_x}{\lambda_6}L_6 + \frac{a\sqrt{\rho}\alpha_s\beta_x}{\lambda_7}L_7+\frac{a\sqrt{\rho}\alpha_s\beta_x}{\lambda_8}L_8\\
    \partial_zB_y & = \frac{\sqrt{\rho}\beta_x}{\lambda_3s_z}L_3 + \frac{\sqrt{\rho}\beta_x}{\lambda_4s_z}L_4 - \frac{a\sqrt{\rho}\alpha_f\beta_y}{\lambda_5}L_5 -\frac{a\sqrt{\rho}\alpha_f\beta_y}{\lambda_6}L_6 + \frac{a\sqrt{\rho}\alpha_s\beta_y}{\lambda_7}L_7 + \frac{a\sqrt{\rho}\alpha_s\beta_y}{\lambda_8}L_8\\
    \partial_zB_z & = \frac{1}{\lambda_1}L_1.
  \end{align}
\end{subequations}

Suppose one has determined an estimate for the incoming characteristics by some means.
Then, Equations \eqref{eq:uprimefroml} could be used to form a simple finite-difference approximation to set primitive variables in the ghost cell layer using the values of the interior cell:
\begin{gather}
  \partial_z\mhdstate \approx \frac{\mhdstate_0-\mhdstate{-1}}{\Delta z} = \Matrix{E}^{-1}\charderivz\\
  \Longrightarrow \mhdstate_{-1}\approx\mhdstate_0 - \Delta z\Matrix{E}^{-1}\charderivz.
\end{gather}

A two dimensional version of this method for setting boundary conditions for a hydrodynamic system was described and implemented in \citet{Thompson:1987a} and \citet{Thompson:1990}.
Our own implementation worked quite well for a variety of one dimensional MHD problems but failed in higher dimensions when coupled to the finite-difference version of the transverse terms as described in the previous Appendix \S\ref{sec:transverse}.
Again, this ultimately led us to implement a finite volume version of characteristic boundaries.
Despite that failure of the finite difference implementation, we have found that the simple link between the characteristic and vertical derivatives is a useful illustrative concept, and thus have included it here.

\section{Details of the simulations}\label{sec:initialconditions}
\subsection{Description of the spheromak initial condition}\label{sec:spheromak}
We define the spheromak following \citet{Rosenbluth:1979}.
In normalized units, the magnetic field $\vect{B}$ is a linear force-free field satisfying
\begin{equation}
  \vect{j} = \kappa\vect{B},
\end{equation}
where $\vect{j}=\del\times\vect{B}$ is the current density and $\kappa$ is a scalar constant.
Taking the curl of the current and using $\del\cdot\vect{B}=0$ we find
\begin{gather}
  \del\times\vect{j} = -\del^2\vect{B} = \del\times(\kappa\vect{B}) = \kappa\del\times\vect{B} = \kappa^2\vect{B},
  \intertext{so that}
  \del^2\vect{B}+\kappa^2\vect{B} = \vect{0}\label{eq:vectHelmholtz}.
\end{gather}
The magnetic field therefore satisfies the vector Helmholtz equation.
The magnetic field $\vect{B}$ can be expressed in terms of the vector potential $\vect{A}$ through $\vect{B} = \curl\vect{A}$.
We use a vector potential written in terms of a scalar field $\psi$,
\begin{gather}
  \vect{A} = \vect{r}\times\frac{\vect{\del}\psi}{\kappa} - \vect{r}\psi,
\end{gather}
so that the resulting magnetic field is
\begin{equation}
  \vect{B} = \vect{\del}\times\vect{A} = \vect{\del}\times\Bigl(\vect{r}\times\frac{\vect{\del}\psi}{\kappa}\Bigr) - \vect{r}\times\vect{\del}\psi.\label{eq:spheromakB}
\end{equation}
The scalar field $\psi$ solves the scalar Helmholtz equation in spherical coordinates \citep[][\S13.1, p.1766]{Morse:1953b},
\begin{gather}
  \del^2\psi + \kappa^2\psi = 0\label{eq:scalarHelmholtz},
\end{gather}
and has solutions
\begin{equation}
  \psi_m^n = j_m(\kappa r)P_m^n(\cos\theta)e^{in\phi},\label{eq:psisoln}
\end{equation}
where $j_m(\kappa r)$ are the spherical Bessel functions of the first kind \citep[\S10.1.1]{Abramowitz:1964} and $P_m^n(\cos\theta)$ are the associated Legendre functions of the first kind \citep[\S8.1.1]{Abramowitz:1964}.
Here, the spherical coordinates $(r,\theta,\phi)$ are defined by $r^2=x^2+y^2+z^2$, the polar angle $\theta = \arctan\frac{y^2+z^2}{x}$, and the azimuthal angle $\phi=\arctan\frac{z}{y}$.
Thus, $x=r\cos\theta,\ y = r\sin\theta\cos\phi$, and $z=r\sin\theta\sin\phi$.
These differ from the more typical spherical coordinates, which set $\zhat$ as the axis of symmetry, by the cyclic permutation $(x\rightarrow y,y\rightarrow z,z\rightarrow x)$.
This choice orients the spheromak so that expansion in $\zhat$ mimics some properties of emerging active regions.
These solutions for the vector Helmholtz equation can be verified by direct substitution of Equations \eqref{eq:psisoln} and \eqref{eq:scalarHelmholtz} into \eqref{eq:spheromakB} and \eqref{eq:vectHelmholtz}.

Our simulation is initialized using the ground state axisymmetric spheromak, defined with $m=1,\ n=0$:
\begin{align}
  \psi_1^0 &= B_0j_1(\kappa r)P_1^0(\cos\theta)\\
  & = B_0\Bigl(\frac{\sin \kappa r}{\kappa^2r^2}-\frac{\cos \kappa r}{\kappa r}\Bigr)\sqrt{\frac{3}{4\pi}}\cos\theta.
\end{align}
$B_0$ is a scalar coefficient that gives the amplitude of the spheromak's magnetic field.
The vector potential is thus
\begin{align}
  \vect{A}_1^0 & = -B_0r\Bigl(\frac{\sin \kappa r}{\kappa^2r^2}-\frac{\cos \kappa r}{\kappa r}\Bigr)\sqrt{\frac{3}{4\pi}}\cos\theta\rhat - \frac{B_0}{\kappa}\Bigl(\frac{\sin \kappa r}{\kappa^2r^2}-\frac{\cos \kappa r}{\kappa r}\Bigr)\sqrt{\frac{3}{4\pi}}\sin\theta\phihat.
\end{align}
Transformed to the Cartesian coordinates used in the \lare{} MHD code, (i.e., $\vect{A}\rightarrow\{A_x,A_y,A_z\}$) the vector potential is
\begin{align}
  \vect{A}_1^0(x,y,z) = B_0\sqrt{\frac{3}{4\pi}}\Bigl(\frac{\cos \kappa r}{\kappa^2r^2}-\frac{\sin \kappa r}{\kappa^3r^3}\Bigr)\bigl\{\kappa x^2,\kappa xy-z,\kappa xz+y\bigr\}.\label{eq:spheromakAcart}
\end{align}

The spheromak solution contains a flux surface at the first zero of the Bessel function $j_1(\kappa r)$, where $\kappa r \approx 4.4934$, which divides the volume into isolated flux domains.
The radial component of the magnetic field is zero at this surface, and so it is convenient to set $\vect{B}=0$ outside this surface because one can formally do so without generating monopoles in the solution.
We therefore set $\kappa\approx4.4934$ so that the first zero of the Bessel function occurs at $r=1$, and then set $\vect{A}= \vect{0}$ outside this surface in our initial condition. 
This choice of boundary conditions for the spheromak solution creates a closed magnetic subvolume surrounded by a field-free plasma with a tangential discontinuity (current sheet) across the $r=1$ flux boundary.
Hereafter we refer to the magnetized portion of the plasma, initially at $r\leq 1$ as ``the spheromak,'' though technically the classical spheromak described by \citet{Rosenbluth:1979} has a vacuum magnetic field that extends to infinity and a magnetized plasma confined within the first zero of the Bessel function.

Numerically, each component of the vector potential is defined at the corresponding edges of a computational cell (e.g., $A_x$ is defined on the for $x$-oriented edges in a cell).
The magnetic field is initialized using the finite--difference version of $\vect{B}=\del\times\vect{A}$, which gives the magnetic field magnitude in the direction normal to each cell face.
\lare{}'s finite difference scheme then ensures that the solenoidal condition $\del\cdot\vect{B}=0$ is satisfied in the initial condition to numerical roundoff, and \lare{}'s flux-constrained-transport scheme ensures that the initial distribution of $\del\cdot\vect{B}$ is constant in time to numerical roundoff, i.e., at roughly $\bigo(10^{-14})$.
However, \lare{}'s Cartesian geometry means that the above precision does not hold at the $r=1$ spherical edge of the spheromak because we set the vector potential to zero outside of this surface.
To remedy the issue we introduce a smoothing function $f(r;e)$ that leaves the spheromak solution unchanged for $r < 1-e$ and has continuous zeroth and first derivatives for $1-e < r < 1+e$:
\begin{gather}
  f(r;e) = \begin{cases}
  1 & , r \le 1-e\\
  \frac{(1+e-r)^2(-1+2e + r)}{4 e^3} & , 1-e < r < 1+e\\
  0 &, 1+e \le r
  \end{cases}\label{eq:smoothing}
\end{gather}
We then let $\vect{A} \rightarrow f(r)\vect{A}$.
The resulting magnetic field is modified near the spheromak boundary to be $\vect{B}=\del\times\vect{A} = f\del\times\vect{A} + \del f\times\vect{A}$.
The effect of the smoothing is to add a small perturbation to the \thetahat{} component of $\vect{B}$ near the $r=1$ boundary of the spheromak, and drop both $\vect{A}$ and $\vect{B}$ smoothly to $\vect{0}$ for $r>1+e$.
We have found that a width of $e=3$ simulation cells is sufficient to ensure $\nabla\cdot\vect{B}=0$ to numerical roundoff everywhere, and again, \lare{}'s flux-constrained-transport scheme maintains this level of precision for the remainder of every simulation.

The initial magnetic configuration is visualized in the top left panel of \figref{fig:sideview} by tracing field lines initiated at two sets of seed points.
The first set of seed points are located near the toroidal axis of the spheromak ($x\approx 0,\ \sqrt{y^2+z^2}\approx 0.6$), and the second set near its symmetry axis ($\abs{x}<1,\ \sqrt{y^2+z^2}\approx 0$). 
The field lines near the toroidal axis loop dominantly in the $y-z$ plane while orbiting that axis at a constant minor radius; these form the toroidal core of the spheromak. 
Field lines initiated near the symmetry axis of the spheromak (i.e., along the $x$ axis) instead form loops primarily in planes perpendicular to the axis of the toridal core, closing along the spheromak's outer $r=1$) surface.
Intermediate field lines vary smoothly between these two configurations.
The later times in Figure~\ref{fig:sideview} also show field lines traced from a third set of seed points that have $z=1.25$ (just above the driving plane at $z_b=1.09$; see next section) and are initiated within a circle of radius $1$ centered on $x=y=0$.  
To ease visual comparison between different simulations, the top-down perspective of \figref{fig:topview} only shows this third set of field lines, and we chose the z=1.25 plane for initialization as opposed to the driving plane so that one field line passing through $(x,y)=(0,0)$ would be visible in \figref{fig:topview}.
The location of the seed points are constant in time, so the visualized field lines at different times highlight diverse regions of the spheromak (a more involved calculation, with the seed points advected with the fluid flow, could also be performed, but we have not undertaken that complication here).

Because we set the field outside the $r=1$ flux surface to zero, our spheromak is initially unbalanced and will expand.
However, we found that expansion to be rather minimal even for relative low background pressures.
Therefore, in order to achieve substantial expansion, we embed the spheromak in an initially non-uniform gas pressure according to
\begin{equation}
    P(x,y,z) = P_{\rm 0} + P_{\rm 1}\left(\frac{B^2}{1+r}\right),\label{eq:pressureperturbation}
\end{equation}
with $P_{\rm 0}$ the background gas pressure and $P_{\rm 1}$ perturbation amplitude of the increased gas pressure internal to the spheromak.
While any gas pressure increase from the outside to the inside of the spheromak will cause some expansion, the additional dependence on magnetic field strength and radial distance from the center of the spheromak encourages a gradual expansion (i.e., keeping $\abs{\vect{v}}\lesssim a$, the fastest propagation speed in the surrounding field-free plasma).
The resulting ground truth simulation then produces a spheromak that expands from the initial condition to a new, larger configuration that is close to equilibrium.

\subsection{Simulation setup, initial condition, and overview}\label{sec:normalization}
For the \emph{GT} simulation we set $N_x = N_y = 256$ and $N_z = 192$ cells. 
The dimensionless domain lengths are $L_x = L_y = 8.0$ and $L_z = 6.0$ and the simulation extends to $\pm4.0$ in the $x$ and $y$ directions, and has $-2.0\leq z\leq4.0$ in the $z$ direction, with all domain boundaries defined at cell faces.
The boundary conditions on the \emph{GT} simulation are set to be periodic in $x$, $y$, and $z$.
The spheromak is centered on the origin, has a radius of $1.0$, and has its symmetry axis in the \xhat{} direction.

The normalization coefficients for the MHD equations are given in SI units as $\lnorm = 1\unit{m}$, $\rnorm = 1\unit{kg}\unit{m}^{-3}$, and $\bnorm = \sqrt{4\pi\times 10^{-7}}T$, i.e., the latter is scaled by the square root of the amplitude of the permeability of free space in SI units.
As described in the text following Equation~\ref{eq:mhd}, all other normalization constants can be derived from these three.
For instance, the normalizing velocity is $\vnorm=1\unit{m}\unit{s}^{-1}$.
Then, in normalized units, the initial condition has a uniform density of $\rho_0=0.1$, a background pressure of $P_{\rm 0} = 6.667\times10^{-3}$ and perturbation amplitude $P_{\rm 1}=1.0$ in \eqref{eq:pressureperturbation}, and a spheromak magnetic field amplitude of $B_0 = 2.0$ in \eqref{eq:spheromakAcart}.
These initial conditions make the total Alfv\'en crossing time across the initial spheromak about $t=2.25$.

The \emph{Data-Driven} simulations span $-2.0\leq x,y\leq 2.0$ and $1.125<z<4.0$ (again corresponding to cell faces).  
The resolution is the same as the \emph{GT} simulation, so both the \emph{Driving} and \emph{Data-Driven} simulations have $N_x = N_y = 128$ and the latter has $N_z = 92$ cells (recall the \emph{Driving} simulation has $N_z=4$, always). 
From the \emph{GT} simulation we extract the cell-center averaged data from the $z=z_b=1.109375$ plane (i.e., the $100$th $z-$cell of the \emph{GT} simulation) at every Courant step, and use this as input to the \emph{Driving} (and therefore \emph{Data-Driven}) simulations.

\section{Comparison to potential and nonlinear force free  extrapolations}
\begin{figure}
    \centering    \includegraphics[width = \textwidth]{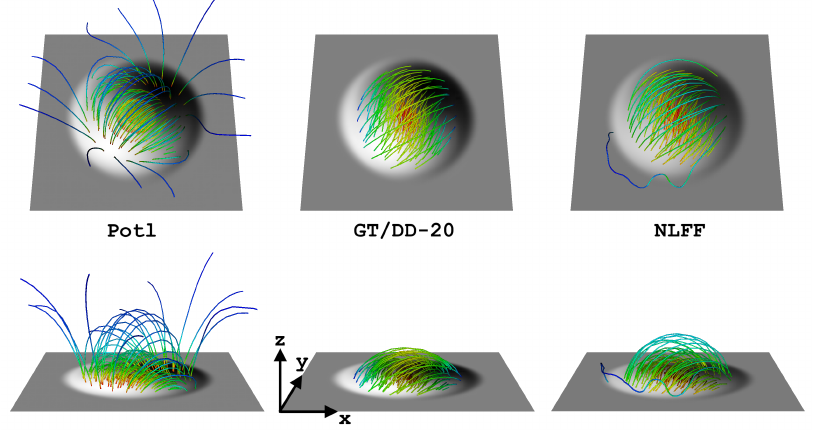}
    \caption{Comparison between a potential field extrapolation (left column), the result of data driving using a driving cadence of $\dtdata=20\dtsim$ (center column), and a nonlinear force free field extrapolation using the optimization code provided by Thomas Wiegelmann (right).  The top and bottom rows show two different perspectives.  The extrapolation time $t=1.55t_N$ is at courant step 380, near the end of the spheromak's expansion.}
    \label{fig:compare_potential_dd_nlff}
\end{figure}

\begin{figure}
    \centering    \includegraphics[width = \textwidth]{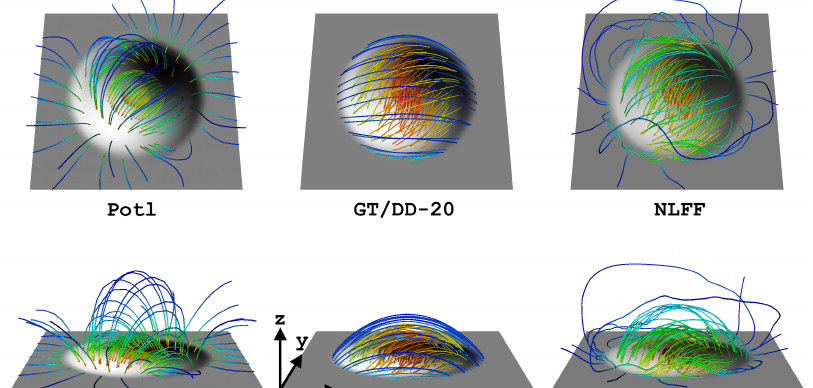}
    \caption{Same as \figref{fig:compare_potential_dd_nlff}, but including field lines initiated near the edge of the emerged portion of the spheromak.}
    \label{fig:compare_potential_dd_nlff2}
\end{figure}
Any single-time extrapolation of the magnetic field will leave out important dynamic information that, in general, will have a major effect on the state of the corona.
Indeed, that is one major motivation to develop data-driven boundary conditions.
Nonetheless, static extrapolations are the simplest possible reconstructions of the coronal magnetic field that can be made, require the fewest computational resources, and are therefore the most often employed method to model the coronal field and interpret coronal observations.  
For that reason we include a comparison of two such extrapolation methods to our MHD based \emph{Data-Driven} solution.
Note that both extrapolations use the limit in which electromagnetic forces are dominant and material forces are negligible, and hence do not provide any information about the plasma quantities.
We therefore must neglect those quantities in the following discussion.

In Figures~\ref{fig:compare_potential_dd_nlff} and \ref{fig:compare_potential_dd_nlff2} we compare a potential field extrapolation (left column) and a nonlinear force free field extrapolation (right column) to the magnetic field resulting from the \emph{Data-Driven} simulation using $\dtdata=20\dtsim$ (center).
The two Figures have the same perspectives but highlight field lines closer to the center of the spheromak (Figure \ref{fig:compare_potential_dd_nlff}) and further towards its edge (Figure \ref{fig:compare_potential_dd_nlff2}).
Recall from the discussion of Figures \ref{fig:sideview} and \ref{fig:topview} that the magnetic field produced by the \emph{Data-Driven} simulation with 20\dtsim{} is nearly indistinguishable from the \emph{GT} field, so we have not included the latter here.
The seed points of the field lines are the same for all panels, so differences between the left or right columns relative to the center column represent errors in the extrapolations.

Both the potential field and the nonlinear force free field were generated using the optimization code provided by Thomas Wiegelmann, and described in \citet{Wiegelmann:2004}, \citet{Wiegelmann:2010}, and \citet{Wiegelmann:2012}.
The potential field extrapolation uses the normal field component $B_z$ in the driving layer as a bottom boundary condition, while the optimization-based nonlinear force free field extrapolation also incorporates the horizontal fields $B_x$ and $B_y$ in the boundary layer.
We did not attempt an exhaustive search of code parameters in order to find the best fit between the field in the \emph{GT} simulation and the results of the extrapolations.
This is how extrapolation codes must often be used in practice when comparison to a ground truth is not possible, and side boundaries can typically only be modeled as periodic, perfectly conducting, or perhaps some asymptotic approach.

This potential field was computed using a version of FFT approach of \citet{Alissandrakis:1981} included in the optimization code\footnote{We do note that a simple Green's function method, also included in Wiegelmann's optimization code, produced similar results. Indeed, \citet{Alissandrakis:1981} notes the equivalence of the approaches when the boundary conditions are properly specified)}.
As such, the boundary conditions assume periodicity in the horizontal directions and exponential decay in the vertical direction.
This produces a field that does not close locally, even though the region is in fact flux balanced and magnetically isolated: the \emph{GT} solution contains a sheath current sheet inside the domain that fully shields the field.
Instead, erroneous magnetic connections outside the domain can be seen in the left hand columns of the figures where field lines initiated near the edge of spheromak in the potential field extrapolation run into the boundary of the computational box.
In the \emph{GT} simulation (or the nearly indistinguishable \emph{Data-Driven} field shown here), the field never reaches the sides of the computational box in any direction, but instead is confined to a small subregion of the full numerical domain.
This is because, dynamically, the spheromak expands into a field free region and includes tangential current sheet bounding the edge of the magnetic spheromak.
At the end of the \emph{GT} simulation, the magnetic field embedded in the blast wave has expanded and the plasma density and pressure have developed gradients such that the system reaches a roughly equilibrium state.
Neither of the extrapolations, which are fundamentally static solutions, can incorporate knowledge of this dynamic evolution.

We therefore see that the potential field volume-filling, as required by the assumption that no currents exist with the domain (which would be required to shield the sources of the field that is imposed at the lower boundary).
The NLFF field also fills the volume, but qualitatively less so in the following sense: the magnitude of $B$ drops off much more rapidly in the NLFF field extrapolation compared to the potential field solution for the given boundary conditions in which the field amplitude goes to zero as $z\rightarrow\infty$. 
This is apparent in the more confined loops for the NLFF field in the right columns of the Figures, especially for those field lines initiated near the edge of the spheromak and shown in the right columns of Figure \ref{fig:compare_potential_dd_nlff2}.
In this sense, the NLFF extrapolation does produce a qualitatively better approximation to the true solution compared to the potential extrapolation, at least in this case \citep[see, however, ][for a discussion of applicability of NLFF field extrapolations to the solar corona in general and the metrics by which they may be assessed]{DeRosa:2009}.
This is, perhaps, expected when the true solution contains currents throughout the solution volume, which are allowed in a NLFF field, whereas a potential field is current-free in the solution volume by definition.

The reality is that the final state of the \emph{GT} simulation and/or \emph{Data-Driven} simulation is neither current free nor force free: while the spheromak in the initial condition is a \emph{linear} force free field, it evolves away from that configuration due to being embedded in the blast wave, and the final state includes non-zero Lorentz forces that are approximately balanced by pressure gradient forces.
This fact severely compromises the NLFF extrapolation's ability to reproduce the ground truth field \citep{Peter:2015}.
The NLFF extrapolation cannot represent this type of force-balance, which partially explains why the NLFF extrapolation produces a greater expansion of field compared to the \emph{GT} solution (or $\dtdata=20\dtsim{}$ \emph{Data-Driven} solution shown in the figures).
Another aspect is that both the potential and the NLFF field extrapolations assume a static configuration and therefore have no notion of the speed at which real information can physically propagate through a system, information that includes the propagation of the shielding current sheet the defines the boundary between the expanding spheromak and the surrounding plasma.
Our method of boundary driving allows the \emph{Data-Driven} simulation to self-consistently reproduce the shielding current sheet in the volume away from the boundary, outside of which the solution remains field-free.
This holds even in the case where $\dtdata = 100\dtsim$, far exceeding the critical driving cadence of $\dtdata\approx40\dtsim$, and when other aspects of the driven solution are severely compromised.
Instead, the NLFF extrapolation produces a field that is substantially more ``puffed up'' than the true solution, and always contains erroneous magnetic connectivity to the external universe.
This demonstrates one way in which the dynamical constraint imposed by the observed temporal evolution of the lower boundary, coupled to a true dynamical model of the full system's evolution, is key to allowing the \emph{Data-Driven} simulation to produce a more faithful representation of the field of the expanded spheromak, including the outer edge of the emerged field's region of influence.
\end{document}